\documentclass[useAMS,usenatbib]{mn2e}
\usepackage{epsfig}
\usepackage{graphicx}
\usepackage{times}
\usepackage{color}

\def\cm3{cm$^{-3}$}
\def\kms{km~s$^{-1}$}
\def\msunyr{M$_{\odot}$\,yr$^{-1}$}
\def\lsun{L$_{\odot}$}
\def\rsun{R$_{\odot}$}

\def\msun{M$_{\odot}$}

\def\one{{\,\sc i}}
\def\two{{\,\sc ii}}

\def\beq{\begin{equation}}
\def\eeq{\end{equation}}

\def\lesssim{\mathrel{\hbox{\rlap{\hbox{\lower4pt\hbox{$\sim$}}}\hbox{$<$}}}}
\def\gtrsim{\mathrel{\hbox{\rlap{\hbox{\lower4pt\hbox{$\sim$}}}\hbox{$>$}}}}

\def\aj{AJ}
\def\pasp{PASP}
\def\apj{ApJ}

\def\apjl{ApJL}
\def\aap{A\&A}
\def\araa{ARA\&A}
\def\aapr{A\&ARv}

\def\mnras{MNRAS}
\def\nat{Nature}

\def\physrep{Phys.~Rep.}
\def\gray{$\gamma$-ray}
\def\grays{$\gamma$-rays}
\def\isoni{$^{56}{\rm Ni}$}
\def\isoco{$^{56}{\rm Co}$}

\title[Radiative transfer of SN IIb/Ib/Ic ejecta]{
Core-collapse explosions of Wolf-Rayet stars and the connection to type IIb/Ib/Ic supernovae}
\author[Luc Dessart et al.]{\vspace{0.3cm} Luc Dessart,$^1$\thanks{email: Luc.Dessart@oamp.fr} D. John Hillier,$^{2}$ Eli Livne,$^3$
Sung-Chul Yoon,$^{4,5}$ Stan Woosley,$^5$ \\
\vspace{0.2cm}
{\LARGE \rm Roni Waldman,$^{3}$ and Norbert Langer$^4$ }\\
  $^1$: Laboratoire d'Astrophysique de Marseille, Universit\'e de Provence, CNRS,
            38 rue Fr\'ed\'eric Joliot-Curie, F-13388 Marseille Cedex 13, France \\
  $^2$: Department of Physics and Astronomy, University of Pittsburgh, USA \\
  $^3$: Racah Institute of Physics, The Hebrew University, Jerusalem, Israel \\
  $^4$: Argelander Institute for Astronomy, University of Bonn, Auf dem H\"{u}gel 71, D-53121, Germany\\
  $^5$: Department of Astronomy and Astrophysics, University of California, Santa Cruz, CA 95064, USA}

\date{Accepted . Received }

\pagerange{\pageref{firstpage}--\pageref{lastpage}} \pubyear{2010}

\begin{document}

\maketitle

\label{firstpage}

\begin{abstract}

We present non-LTE time-dependent radiative-transfer simulations of supernova (SN) IIb/Ib/Ic spectra
and light curves, based on $\sim$\,10$^{51}$\,erg piston-driven ejecta,
with and without \isoni, produced from single and binary
Wolf-Rayet (W-R) stars evolved at solar and sub-solar metallicities.
Our bolometric light curves show a 10-day long post-breakout plateau with a luminosity of 1-5$\times$10$^7$\,\lsun,
visually brighter by $\gtrsim$10\,mag than the progenitor W-R star.
In our \isoni-rich models, with $\sim$3\,\msun\ ejecta masses, this plateau precedes a 20 to 30-day long
re-brightening phase initiated by the outward-diffusing heat wave powered by radioactive decay at depth.
A larger ejecta mass or a  deeper \isoni\ location increases the heat diffusion time and acts to both delay and
broaden the light-curve peak. Discriminating between the two effects requires spectroscopic modelling.
In low ejecta-mass models with moderate mixing, \gray\ leakage
starts as early as $\sim$50\,d after explosion and causes the nebular luminosity to steeply decline by $\sim$0.02\,mag/d.
Such signatures, which are observed in standard SNe IIb/Ib/Ic, are consistent with low-mass progenitors derived from
a binary-star population. We propose that the majority of stars with an initial mass $\lesssim$20\,\msun\ yield SNe II-P
if ``effectively" single, SNe IIb/Ib/Ic if part of a close binary system, and SN-less black holes if more massive.
Our ejecta, with outer hydrogen mass fractions as low as $\gtrsim$0.01 and a total hydrogen mass of
$\gtrsim$0.001\,\msun,  yield the characteristic SN IIb spectral morphology at early times.
However at later times,  $\sim$15\,d after  the explosion, only H$\alpha$ may remain as a weak absorption feature.
Our binary models, characterised by helium surface mass fractions of $\gtrsim$0.85, systematically show He\one\ lines
during the post-breakout plateau, irrespective of the \isoni\ abundance.
Synthetic spectra show a strong sensitivity to metallicity, which offers the possibility to constrain it directly from SN
spectroscopic modelling.
\end{abstract}


\section{Introduction}

  Unlike numerous Type II SNe with a progenitor identification on pre-explosion images (see \citealt{leonard_10} for a
  recent review), SNe IIb/Ib/Ic can currently be constrained only through the analysis of their light.  SNe IIb/Ib/Ic events, understood as
  successful explosions following  the core collapse of a H-poor/H-deficient progenitor star, represent a number of
  challenges for modern astrophysics. The traditional perspective was that, showing no hydrogen-line features in their spectra, their
  progenitor massive stars had lost their hydrogen envelope prior to explosion, in the form of a radiatively-driven wind \citep{cak}.
  During the star's evolution, mass loss progressively peels off the massive-star envelope leaving only a residual hydrogen shell.
  Eventually this is also peeled away leaving a
  shell whose composition is dominated by helium and nitrogen (WN star; \citealt{crowther_etal_95}).
  As evolution and peeling continue, the surface then becomes helium and carbon dominated (yielding a WC star;
  \citealt{DCH00_WC_neon,crowther_etal_02}),
  and then carbon and oxygen dominated (yielding a WO star; \citealt{conti_76,maeder_meynet_94,kingsburgh_etal_95}).
   Owing to the stiff dependence of mass loss on luminosity/mass, low-mass massive stars lose little mass and die as Type II SNe.  Consequently SNe IIb/Ib/Ic would have to come from higher-mass ($M\lesssim$\,30\,\msun) massive stars.

   From the above  simplistic argument, one can picture a sequence of increasing main-sequence mass as we go through progenitors of
   SNe II-P, II-L, IIb, Ib, and Ic \citep{heger_etal_03,crowther_07,georgy_etal_09}.  Recent revisions downwards of massive
   star steady-state mass-loss rates may challenge this scenario \citep{bouret_etal_05}. Although still quite speculative at present,
   transient, and perhaps recurrent, phases of intense mass loss (e.g., of the eruptive kind seen in $\eta$ Car) may
   be a viable alternative to steady-state mass loss (see,
   for example, \citealt{langer_etal_94,maeder_meynet_00,shaviv_00,owocki_etal_04,guzik_05,DLW10a}).
   A second issue is the viability of the explosion mechanism
   for stars of increasing mass, since their higher-mass cores are increasingly more bound, exposing the newly-formed
   SN shock to a tremendous accretion rate \citep{burrows_etal_07a}.
   Finally, we need to understand what differentiates the progenitors of SNe IIb, Ib and Ic, as well as the progenitors
   of Type Ic hypernovae which might be associated with a \gray-burst signal (for a discussion,
   see e.g.,\citealt{woosley_bloom_06,fryer_etal_07}).

   Using radiation-hydrodynamics simulations, \citet{EW88} were the first to perform a comprehensive study
   of hydrogen-less massive-star cores, representative of hydrogen-deficient W-R stars. Based on the available
   observations of SNe Ib at that time, they concluded that ejecta from such W-R progenitor models could reproduce
   the general SN-Ib light curve morphology if powered by decay energy from $\sim$0.1\,\msun\ of \isoni.
   However, the generally narrow peak of their light curves supported only 4-7\,\msun\ ejecta, corresponding to 15--25\,\msun\ progenitor
   main-sequence masses. Additional SN-Ib light curves  acquired since the study of  \citet{EW88} show a similar behaviour,
   and thus strengthen the notion that the bulk of SNe Ib are associated with relatively
   low-mass ejecta \citep{richardson_etal_02,richardson_etal_06,drout_etal_10}.

   Since the scenario invoking higher-mass massive stars appears not to be viable,
   binary-star evolution channel is favored for the production of SNe Ib/c, as well as
   SNe IIb \citep{utrobin_94,woosley_etal_94,young_etal_95, blinnikov_etal_98,fryer_etal_07}.
   A similar conclusion emerges from independent considerations based on the observed SN Ib/c rate \citep{smith_etal_10}.

   In addition to single-star evolution models of W-R stars \citep{langer_etal_94,WHW02}, we now have physically-consistent
   predictions for the binary-star counterpart (\citealt{yoon_etal_10}; see also preliminary explorations
   of \citealt{woosley_etal_95} and \citealt{nomoto_etal_95}).
   The critical ingredient of these scenarios is that
   a star may lose mass through mass transfer to a companion. Consequently, the mass-loss/luminosity scaling that inhibits
   mass loss in low-mass massive stars does not apply, allowing much lower mass progenitors to explode
   as ``hydrogen-less" cores.

   The best observations and tailored analyses exist for only a few objects, usually the brightest or
   the weirdest, and thus not representative of the SN IIb/Ib/Ic population.
   To cite a few, this includes the Type IIb SN 1993J, associated with a small ejecta mass resulting from the explosion
   of a rather low-mass progenitor star in a binary system \citep{WEW94_SN1993A,young_etal_95,blinnikov_etal_98}, with a
   residual hydrogen envelope \citep{swartz_etal_93b,baron_etal_95}. The Type Ic SN 1994I was also
   the focus of numerous studies, suggesting a low-mass ejecta, potentially non-deficient in helium
   despite its classification \citep{wheeler_etal_94,millard_etal_99,sauer_etal_06}.
   One exception is SN 2008D, which appears as a very standard Type Ib SN but with a shock breakout
   detection, hence a very well-defined time of explosion. In this object, a short post-breakout plateau
   is observed before the SN re-brightens merely 4\,d after explosion and peaks 1\,mag brighter
   about 20\,d later \citep{soderberg_etal_08,chevalier_fransson_08,modjaz_etal_09}.

   The helium abundance is thought to distinguish SNe Ib and Ic, although the He\one-line signatures
   needed to make this assessment can be influenced by numerous complications.  Due to the
   high excitation energy needed, He\one-lines may not be excited, and thus
   helium could be present but without associated spectral signatures.
   In addition, the presence or absence of He\one\ lines is conditioned not just by the composition
   but also by non-thermal excitation/ionization processes born out of \gray-emission from radioactive decay
   of \isoni\ and \isoco\ isotopes \citep{lucy_91,swartz_91,KF92,swartz_etal_93a,swartz_etal_95,KF98a,KF98b}.
     He\one\ lines may therefore be affected by the efficiency of outward mixing
   of \isoni. Unfortunately, the mixing characteristics are adjusted for convenience rather than computed from first principles.
   Further, light-curve and spectral calculations are performed with separate codes that make distinct
   approximations and thus global physical consistency is lacking in current modelling of SNe IIb/Ib/Ic.

   Many SNe Ib may have traces of hydrogen at their surface \citep{deng_etal_00,branch_etal_02,parrent_etal_07},
   and this may even apply to SNe Ic \citep{jeffery_etal_91,branch_etal_06}. While an interesting possibility,
   such inferences are always jeopardised by the weakness of the H\one\ features at the heart of the debate (which
   suffer from line overlap with C\two\ and Si\two\ lines; see, e.g., \citealt{ketchum_etal_08})
   and the shortcomings of the radiative-transfer approach used (e.g., LTE versus non-LTE, time-dependent effects
   on the ionisation, ejecta structure and composition). As emphasised here (see also \citealt{james_baron_10}), obtaining
   early-time spectra and light curves of SNe IIb/Ib/Ic is the only way to settle this and the He abundance issue.

   Here, we present non-LTE time-dependent calculations of SN ejecta based on piston-driven explosions
   of WN and WC/WO progenitor stars.\footnote{W-R, WN, WC, and WO are spectroscopic designations. Here, for convenience,
   we use the same designations for stars with the typical surface composition of a WN star etc. As some of the progenitors have
  very low final masses and large radii, but relatively low mass-loss rates, emission lines in their spectra may
  be unusually weak.}
   These SNe IIb/Ib/Ic calculations have a higher level of consistency than earlier calculations --- they combine
   stellar-evolutionary models of the progenitors, hydrodynamics calculations of the explosion (albeit
   artificial), and non-LTE time-dependent radiative transfer of the full ejecta yielding simultaneously multi-color
   light curves and spectra.
   We discuss the gas and radiative properties of such SN ejecta and what determines their
   classification as SN IIb, Ib, or Ic.  Our simulations
   ignore non-thermal electrons and do not include any mixing induced by the explosion,\footnote{In practice,
   the sharp \isoni-distribution left behind by the explosion is smoothed in our models by numerical diffusion
   as we remap the ejecta at the start of each step in our time sequences.
   This introduces a very moderate mixing, typically over a velocity-width of $\sim$500\,\kms.}
   and thus set a lower threshold on the expected strength of H{\sc i} and He{\sc i} lines.
   More importantly, the calculations show that under certain conditions, which
   we detail,  H{\sc i} and He{\sc i} lines can be seen even in the absence of any unstable nuclei and high-energy electrons.

   In Section~\ref{sect_setup}, we give a brief overview of the models selected from the comprehensive study
   of \citet{yoon_etal_10}, including how we proceeded from such progenitors to make SN ejecta.
   As one aspect of
   this work is to understand how the progenitor properties effect the H and He lines observed in SN spectra, we
   discuss the difference in progenitor properties.
    We then present our results, discussing simultaneously the properties of the gas (ejecta) and of the emergent
   radiation (spectra and light curves). The advantage  of our approach is that it {\it simultaneously} yields
   synthetic light curves and spectra, allowing a direct assessment of spectra for a given light-curve phase.
   Our synthetic bolometric light curves and how they are affected by variations in photospheric
   conditions are described in Section~\ref{sect_lbol}.
   We then present our synthetic spectra, addressing in turn the potential signatures of hydrogen (Section~\ref{sect_h}),
   helium (Section~\ref{sect_he}), CNO elements (Section~\ref{sect_cno}), intermediate-mass elements
   (IMEs; Section~\ref{sect_ime}), and iron-group elements  (Section~\ref{sect_z}). A crucial question
   which we address is when will the resulting SN be classified as Type Ib, or Ic, or IIb.
   Section~\ref{sect_disc} is devoted to a discussion of our results, which we confront to observations. This study is
   not a review so we merely select a few well-observed cases to illustrate our findings.
   In Section~\ref{sect_concl}, we present our conclusions and future goals.

\begin{table*}
\begin{center}
\caption[]{Progenitor-model properties, including global characteristics and surface mass fractions of important species.
We include the progenitor-envelope binding energy outside of 1.4 (top three models) and 2.0\,\msun\ (the rest of the
models). Numbers in parenthesis refer to powers of ten. Ages generally refer to the evolutionary time since the main sequence.
For model  Smi60mf7p08z1, it is instead the time since the onset of core-helium burning. [See text for discussion.]
\label{tab_presn}}
\begin{tabular}{l@{\hspace{1.6mm}}c@{\hspace{1.6mm}}c@{\hspace{1.6mm}}c@{\hspace{1.6mm}}
c@{\hspace{1.6mm}}c@{\hspace{1.6mm}}c@{\hspace{1.6mm}}c@{\hspace{1.6mm}}c@{\hspace{1.6mm}}c@{\hspace{1.6mm}}
c@{\hspace{1.6mm}}c@{\hspace{1.6mm}}c@{\hspace{1.6mm}}c@{\hspace{1.6mm}}c@{\hspace{1.6mm}}c@{\hspace{1.6mm}}}
\hline
 Model    &  $M_{\rm i}$  & $Z$                 & $M_{\rm f}$ &  $R_{\ast}$ & Age  & $E_{\rm b}$  & $X_{\rm H}$    &$X_{\rm He}$ & $ X_{\rm C}$ &$ X_{\rm N}$ & $X_{\rm O}$& $X_{\rm Ne}$& $X_{\rm Mg}$ & $X_{\rm Si}$  & $X_{\rm Fe}$    \\
                &   [\msun]                   & [$Z_{\odot}$] & [\msun]        &  [\rsun]          & Myr  &            [B]  & \multicolumn{9}{c}{}  \\
\hline
Bmi18mf3p79z1                &     18.0     &  1.0       &   3.79    & 10.0    & 9.8  & 0.20   &  4.27(-2) &     9.38(-1) &     2.02(-4) &     1.33(-2) &  3.76(-4) & 1.85(-3)   &    7.25(-4)  &     7.34(-4) &     1.36(-3) \\
Bmi18mf4p41z1                &     18.0     &  1.0       &   4.41    & 12.3    & 9.5  & 0.32   &  1.25(-1) &     8.55(-1) &     1.48(-4) &     1.33(-2) &  4.53(-4) & 1.85(-3)   &    7.25(-4)  &     7.34(-4) &     1.36(-3) \\
Bmi25mf5p09z1                &     25.0     &  1.0       &   5.09    & 4.35    & 7.5  & 0.61   &  6.33(-5) &     9.81(-1) &     2.92(-4) &     1.33(-2) &  3.21(-4) & 1.85(-3)   &    7.23(-4)  &     7.34(-4) &     1.36(-3) \\
Bmi25mf6p49z1                &    25.0    &  1.0         &   6.49    & 3.64    & 6.8  & 0.58   &   0.0         & 9.81(-1)     &   4.23(-4)   &  1.31(-2)    & 3.11(-4)  &  1.69(-3)  &   6.68(-4)    &    7.37(-4) &   1.36(-3)  \\
Smi60mf7p08z1                &    60.0    &  1.0        &   7.08    & 0.45     & 0.5     & 0.86   &   0.0         & 1.30(-1)     &   4.77(-1)   &       0.0       & 3.45(-1)  &  3.42(-2)  &   1.07(-3)    &    8.25(-4) &   1.36(-3)  \\
Bmi25mf7p3z0p2             &    25.0    &  0.2        &   7.30     & 2.10      & 7.3  & 0.46   &   0.0        & 9.99(-1)      &   6.01(-5)   &  2.6(-3)      & 5.66(-5)  &  3.23(-4)  &   1.23(-4)    &    1.49(-4) &   2.72(-4)   \\
Smi25mf18p3z0p05        &    25.0    &  0.05      &  18.3     & 1.14      &10.0 & 1.49  &   0.0         & 5.34(-1)     &   1.72(-1)    &  2.96(-3)   & 2.17(-1)  &  5.82(-2)  &   1.57(-2)    &    4.61(-5) &   6.80(-5)  \\
\hline
\end{tabular}
\end{center}
\end{table*}

\begin{table*}
\begin{center}
\caption[]{Same as Table~\ref{tab_presn}, but now showing the explosion/ejecta properties computed by {\sc kepler} and {\sc v1d},
including the total ejecta yields for important species. The last column gives the velocity of the ejecta shell that bounds 99\% of
the total \isoni/\isoco\ mass. Because of the sharp composition boundaries in such unmixed 1D models, these values are altered
by up to 10\% when remapping onto the lower-resolution  grid in {\sc cmfgen}.
[See Table~\ref{tab_presn} and the text for additional details.]}
\label{tab_comp}
\begin{tabular}{l@{\hspace{1.6mm}}c@{\hspace{1.6mm}}c@{\hspace{1.6mm}}c@{\hspace{1.6mm}}c@{\hspace{1.6mm}}c@{\hspace{1.6mm}}c@{\hspace{1.6mm}}c@{\hspace{1.6mm}}
c@{\hspace{1.6mm}}c@{\hspace{1.6mm}}c@{\hspace{1.6mm}}c@{\hspace{1.6mm}}c@{\hspace{1.6mm}}c@{\hspace{1.6mm}}c@{}}
\hline
Model   &    $M_{\rm remnant}$  &    $M_{\rm ejecta}$ & $E_{\rm kin}$ & $M_{\rm H}$ &$M_{\rm He}$ & $M_{\rm C}$  & $M_{\rm N}$  & $M_{\rm O}$ & $M_{\rm Ne}$ & $M_{\rm Mg}$& $M_{\rm Si}$& $M_{\rm Fe}$ & $M_{^{56}{\rm Ni}}$  & $V_{^{56}{\rm Ni}}$   \\
                  &    [\msun]                 &    [\msun]       &   [B]           &   [\msun] &   [\msun]   &    [\msun]    &   [\msun]  &   [\msun]  &   [\msun] &    [\msun]   &[\msun]  & [\msun]    & [\msun]   & \kms\ \\
\hline
Bmi18mf3p79z1               & 1.40 &  2.39 & 1.2 & 1.94(-3) &   1.50(0) &   1.32(-1)     &     1.19(-2) &     3.22(-1) &     9.24(-2) &     4.68(-2) &     7.94(-2) &     8.94(-3) &     1.84(-1)  & 2750 \\
Bmi18mf4p41z1               & 1.50 &  2.91 & 1.2 & 6.70(-3) &   1.64(0) &   1.63(-1)     &     8.83(-3) &     5.36(-1) &     1.47(-1) &     6.99(-2) &     1.05(-1) &     7.91(-3) &     1.70(-1)  & 2510 \\
Bmi25mf5p09z1               & 1.48 &  3.61 & 1.2 & 1.39(-6) &   1.58(0) &   4.31(-1)     &     8.56(-3) &     1.15(0)  &     7.60(-2) &     7.30(-2) &     7.01(-2) &     8.85(-3) &     2.37(-1)  & 1250 \\
Bmi25mf6p49z1              & 2.54 &  3.95  & 1.0 &  0.0      &   1.61(0)  &   3.54(-1)     &    8.08(-3) &   1.53(0) &   7.64(-1)  &    1.84(-1) &      3.36(-2) &   5.92(-3) &  0.0  & 0.0 \\
Smi60mf7p08z1              & 2.88 & 4.20   & 1.0 & 0.0      &   2.25(-1) &   1.19(0)      &    0.0      &   2.65(0) &   7.79(-1)  &    2.10(-1) &      2.24(-2) &   7.21(-3) &  0.0  & 0.0 \\
Bmi25mf7p3z0p2            & 2.33 &  4.97 &  1.0 &  0.0 &   2.07(0)  &   3.98(-1)     &    1.21(-3) &   1.78(0) &   9.25(-1)  &    1.22(-1) &      4.19(-3) &   1.40(-3) &  0.0  &   0.0 \\
Smi25mf18p3z0p05       & 2.51 & 15.79 & 1.0 &  0.0      &   1.32(0)  &   2.03(0)      &    1.45(-3) &   9.11(0) &   2.83(0)   &    4.31(-1) &      5.87(-1) &   1.08(-3) &  0.0  & 0.0 \\
\hline
\end{tabular}
\end{center}
\end{table*}

\section{Pre-SN evolution, explosion, and model setup}
\label{sect_setup}

The radiative-transfer calculations are based on SN ejecta that are
produced in two steps. First, we simulate the evolution of a massive star from
the main sequence until an advanced stage (either the end of neon-core burning or the
formation of a degenerate iron core). Second,  we use radiation hydrodynamics to simulate
the explosion by driving a piston at the base of the SN progenitor envelope.

For the stellar-evolution model inputs, we use a sample of simulations presented by \citet{yoon_etal_10},
who focused on binary-star evolution for the production of W-R stars at solar and 1/5th-solar metallicity
and who used  various primary/secondary masses and a range of orbital periods (typically on the order of a few days).
In their study, the initial (final) mass of the primary was in the range
12--60\,\msun\ (1.4-7.3\,\msun), the final mass range reflecting the complicated mass-loss and mass-transfer history.
These calculations were performed with an optimized nuclear network that included all elements up to $A=30$, and iron.
In a forthcoming study, a larger network containing all elements up to iron will be included.
From this sample, we select a few W-R star models with a final mass in the range 3.79 to 7.3\,\msun, which we name here
Bmi18mf3p79z1, Bmi18mf4p41z1,  Bmi25mf5p09z1, Bmi18mf6p49z1, and Bmi18mf7p3z0p2.

The adopted model nomenclature specifies the evolution channel (B/S for binary-/single-star evolution),
the main-sequence mass (which corresponds to X, given in \msun, in miX; when present, ``p'' introduces the decimals),
the mass at the end of the simulation (which corresponds to X, given in \msun, in mfX; this value
should be very close to that at the time of core collapse in all cases),
the environmental metallicity (which corresponds to X, given in units of the solar value, in zX).
Thus Bmi18mf3p79z1 is a binary model for an 18\,\msun\ main-sequence mass, with a final mass of 3.79\,\msun,
and evolved at solar metallicity.
We also complement this subset with two single-star models.
These include Smi60mf7p08z1, a 20\,\msun\ helium-star model (60\,\msun\ star on the main sequence) evolved at solar
metallicity, and Smi25mf18p3z0p05, a 25\,\msun\ model which evolved chemically homogeneously at a
metallicity of 0.05\,Z$_{\odot}$ \citep{yoon_etal_06}.

For historical reasons, this set of simulations is split into two distinct samples (see below).
Models Bmi18mf3p79z1, Bmi18mf4p41z1, Bmi25mf5p09z1 were mapped
as single stars and evolved with {\sc kepler} \citep{weaver_etal_78} until the formation
of an iron core, before being exploded by means of a piston (using a $\sim$1.4\,\msun\ mass cut)
to yield an asymptotic ejecta kinetic energy of 1.2\,B (where 1\,B = 1\,Bethe = $10^{51}$\,erg).
The explosions produced 0.171, 0.182, and 0.195\,\msun\ of $^{56}$Ni respectively, while
the velocity of the ejecta shell that bounds 99\% of the total \isoni/\isoco\ mass is 2750,
2510, and 1250\,\kms. A more detailed presentation covering the post-helium burning evolution
and explosion of these and other similar models is deferred to Woosley, Kasen, \& Yoon (in preparation).

The 1D Lagrangian {\sc kepler} simulations employ a grid of 500-1000 points to cover the ejecta using
a mass grid, thus resolving and predicting sharp discontinuities in the composition distribution.
In contrast, our radiative-transfer simulations (performed with {\sc cmfgen}, see below),
employ an optical-depth grid which is more relevant for radiation transport
than a mass grid, but the drawback is that sharp composition boundaries cannot be resolved (at least without
special procedures).
As we step in time through our sequence, the remapping procedure on such un-mixed models causes
numerical diffusion of sharp composition discontinuities, which effectively acts as mixing.
We find that this smears the \isoni\ profile typically over 500\,\kms. This also leads to a modest change in
\isoni\ ejecta abundance compared to the {\sc kepler} values, which are now 0.184, 0.170, and 0.237\,\msun\
for models Bmi18mf3p79z1, Bmi18mf4p41z1, and Bmi25mf5p09z1 (these cumulative isotopic yields are preserved
at the 10\% level throughout the time sequence).

The second set of models was used without further evolution, remapped into V1D \citep{livne_93,DLW10a,DLW10b},
and exploded, as in {\sc kepler}, using a piston (with a 2.0\,\msun\ mass cut) to yield an asymptotic ejecta kinetic
energy of 1\,B.\footnote{This slight
difference of 0.2\,B in ejecta-kinetic energy between these two sets of models has only a modest influence on our results.
The difference in piston-mass-cut is of no relevance since we primarily focus on the early-time evolution which is sensitive only to the outer ejecta.
The global properties of our progenitor models (e.g., M$_{\ast}$, R$_{\ast}$, etc.) would not be changed significantly
if they had all been evolved until the formation of a degenerate iron core, since the relevant nuclear timescales are shorter
than, e.g., the Kelvin-Helmholtz timescale.}
Unstable-nuclei abundances were deliberately set to zero in this second set (we do not include any explosive nucleosynthesis
for those V1D simulations). Since these models were not evolved until iron-core collapse,
the composition of the inner 2-3\,\msun\ are not converged.
Our discussion of such \isoni-deficient models will thus be limited to early times, when the photosphere lies well outside this inner region
potentially polluted by \isoni\ in realistic explosions.
This distinction between our two samples also serves to reveal the effect of decay heating on the
light curves and spectra, which can only affect models Bmi18mf3p79z1, Bmi18mf4p41z1, and Bmi25mf5p09z1.
It also permits a discussion of SNe Ib light curves and spectra in which unstable nuclei are absent, which could
occur, for example, with complete fallback of the inner ejecta.

The splitting of the models into two groups primarily arose because
the code of Yoon/Langer does not follow stellar evolution until the formation of a degenerate
and collapsing iron core, and
because at the time {\sc v1d} could not compute explosive nucleosynthesis.
There is also a chronological reason for this split. The work presented in this paper includes computations
started in 2008 (i.e., models Bmi18mf3p79z1,  Bmi18mf4p41z1, and Bmi25mf5p09z1). Later,
we realised that for a discussion of the early-time evolution of SNe IIb/Ib/Ic, the actual core properties are
not directly relevant. We thus augmented
our sample with additional models from \citet{yoon_etal_06,yoon_etal_10}, evolved only until neon-core burning,
and exploded them with {\sc v1d} at this stage of evolution. Our choice of augmenting the sample was made
after we had computed the
light curve and spectra of models Bmi18mf3p79z1, Bmi18mf4p41z1, and Bmi25mf5p09z1. By including these
\isoni-deficient models, we cover a wider range of properties, including higher-mass binary models,
single-star models, solar- as well as sub-solar metallicity models, and a chemically-homogeneous model.

In Tables \ref{tab_presn} and \ref{tab_comp}, we summarise the properties of the progenitors
including the surface (outer ejecta) composition for the dominant species and the total ejecta yields,
in solar masses, for important species.
For each model we also quote the envelope binding energy which is found to be significantly
smaller in binary-star than in single-star
progenitors (see also \citealt{DLW10a}) -- this is a particularly important feature for the explosion mechanism
(see Section~\ref{sect_concl}).
The remnant mass $M_{\rm remnant}$ varies from small ($\gtrsim$1.4\,\msun) for the lower-mass models (evolved
until iron-core collapse and exploded with {\sc kepler}) up to 2.88\,\msun\ for the rest of the sample. The larger
remnant masses may result, in part, from
the flatter density distribution of these model envelopes, a possible artefact of the truncated evolution
which prevented the formation of a denser core surrounded by a steeper density decline. Until we know how these stars explode
the remnant-mass value we obtain may be inaccurate.
At the start of the {\sc cmfgen} simulations, at 1-2\,d after explosion, all these SN ejecta are in homologous expansion.

Since the companion star in our selected binary systems may influence the SN radiation associated with the
successful explosion of the primary star, we also give
the binary system and secondary star properties. Binary-star models  Bmi18mf3p79z1, Bmi18mf4p41z1,
Bmi25mf5p09z1, Bmi18mf6p49z1, and Bmi18mf7p3z0p2
are characterised at the end of neon-core burning (of the primary star) by a systemic orbital period (separation)
of 29.7\,d (120.8\,\rsun), 31.5\,d (125.6\,\rsun), 21.3\,d (110.0\,\rsun), 27.4\,d (127.9\,\rsun), and 13.0\,d (81.1\,\rsun).
In the same order, the corresponding surface radius (stellar mass) of the secondary star is 10.9\,\rsun\ (23.0\,\msun),
9.3\,\rsun\ (22.5\,\msun), 26.0\,\rsun\ (34.3\,\msun), 11.8\,\rsun\ (31.0\,\msun), and 11.3\,\rsun\ (35.4\,\msun).

\begin{figure*}
\epsfig{file=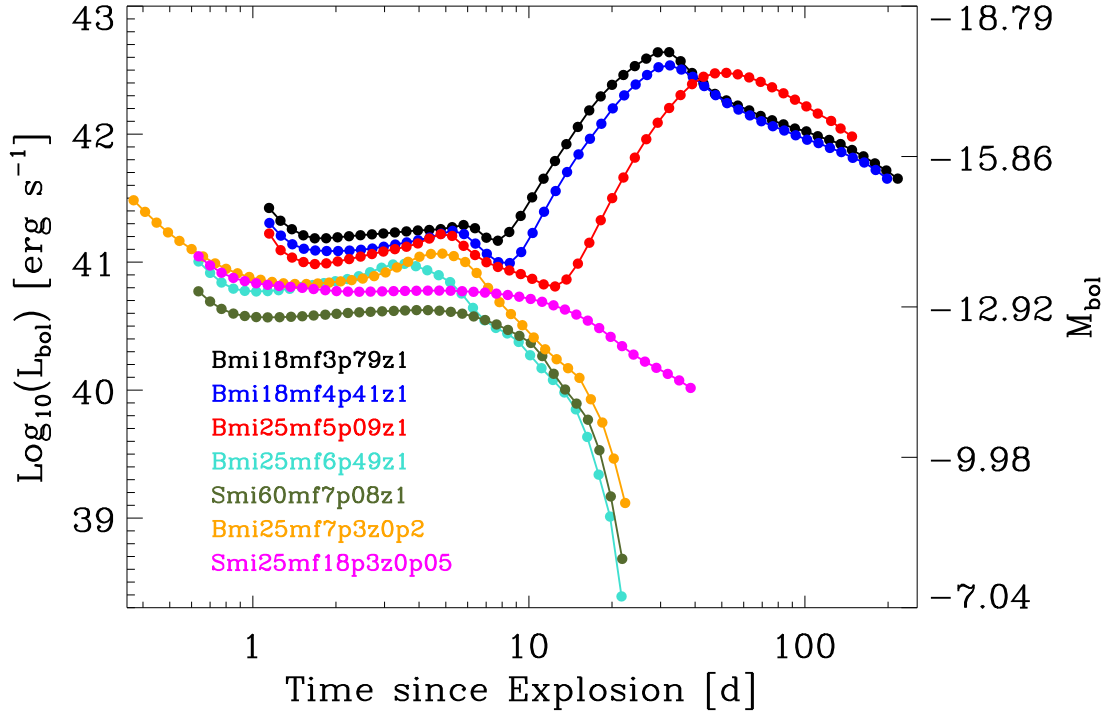,width=15cm}
\caption{Synthetic bolometric light curves computed with our non-LTE time-dependent {\sc cmfgen} simulations
based on binary- and single-star W-R progenitor models (see Tables\ref{tab_presn}-\ref{tab_comp}).
A colour coding is used to differentiate the models, and dots indicate the actual post-explosion times at which the computations are
performed (in a time sequence). Simulations including unstable nuclei (here exclusively produced through the decay of \isoni\ and
\isoco\ isotopes)
brighten about a weak after the onset of a post-breakout plateau, while all other simulations show a precipitous fading as the photosphere
recedes to the deeper helium-deficient ejecta layers of increasing mean-atomic weight and original binding energy.
For reference, we give the bolometric magnitude on the right-hand-side ordinate axis.
\label{fig_lbol_all}
 }
\end{figure*}

Binary-star progenitor models tend to have larger radii than single-star ones, especially if they
have retained a residual hydrogen envelope.
This concerns models Bmi18mf3p79z1 and Bmi18mf4p41z1, which both contain a few 0.001\,\msun\ of hydrogen, with a
typical surface mass fraction of 0.01--0.1.
Model Bmi25mf5p09z1also has some hydrogen at the surface, but with a very low mass fraction of 6.33$\times$10$^{-5}$
and a cumulative mass of only 1.39$\times$10$^{-6}$\,\msun.  After the explosion, such low-mass non-hydrogen
deficient layers travel at speeds $\gtrsim$\,15000\,\kms\ in this model, lie well above the photosphere past 1\,d, and
thus leave no spectral feature (Section~\ref{sect_h}).
The bulk of the envelope mass is lost through mass transfer, but the presence
of hydrogen at the pre-SN stage is subsequently conditioned  primarily by the stellar-wind mass-loss rate.
In practice, the pre-SN model surface hydrogen abundance is higher for lower stellar-wind mass loss rates, which can result
from the adopted mass-loss recipe, or from sub-solar metallicity (see \citealt{yoon_etal_10} for details).

All progenitor models, except Smi25mf18p3z0p05, have a rather small final mass, producing ejecta masses as low as
2.39\,\msun\ in model Bmi18mf3p79z1. The low metallicity single star model, Smi25mf18p3z0p05, produces an ejecta
mass of over 15\,\msun\ ejecta --- over 3 times that of any other model.  All models contain $\sim$1\,\msun\ of helium,
with a surface mass fraction that is close to unity in our binary-star models, but substantially lower
(0.13 and 0.54) in the two single star models.
The model ejecta have a similar chemical stratification (reflecting the analogous progression through H, He, C etc. core burning), but
they show quantitative difference in their envelope/ejecta yields.
Consequently, we identify distinct spectroscopic signatures associated with such abundance variations, which complement
those associated with H and He.

We evolve our hydrodynamical inputs using the non-LTE time-dependent
radiative-transfer code {\sc cmfgen} \citep{HM98_lb}. The approach and setup are analogous to that presented for the Type II-peculiar SN 1987A
\citep{DH10a} and for Type II-Plateau (II-P) SNe \citep{DH11}, and are thus not repeated here.
We assume the SN is free from external disturbances, such as interaction with the pre-SN wind or a companion star,
or irradiation from the newly-born neutron star. Importantly,
in \isoni-rich models, we assume a local deposition of radioactive-decay energy, treated as a pure heating source.
Our treatment of decay energy as a pure heating source
leads to an underestimate of the excitation and ionisation of the gas at the photosphere as soon \grays\ may
reach it \citep{lucy_91,swartz_91,KF92,swartz_etal_93a,KF98a,KF98b,swartz_etal_95}. A corollary is that as long
as there is no post-breakout re-brightening of our SN models, we are confident that the photosphere cannot be influenced
by \grays\ from radioactive decay, and thus our computed spectra are not compromised by this approximation.
Hence, we can discuss the early-phase spectroscopic evolution of the SN when our non-LTE time-dependent
approach is physically accurate, starting as early as $\lesssim$1\,d after shock breakout.

Our Monte-Carlo transport code predicts that \gray\ leakage causes $\sim$0.01\% (10\%) of the decay energy
to escape at $\sim$40\,d ($\sim$85\,d) after explosion in models  Bmi18mf3p79z1 and Bmi18mf4p41z1.
In model Bmi25mf5p09z1, this leakage causes $\sim$0.01\% (10\%) of the decay energy to escape at 75\,d (180\,d)
after explosion (Section~\ref{sect_lbol_ldecay}). Further, since LTE holds below the photosphere, the bolometric
luminosity is insensitive to how the \gray\ energy is downgraded. Thus, our synthetic bolometric light curves are strictly
accurate over at least two months after explosion.

\begin{figure*}
\epsfig{file=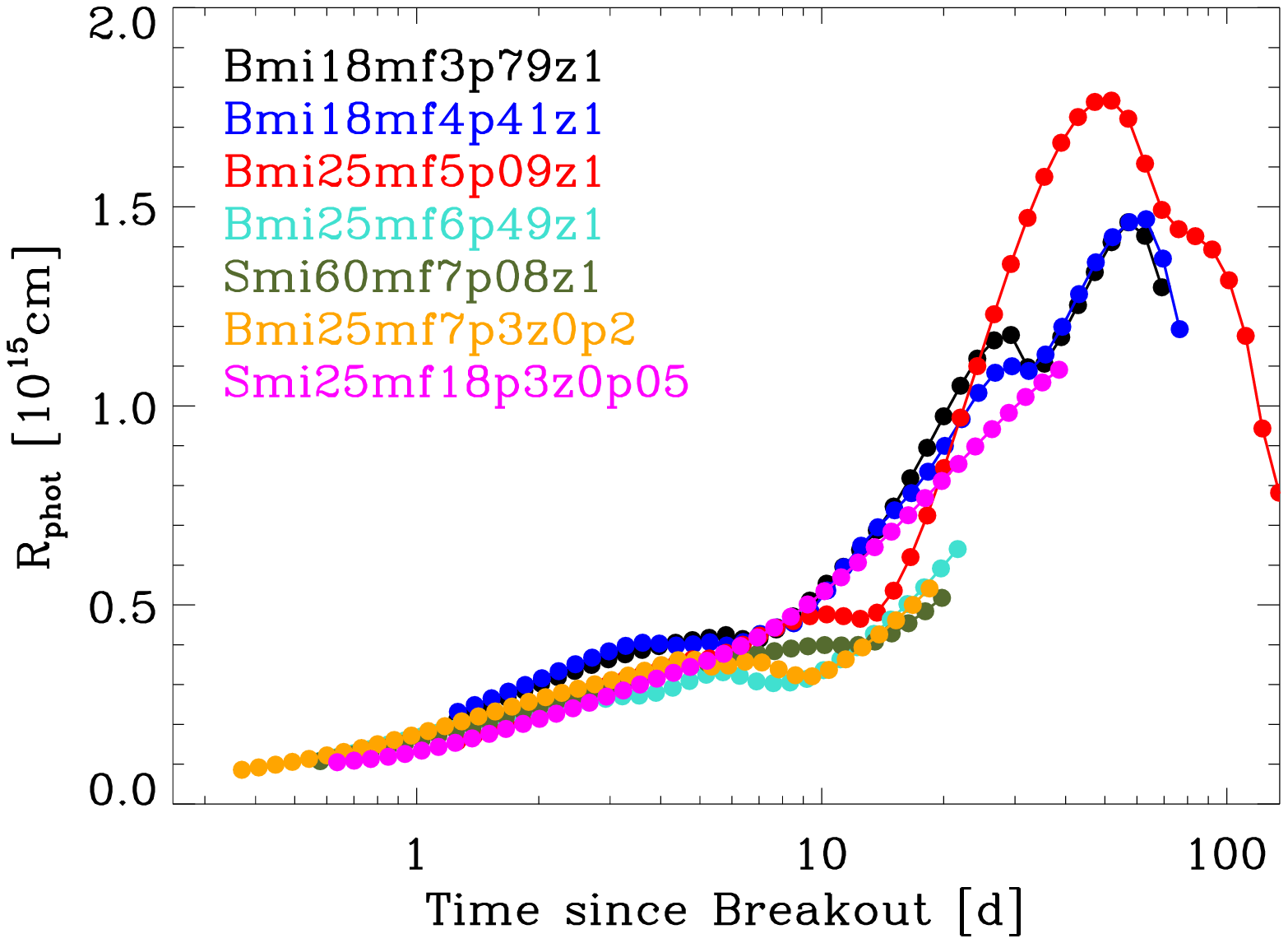,width=8.5cm}
\epsfig{file=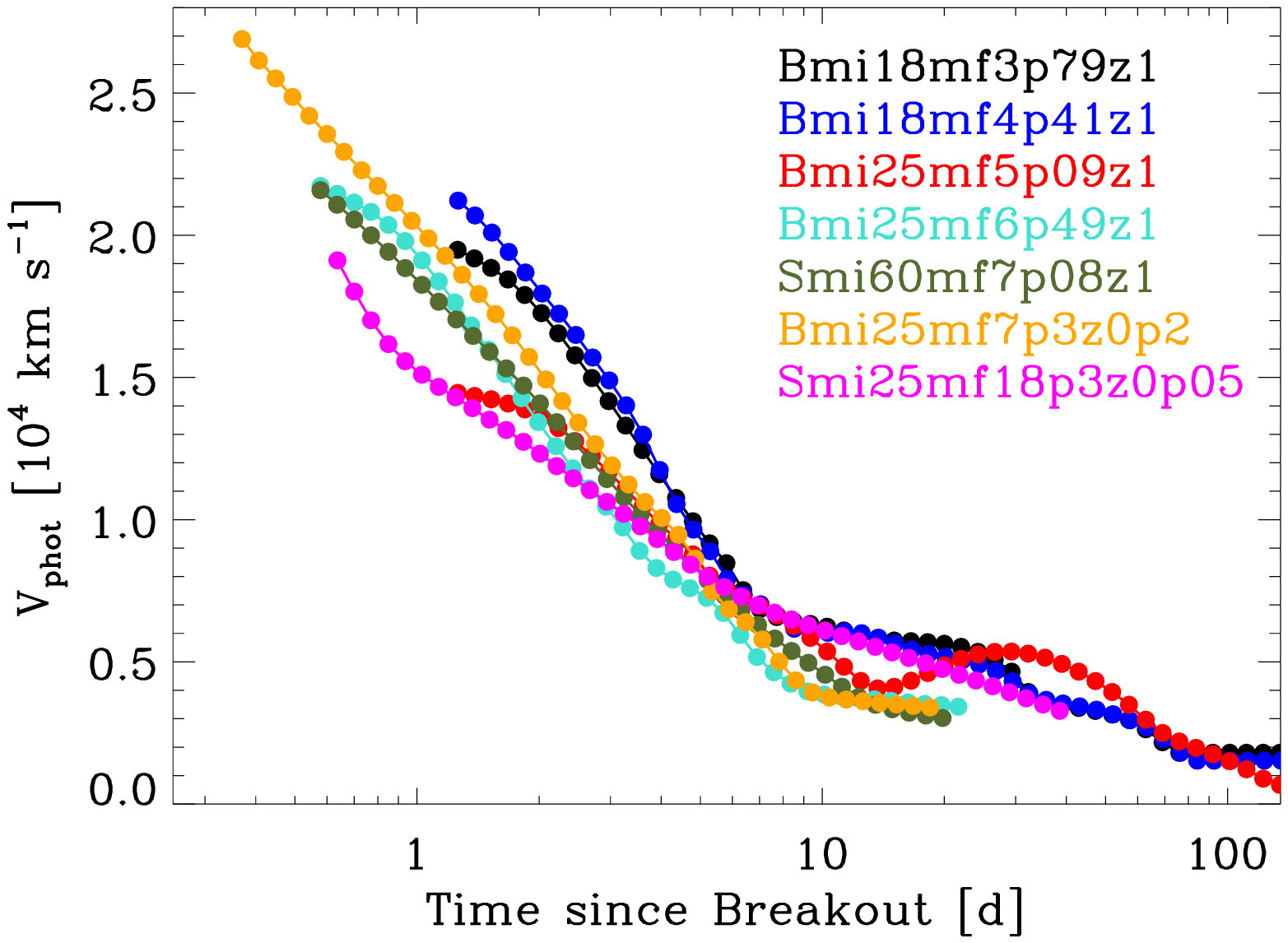,width=8.5cm}
\epsfig{file=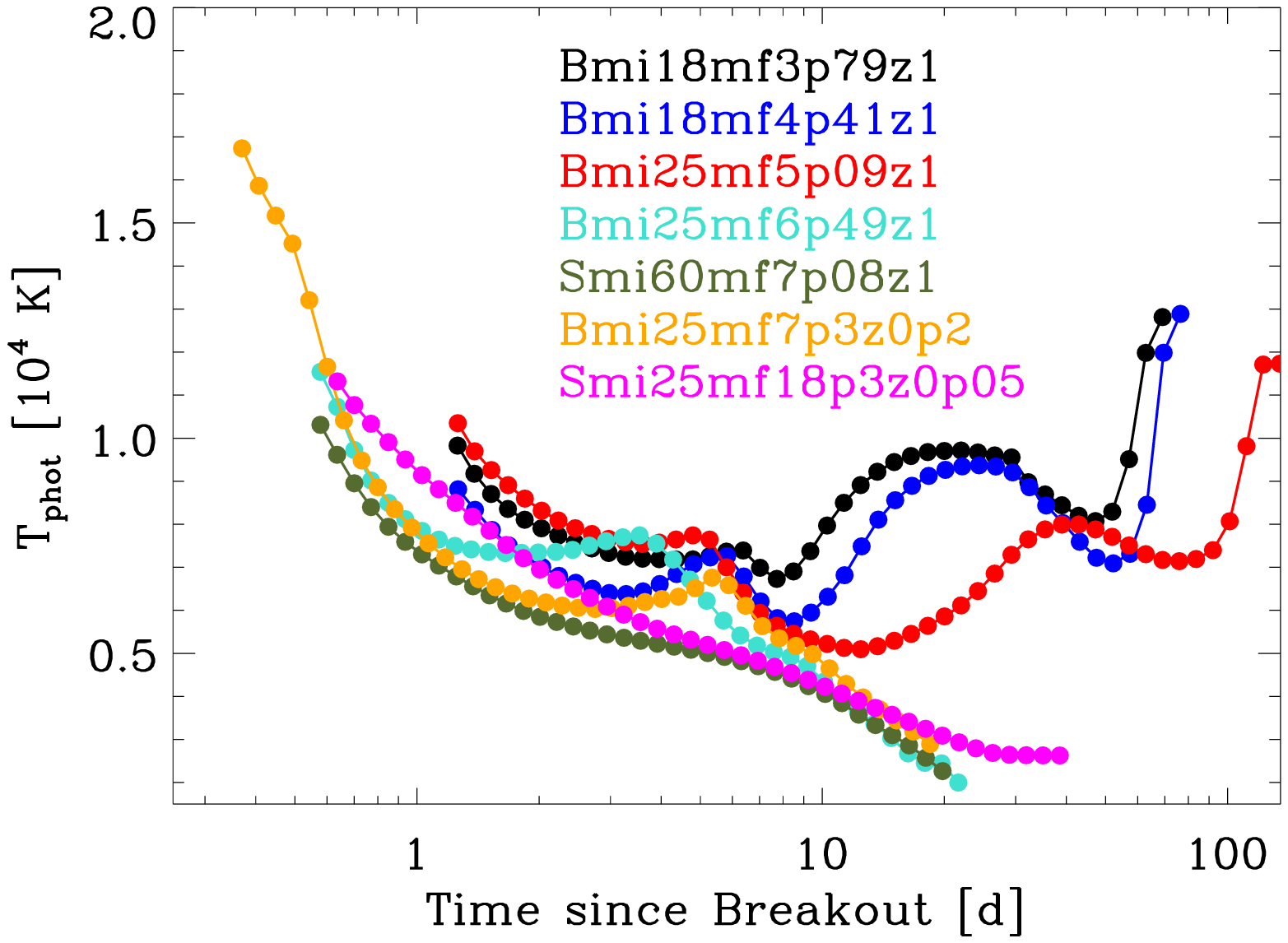,width=8.5cm}
\epsfig{file=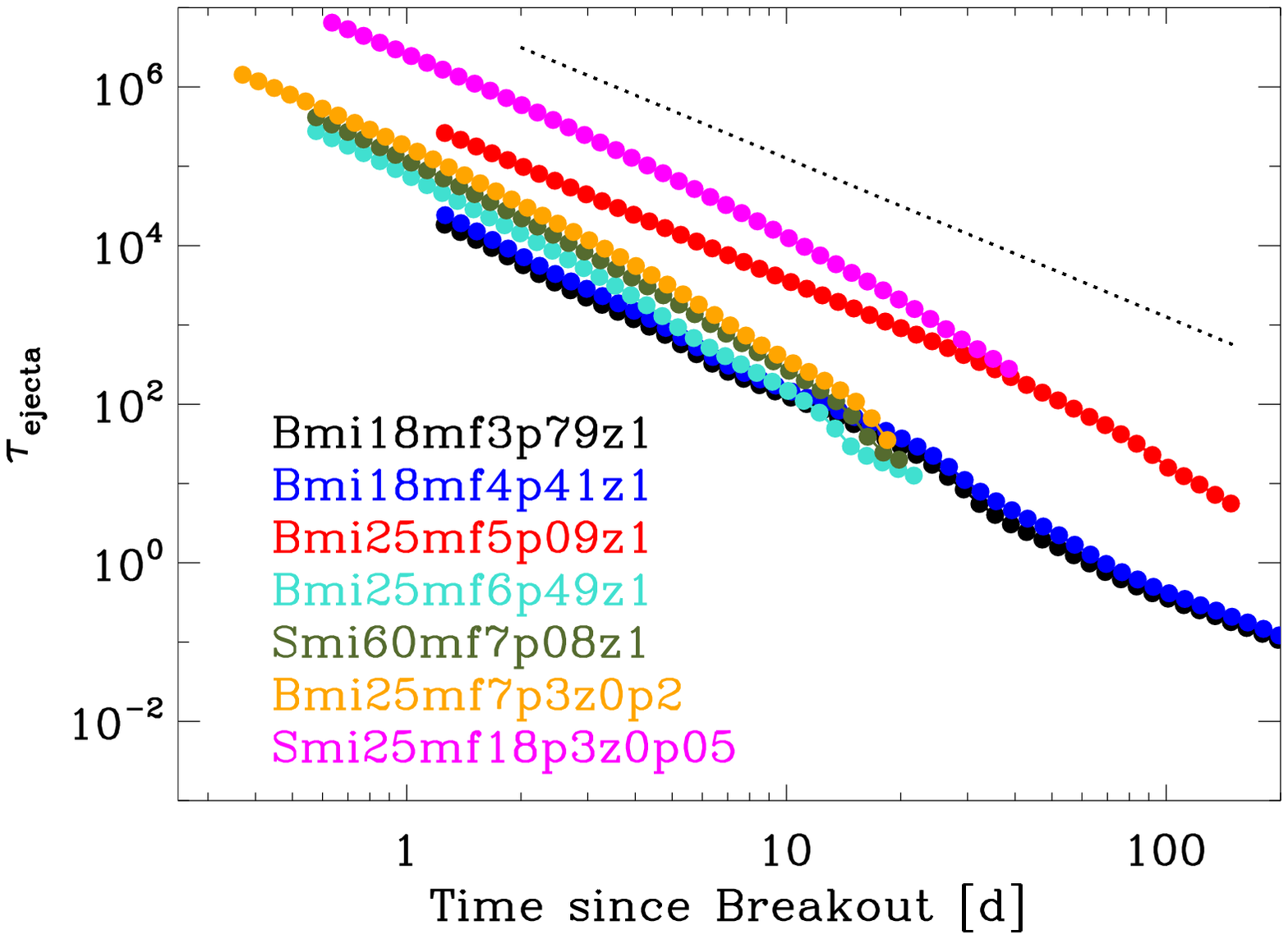,width=8.5cm}
\caption{Same as Fig.~\ref{fig_lbol_all}, but now showing the evolution of the
radius ($R_{\rm phot}$; top left), the velocity ($V_{\rm phot}$; top right), and the temperature
($T_{\rm phot}$; bottom left) at the photosphere. For reference, we also
include the evolution of the total ejecta electron-scattering optical depth
($\tau_{\rm ejecta}$; bottom right; the dotted black line indicates the
1/time$^2$ evolution expected under constant ionization conditions).
\label{fig_phot_all}}
\end{figure*}

\section{Bolometric light curves and evolution of photospheric conditions}
\label{sect_lbol}

   Starting our non-LTE time-dependent simulations at a post-breakout time of $\lesssim$1\,d,
   our synthetic bolometric luminosities initially decline from the peak value reached
   at shock breakout.
   The early post-breakout fading stems from energy losses associated with radiation and expansion (the latter being
   particularly significant for compact progenitor stars), causing a tremendous cooling of the ejecta.
   In our simulations this rapid bolometric fading is largely missed --- we catch mostly the end of it
   at 1--2\,d. The luminosity then levels off at a value of
   1--5$\times$10$^7$\,\lsun\ and initiates a short plateau (Fig.~\ref{fig_lbol_all}).
   This plateau extends in all models over a number of days before the SN either brightens or fades.

   Brightening is initiated in our \isoni-rich models Bmi18mf3p79z1, Bmi18mf4p41z1, and Bmi25mf5p09z1
   as soon as the photosphere ``feels" the influence of the heat wave generated from the decay
   of radioactive nuclei (primarily \isoni\ to \isoco) at greater depths.
   The re-brightening is delayed in model
   Bmi25mf5p09z1 because its ejecta mass is larger and the \isoni-rich layers are located deeper ($\lesssim$1250\,\kms) --
   both increase the diffusion time for the energy released at depth.
   In contrast, sample objects without \isoni\ eventually, and irrevocably, fade at $\sim$10\,d after exploding.
   This time is about the same as the time of re-brightening in the \isoni-rich models, and in some sense, is quite degenerate
   given the range of ejecta masses in our sample.
   The models that re-brighten reach a peak luminosity at about 30-50\,d after shock breakout, before fading to a
   luminosity corresponding to the energy decay rate of $^{56}$Co to $^{56}$Fe (recall that  we currently assume
   full-trapping of \grays\ in these {\sc cmfgen} calculations).

\begin{figure*}
\epsfig{file=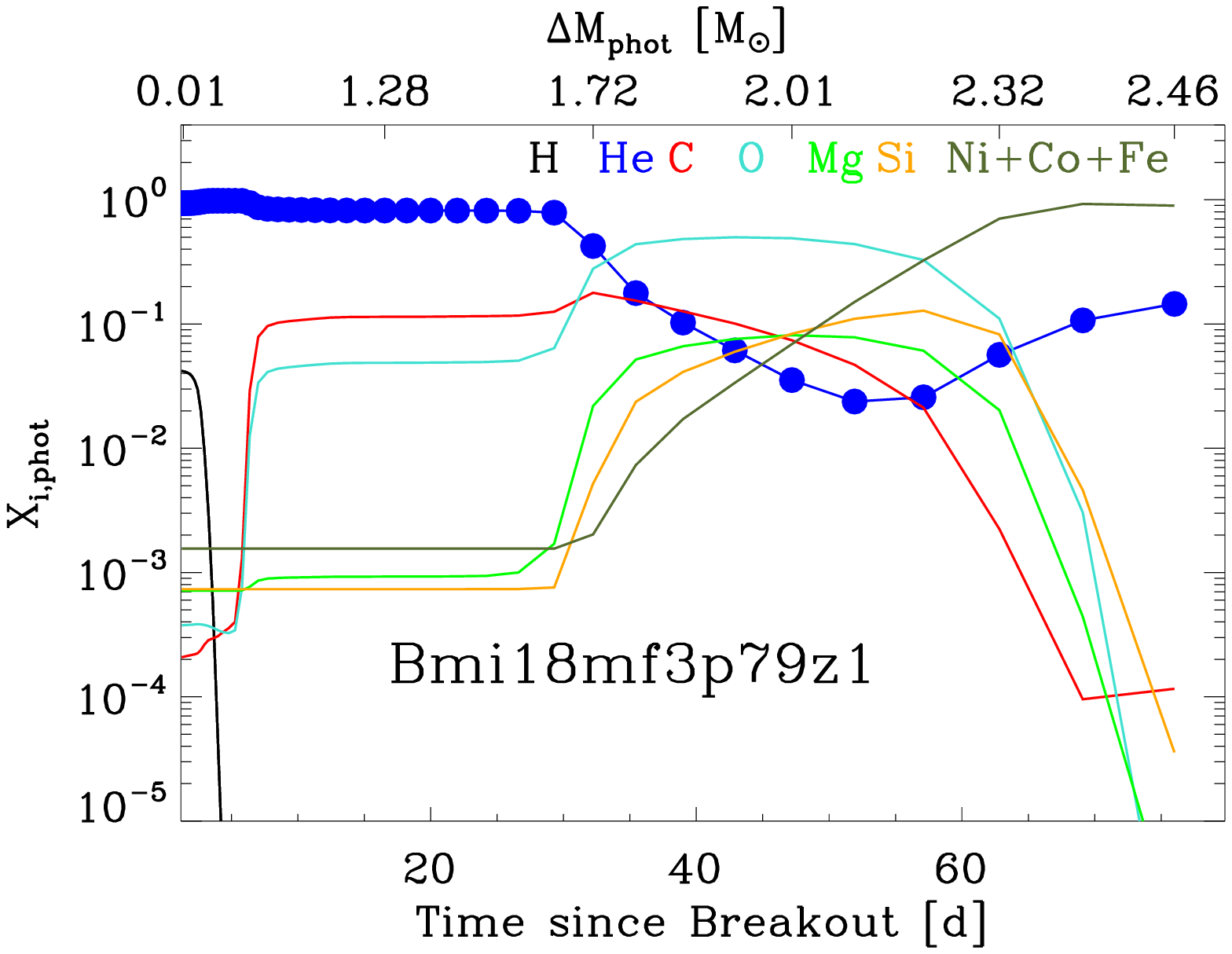,width=8cm}
\epsfig{file=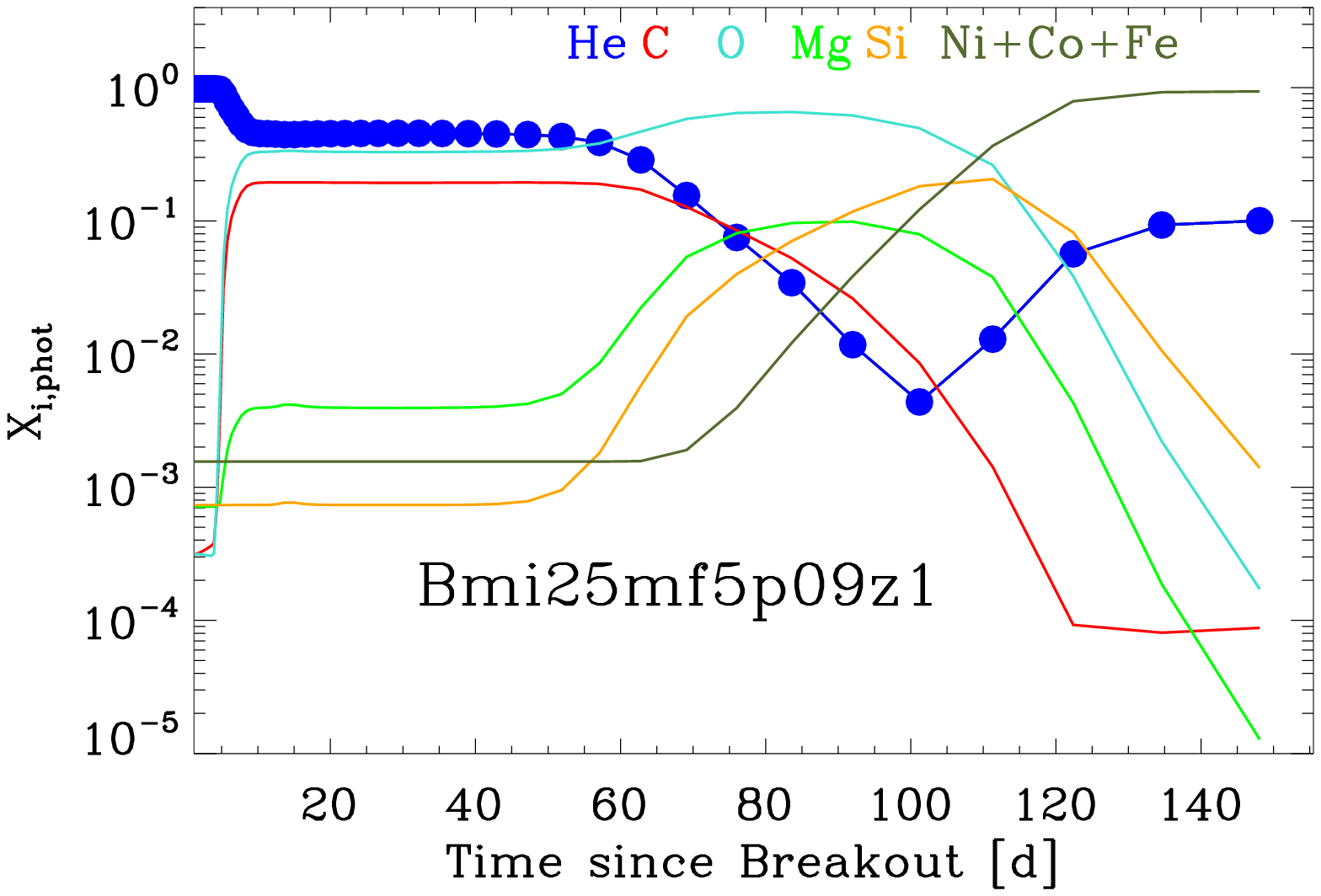,width=8cm}
\epsfig{file=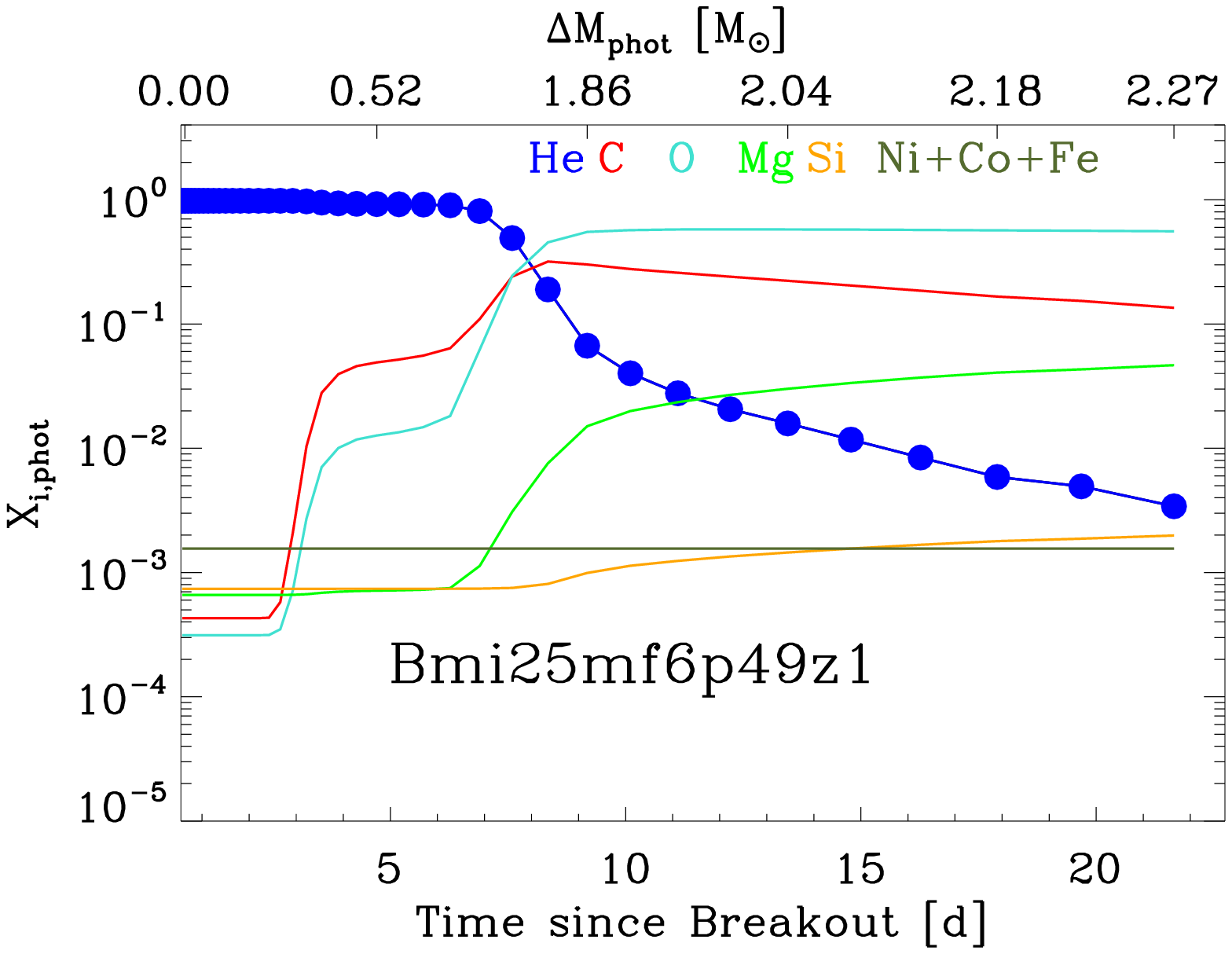,width=8cm}
\epsfig{file=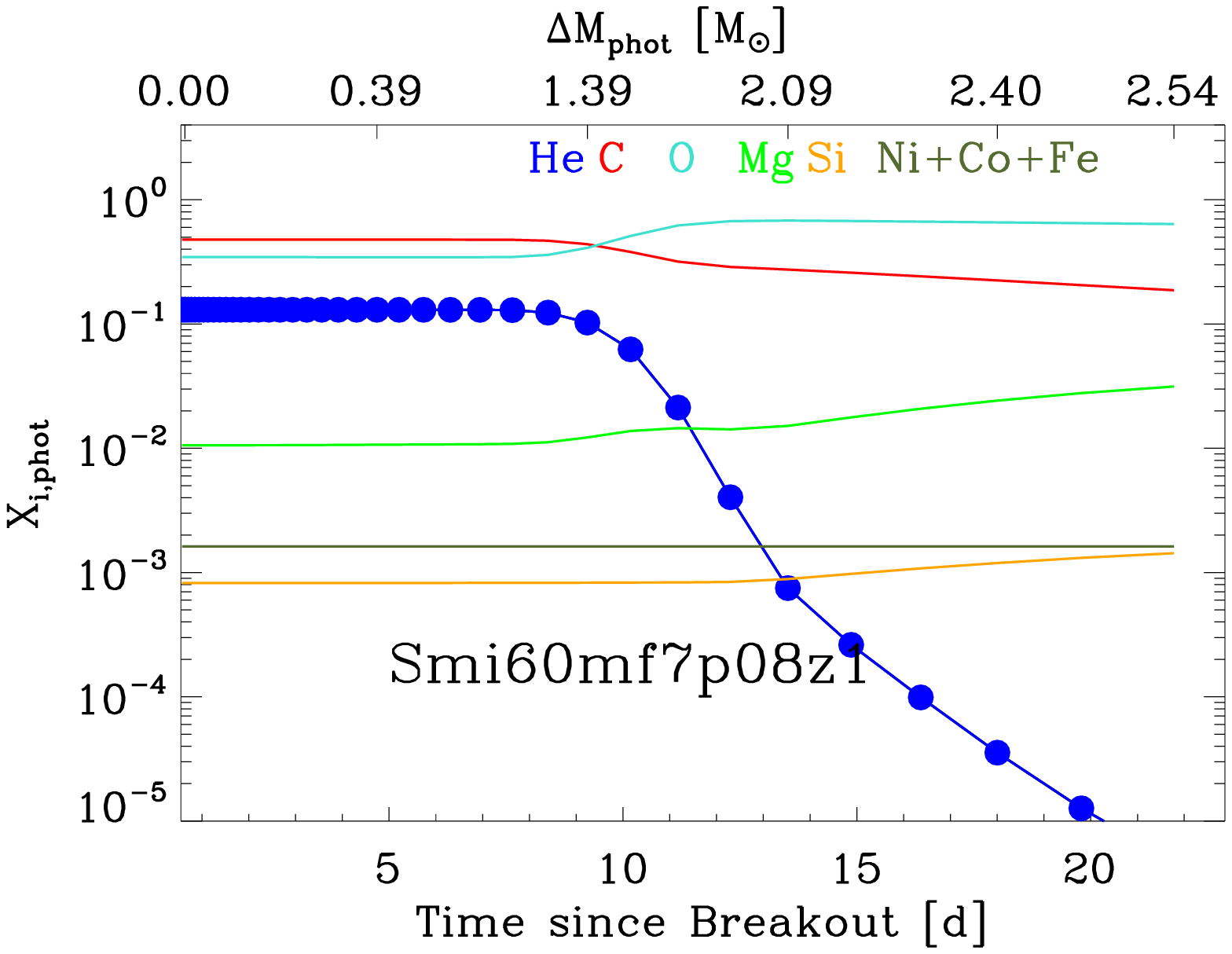,width=8cm}
\epsfig{file=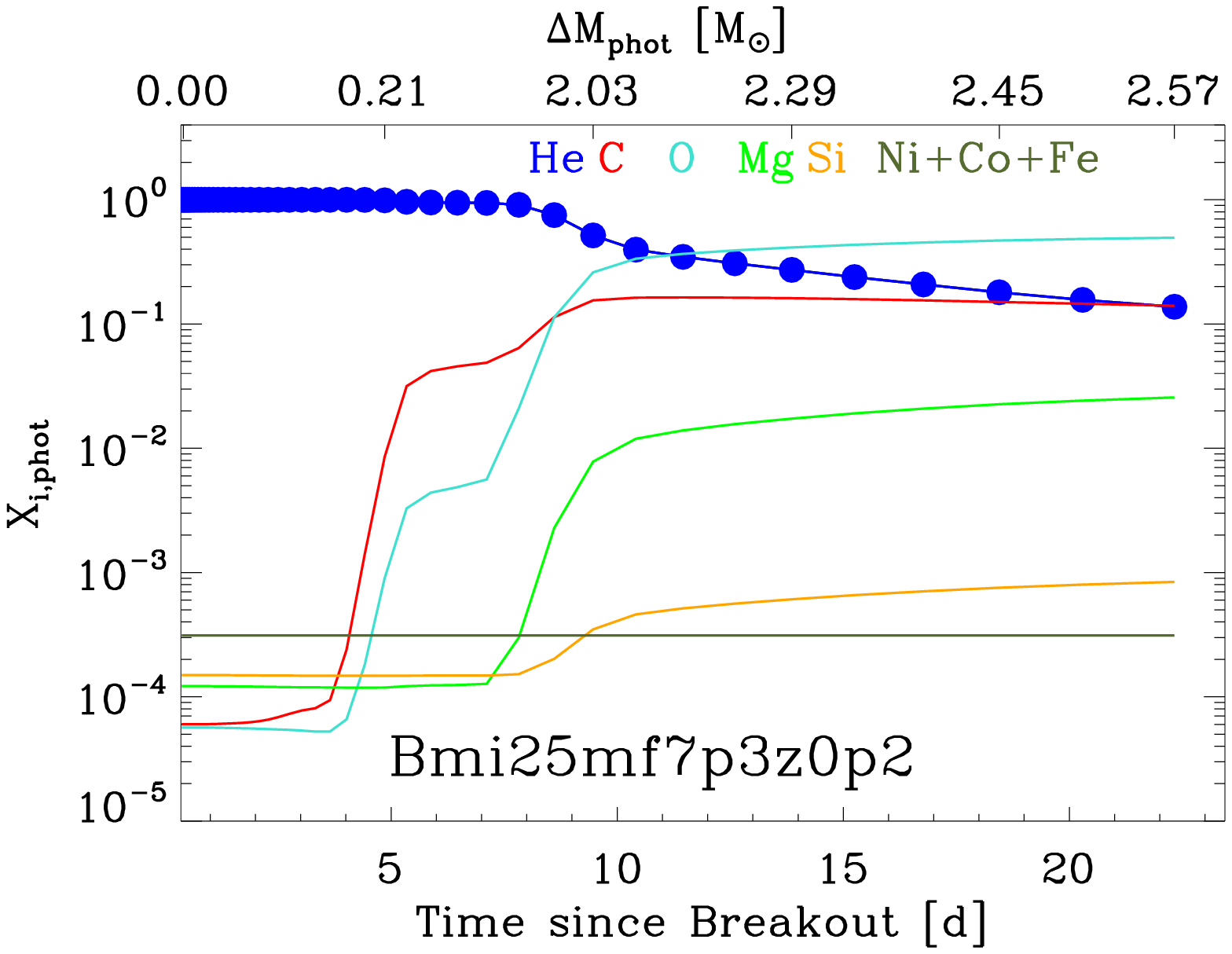,width=8cm}
\epsfig{file=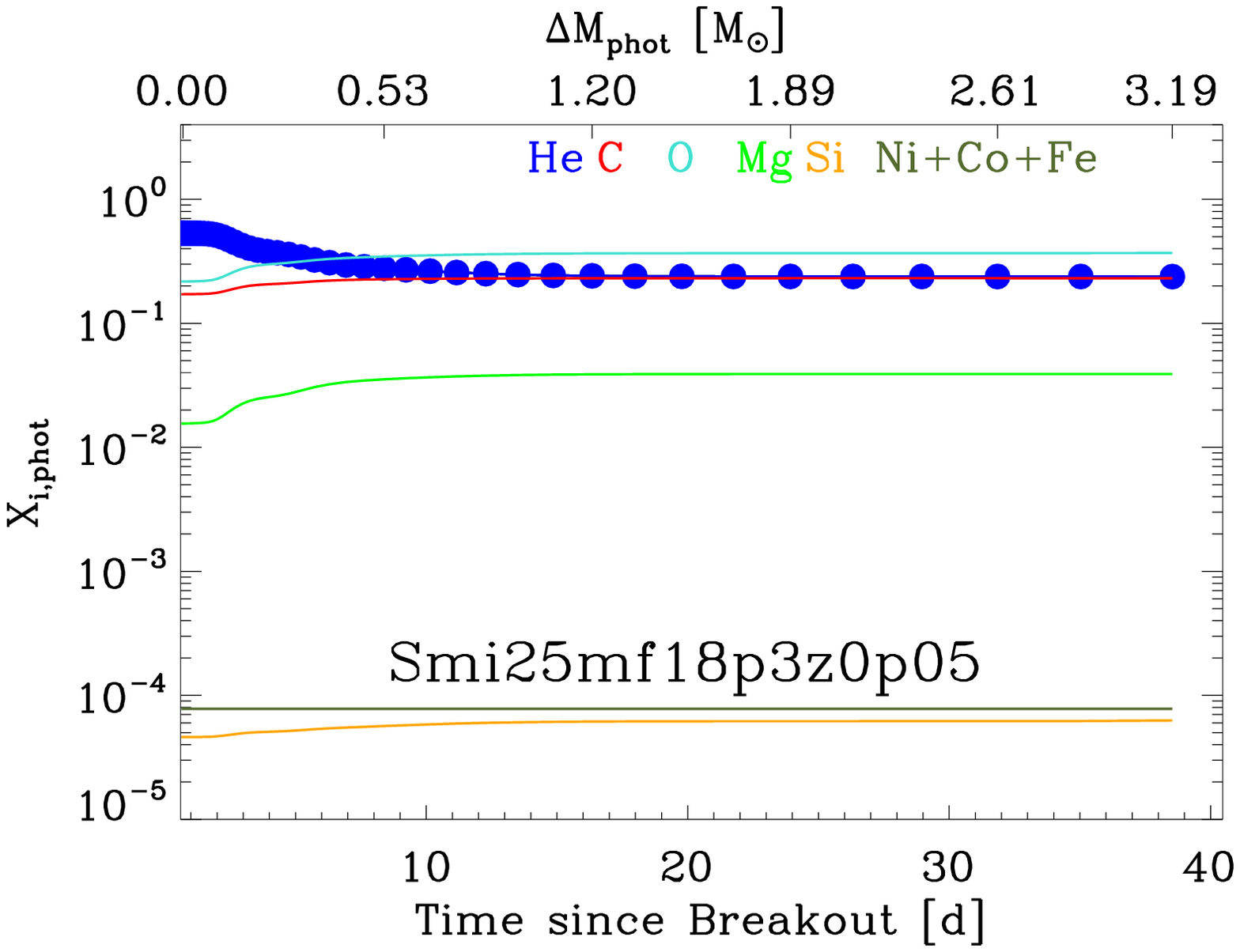,width=8cm}
\caption{
Evolution of the composition (given as mass fractions) at the photosphere for models
Bmi18mf3p79z1 (top left; the time of peak brightness is $\sim$30\,d;
model Bmi18mf4p41z1 has a very similar chemical stratification
and is not shown), Bmi25mf5p09z1 (top right; the time of peak brightness is $\sim$50\,d;
the top axis is not shown since $\Delta M_{\rm phot}$ is a non-monotonic function of time in this model),
Bmi25mf6p49z1 (middle left), Smi60mf7p08z1 (middle right),
Bmi25mf7p3z0p2 (bottom left), and Smi25mf18p3z0p05 (bottom right).
The same colour coding is used in all panels of this figure to differentiate the species represented.
Dots refer to the actual post-explosion times of the radiative-transfer computations (drawn on the curve
corresponding to the helium mass fraction).
Note that the timespan for \isoni-deficient simulations is limited to $\sim$30\,d, while \isoni-rich models (top row)
are evolved all the way to the nebular phase.
\label{fig_phot_comp}}
\end{figure*}

   This general light-curve morphology, discussed in a similar context by \citet{EW88}, is well known
and generally reflects the evolution of the photospheric properties, which we show in
Figs.~\ref{fig_phot_all}--\ref{fig_phot_comp}.
The modest increase in radius combined with the large decrease in temperature at the photosphere
cause the luminosity to drop after breakout. The plateau arises when the temperature decrease at the photosphere slows down,
which is associated with the recombination of ejecta layers to their neutral state (primarily He, but also CNO
elements in such W-R progenitor stars). \citet{EW88} obtain a similar early post-breakout plateau in all their simulations.
For reference, we also show in the bottom-right panel of Fig.~\ref{fig_phot_all} the time evolution of the ejecta electron-scattering
optical  depth, which tend to drop faster than 1/time$^2$ due to changes in ionization. This figure also serves to identify the time at which
each of the ejecta becomes optically thin.

\begin{figure}
\epsfig{file=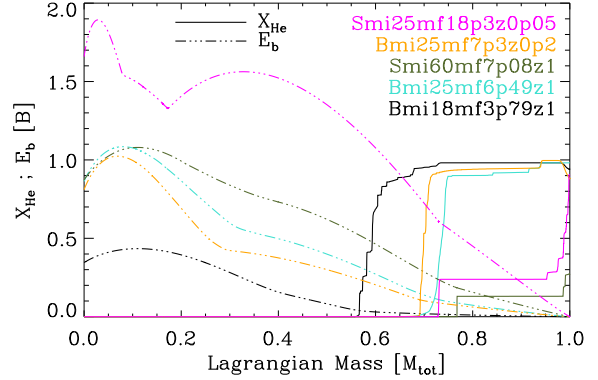,width=8.5cm}
\caption{Illustration of the pre-SN envelope helium mass fraction (solid line) and binding energy (broken line),
as a function of the Lagrangian mass (in units of the total ejecta mass) for our models
Bmi18mf3p79z1 (black), Bmi25mf6p49z1 (turquoise), Smi60mf7p08z1 (olive), Bmi25mf7p3z0p2 (orange), and
Smi25mf18p3z0p05 (magenta). Models with brighter post-breakout plateau luminosities are those
whose progenitors have the least-bound outer envelopes, which also tend to be more helium-rich
and stem from binary-star evolution.
\label{fig_eb}}
\end{figure}

Free of any decay heating at early times, the evolution of the outer ejecta is essentially
adiabatic (modulo radiative cooling near the photosphere), and thus ejecta internal energy is lost
(and ionization state reduced) to $pdV$ work primarily.
What primarily determines the plateau brightness is therefore the amount of energy initially deposited by the shock and
the size of the progenitor envelope.
The former is modulated by the differential rate of ejecta expansion between models. Given the
uniform ejecta kinetic energy in our sample (1 or 1.2\,B),
and the rough inverse correlation of size with ejecta mass for our limited set of models,
models with smaller ejecta mass tend to have a larger plateau brightness
(e.g., model Bmi18mf3p79z1 has the largest kinetic energy per unit ejecta mass and the largest post-breakout plateau brightness).
The latter is connected to the progenitor-envelope binding energy, which we show for a few models
in Fig.~\ref{fig_eb}. Objects that have brighter post-breakout plateaus are those with the least bound outer envelopes.
Because of stellar evolution, these also tend to be more helium rich (greater enrichments in C/O testify for a greater proximity
to the more tightly bound and hotter regions of the progenitor core where nuclear reactions take place).
Consequently, ejecta produced both from a larger mass and a more compact progenitor star are the faintest of all during the short early-time plateau.

  The early-time plateau is not always strictly flat but instead may slant up or down.
  This seems to correlate directly to the depth variation of the ejecta helium abundance (and the associated recombination energy)
  and of the corresponding progenitor-envelope binding energy.
  If the helium mass fraction is constant with depth (or if the
  progenitor-envelope binding energy varies little with depth), the SN luminosity
  is constant or increases, while if the helium mass fraction decreases with depth (or if the
  progenitor-envelope binding energy increases steeply with depth), the luminosity ebbs. The main kinks in helium mass fraction
  shown in Fig.~\ref{fig_phot_comp} thus correspond to points of inflection in the light curves of Fig.~\ref{fig_lbol_all}.
  In all \isoni-deficient models, the precipitous fading at $\sim$10\,d coincides with the sudden decrease of the
  helium mass fraction at the photosphere, and its recession to the oxygen-rich, originally more tightly-bound,
  ejecta layers. We find that the plateau is shorter for outer ejecta having a larger mean-atomic weight,
  because of the lower effective opacity and the greater original binding-energy (in absolute terms) of the corresponding layers.

   At the time of re-brightening in \isoni-rich models, the heat wave from decay energy at depth causes the
   photospheric temperature to reverse its decrease and to rise.
   This rise causes a modest change to the ionisation, but helium remains essentially neutral at the photosphere
   (at later times, non-thermal ionisation, which we neglect, could alter this ionisation state).
   However, the recombination wave is slowed down and the
   photosphere recedes much more slowly in velocity space in models Bmi18mf3p79z1 and Bmi18mf4p40z1
   (Fig.~\ref{fig_phot_all}).  During the rise to peak
   brightness, the photospheric velocity decreases by no more than a few percent. Because velocity relates to mass in SN ejecta,
   this implies that the photosphere probes the same region of the ejecta during that time.
   In model Bmi25mf5p09z1, the influence of the delayed heat wave is much stronger and causes the photosphere
   to move out in velocity from 4000 to 5400\,\kms\ (equivalent to an outward shift in mass of 0.6\,\msun).

   At the onset of re-brightening the photosphere has receded through the very rich He-rich shell. While as much
   as $\lesssim$\,50\% of the ejecta mass may have passed through the photosphere, the photosphere
   is still well outside of the \isoni-rich regions. The re-brightening, and rise to peak, thus stems from the heat wave triggered
    by radioactive decay heating at greater depth.  Because the  \grays\  are subject to a large optical depth at such early times,
    this diffusing heat wave will always reach the photosphere before direct heating by \grays\ can occur. More generally,
    diffusion of heat from the site of radioactive decay seems to be the cause of post-breakout re-brightening
    in all SN II-pec/IIb/Ib/Ic progenitors. As the re-brigthening phase
   continues, the photosphere recedes through the He-C-O region which is still He dominant, although in Bmi25mf5p09z1 He, C, and O
   have comparable mass-fractions. At the time of peak brightness the photosphere is at the base of this He-C-O shell, and is
   entering the oxygen-rich core.

  Interestingly, the light curve of the \isoni-rich model Bmi25mf5p09z1 qualitatively follows the early evolution of
  \isoni-deficient  models, with the onset of a fading at 5--7\,d as the photosphere leaves the helium-rich regions
  of the ejecta. However, this fading only lasts about a week before the heat wave reaches the photosphere and
  triggers a re-brightening.  For yet a larger mass ejecta (with a similar \isoni\ distribution and mass), or for a
  deeper location of \isoni-rich material, this temporary fading could last longer.

   In \isoni-rich models, the time to peak, the width of the peak, and the peak luminosity are conditioned
   by the ejecta mass, as well as the \isoni\ mass and its distribution.
   A larger \isoni\ mass enhances the peak luminosity. Modulations of
   the \isoni\ distribution in mass/velocity space, as well ejecta mass (for a fixed kinetic energy)
   alter the width of, and the rise time to, the peak, which increase with \isoni\ mass (amount of heating) or
   ejecta mass (diffusion time).
   The 2.39\,\msun\ ejecta endowed with 0.184\,\msun\ of \isoni\  cause model Bmi18mf3p79z1 to peak earlier ($\sim$30\,d)
   at a larger value and faint faster after peak.
   The 3.61\,\msun\ ejecta endowed with 0.237\,\msun\ of \isoni\  cause model Bmi25mf5p09z1 to peak later ($\sim$50\,d)
   at a smaller value and fade slower after peak.
   Model Bmi18mf4p41z1 behaves analogously to Bmi18mf3p79z1.

   As discussed above, the influence of decay energy (via the energy diffusion wave) only kicks in at about 10 days in our \isoni-rich
   models (the photosphere hits the \isoni-rich layers not earlier than 30\,d in our models; Fig.~\ref{fig_phot_comp}).
   From this property, and prior to $\sim$10\,d, we can safely discuss the spectroscopic signatures in these simulations
   without any consideration of decay energy, gamma-rays, and non-thermal electrons, since these do not influence, either directly
   or indirectly, the emergent radiation. This also validates the use of progenitor models that were not evolved all the way
   to the formation and collapse of the iron core --- at early times only the outer ejecta are probed.

   As the spectral evolution  is gradual over the first 10\,d after explosion, we  discuss it at two representative times --- 1.5 and 7.0\,d.
   A montage of the spectra for all simulations
  is shown in Figs.~\ref{fig_spec1}--\ref{fig_spec2} (where we plot  $\lambda^2 \times F_{\lambda}$ for better visibility).
  From bottom to top, we stack the spectra in order of increasing ejecta mass.
  We have compiled numerous simulation results in these figures. Besides the full non-LTE time-dependent spectrum,
  we also show as coloured lines the spectrum obtained when the bound-bound transitions of a selected
  species are omitted (for example, the red line shows the spectrum resulting from the neglect of H\one-bound-bound transitions).
  This does not give the line flux associated with a given species, but clearly marks the spectral regions that it affects.
  Furthermore, below each spectrum, we draw tick marks for all lines with an equivalent widths EW greater than 10\,\AA, with a thickness
  that scales with $\log$(EW). For this figure, we address the contributions from the following important species:
  H\one, He\one, C\one, N\one, O\one, Na\one, Mg\two, Ca\two, Si\two, Ti\two, and Fe\two.

  At $\sim$1.5\,d after explosion (Fig.~\ref{fig_spec1}), our synthetic spectra show a similar slope, with the peak of
  the spectral energy distribution (SED) at $\sim$5000\,\AA.
  Line features are more numerous and pronounced in the blue part of the spectrum, although a few broad features
  are also present in the red part of the spectrum. As indicated in this montage, we identify lines of H\one\ (models Bmi18mf3p79z1 and
  Bmi18mf4p41z1),  of He\one\ (all models except Smi60mf7p08z1 and Smi25mf18p3z0p05), and of once-ionised CNO elements.
  The standard Ca\two\,H\&K and 8500\,\AA-multiplet lines are present, although with a strength that varies depending on the
  Ca ionisation state.
  A forest of lines from Ti\two\ and Fe\two\ are present in all models apart from the low-metallicity models
  Bmi25mf7p3z0p2 and Smi25mf18p3z0p05.
  Nearly all the lines we predict in model Bmi25mf7p3z0p2  are from He\one.
  In contrast, model Bmi25mf5p09z1 essentially shows  just Fe\two\ lines.
  At 7.0\,d after explosion (Fig.~\ref{fig_spec2}), the SEDs are significantly reddened and line blanketing effects are stronger,
  primarily because of Fe\two.

\begin{figure}
\epsfig{file=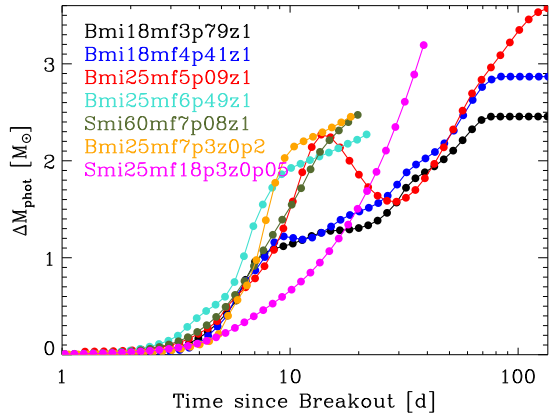,width=8.5cm}
\caption{Same as Fig.~\ref{fig_phot_all}, but now showing the evolution of the mass above the photosphere ($\Delta M_{\rm phot}$).
Notice the sizeable reversal of the photosphere trajectory in velocity/mass for model Bmi25mf5p09z1.
\label{fig_mphot}}
\end{figure}

\section{Spectral properties during the post-breakout plateau}
\label{sect_spec}

  Overall, the different spectral properties are conditioned by the differences in composition between models, which reflect
  the differing evolutionary paths  and
   environmental  metallicity. The composition is also important
  because ejecta elements have a range of ionisation potentials.
  It is about 11.3-13.6\,eV for H\one, C\one, and O\one, but it is 24.6\,eV for He\one, thus about twice as high.
  Due to the high ionisation potential of He it may not contribute to the spectrum, and thus its presence can be
  hidden.
  Furthermore, the ionisation state  of the photosphere is time dependent and conditioned by complex
  collisional/radiative processes, requiring detailed radiative-transfer simulations.
  Below, we present our results for lines associated with the main elements of interest: H, He, CNO,
  intermediate-mass elements (IMEs), and iron-group elements.

\begin{figure*}
\epsfig{file=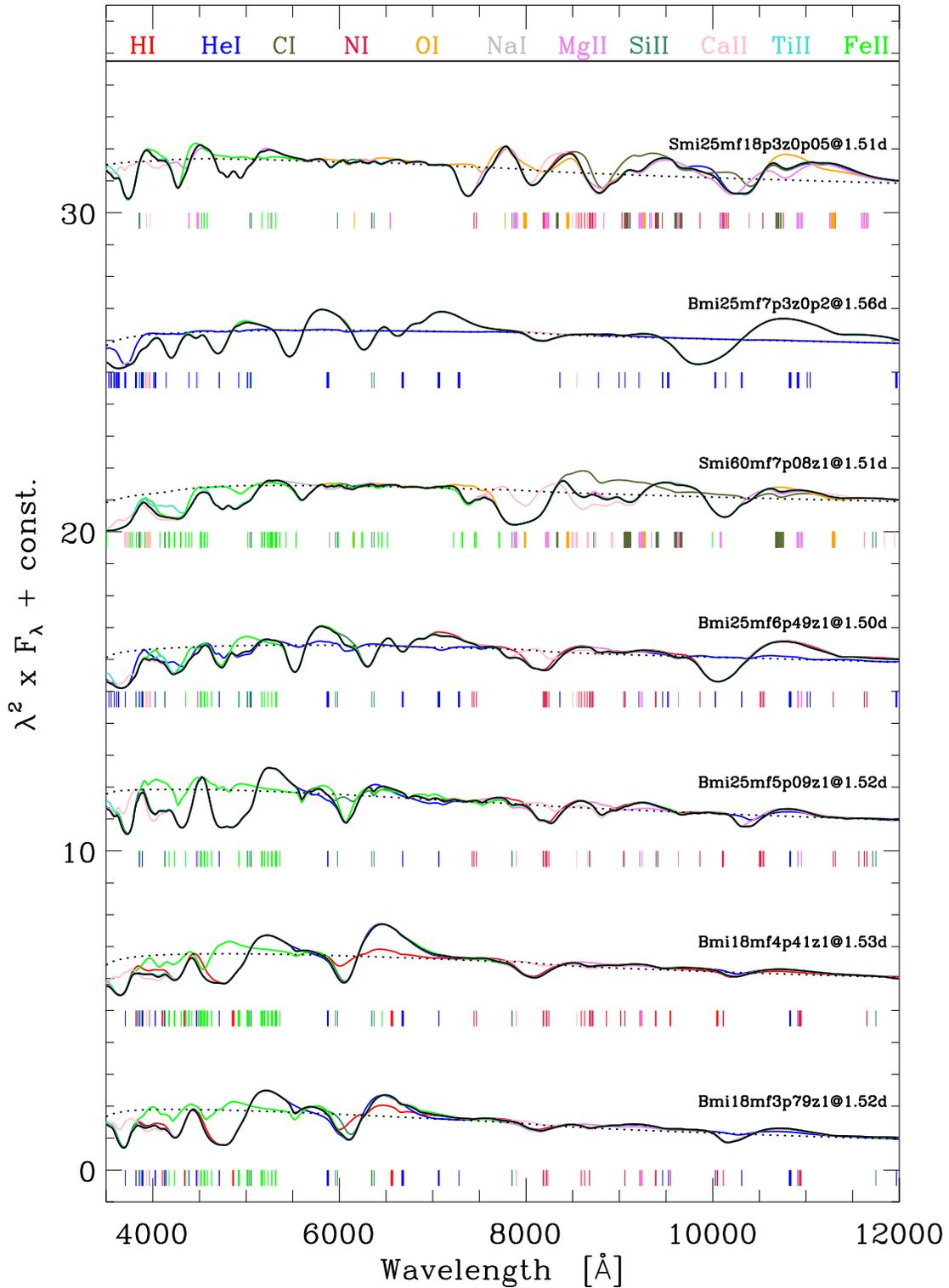,width=15.5cm}
\caption{Scaled synthetic spectra (black) for each of our simulations at $\sim$1.5\,d after explosion (we show
the quantity $\lambda^2 \times F_{\lambda}$ to better reveal the weaker features at longer wavlengths).
At each epoch, we overplot the continuum-only synthetic flux (black dotted line; the same vertical scaling
is applied), as well as the synthetic spectra obtained when bound-bound transitions of a given ion are omitted (see
colour-coding at top). This highlights the relative importance of lines, and their associated ions, in different spectral regions.
Coloured tickmarks indicate the ion producing a line with a Sobolev equivalent width of 10\,\AA\ (in absolute value),
and with a thickness that reflects the magnitude of this equivalent width on a logarithmic scale.
\label{fig_spec1}
 }
\end{figure*}

\begin{figure*}
\epsfig{file=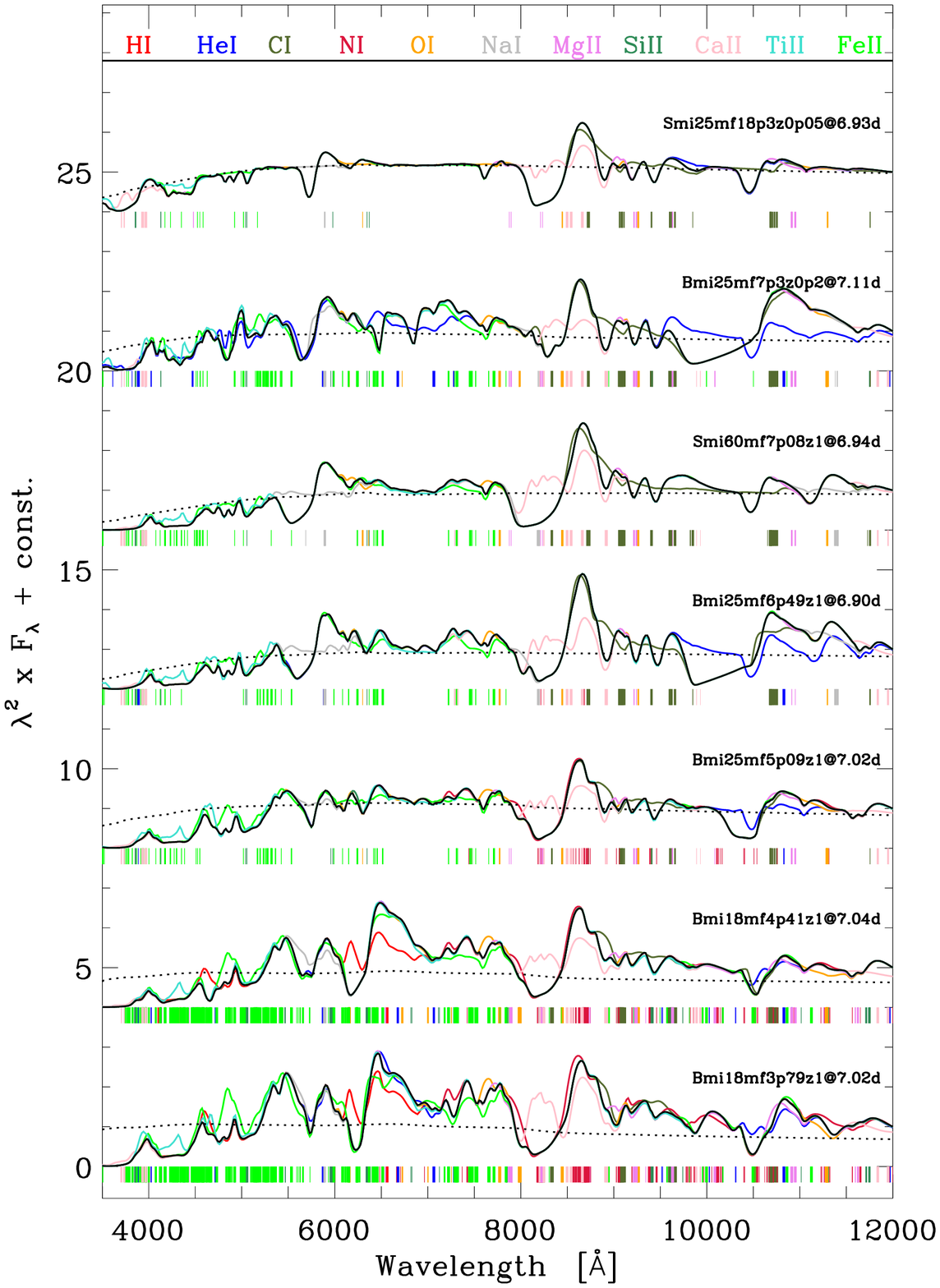,width=16cm}
\caption{Same as Fig.~\ref{fig_spec1}, but now at $\sim$7.0\,d after explosion.
\label{fig_spec2}
 }
\end{figure*}

\subsection{Hydrogen}
\label{sect_h}

    In our model set, only models Bmi18mf3p79z1, Bmi18mf4p41z1, and Bmi25mf5p09z1 contain
    hydrogen, with a maximum mass fraction of 0.043, 0.125, and 6.3$\times$10$^{-5}$, and
    cumulative masses of 1.94$\times$10$^{-3}$, 6.7$\times$10$^{-3}$, and 1.39$\times$10$^{-6}$\,\msun.
    This hydrogen is typically present above $\sim$10000\,\kms.
    In the corresponding regions, the composition is 85-98\% helium, with an integrated mass of $\sim$1.5\,\msun.
    In model Bmi25mf5p09z1, the hydrogen mass fraction of $\sim$10$^{-4}$\,\msun\ is too small to produce
    any feature, but in the other two models, synthetic spectra clearly show H$\alpha$.
    Other Balmer lines are weak and overlap with metal lines which can compromise their
    definite identification (for example, H$\beta$ merely broadens a broad feature due to iron-group elements).
    At 7.0\,d after explosion, H$\alpha$ becomes the second strongest/broadest line
    in the optical spectra of models Bmi18mf3p79z1 and Bmi18mf4p41z1. It is stronger and broader in the latter
    model due to its larger hydrogen mass fraction. Interestingly, our early-time synthetic spectra
     for models Bmi18mf3p79z1/Bmi18mf4p41z1 look very similar to those of Bmi25mf5p09z1 if we were to
     simply remove the H$\alpha$ line from the former models. This suggests that such low amounts of hydrogen do
     not influence sizeably the radiative-transfer solution, but can alter the SN classification.

\begin{figure}
\epsfig{file=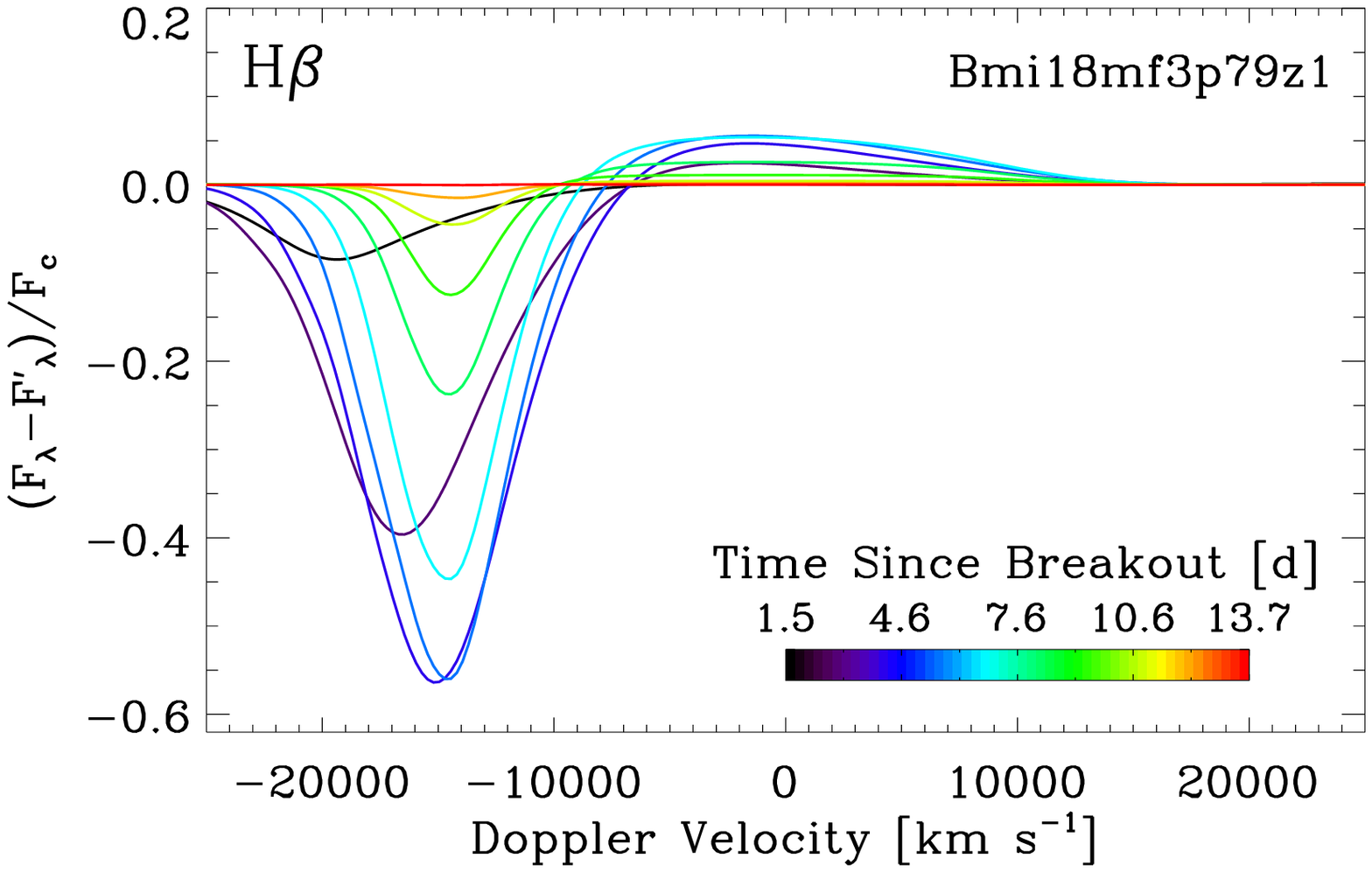,width=8.5cm}
\epsfig{file=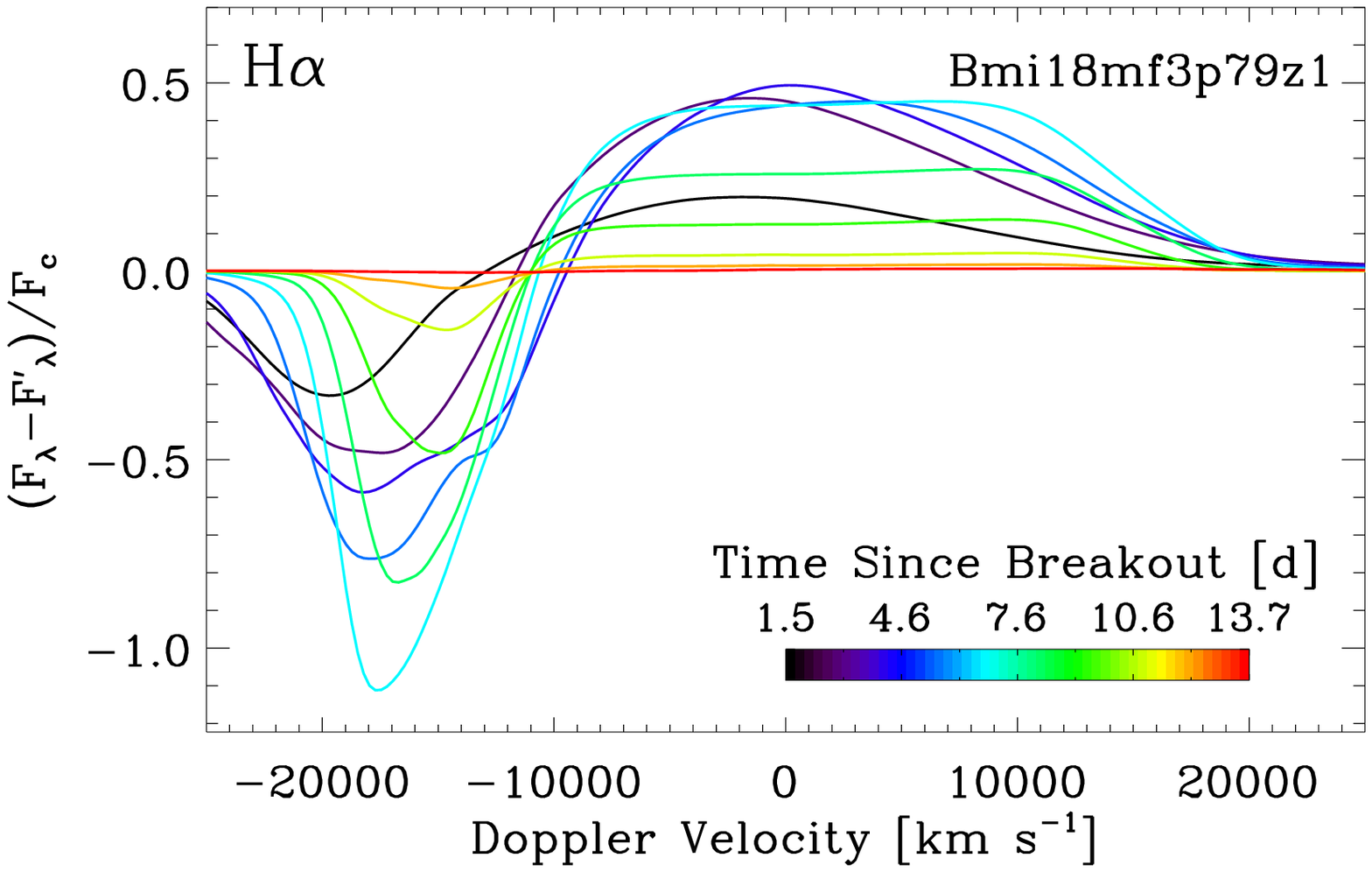,width=8.5cm}
\epsfig{file=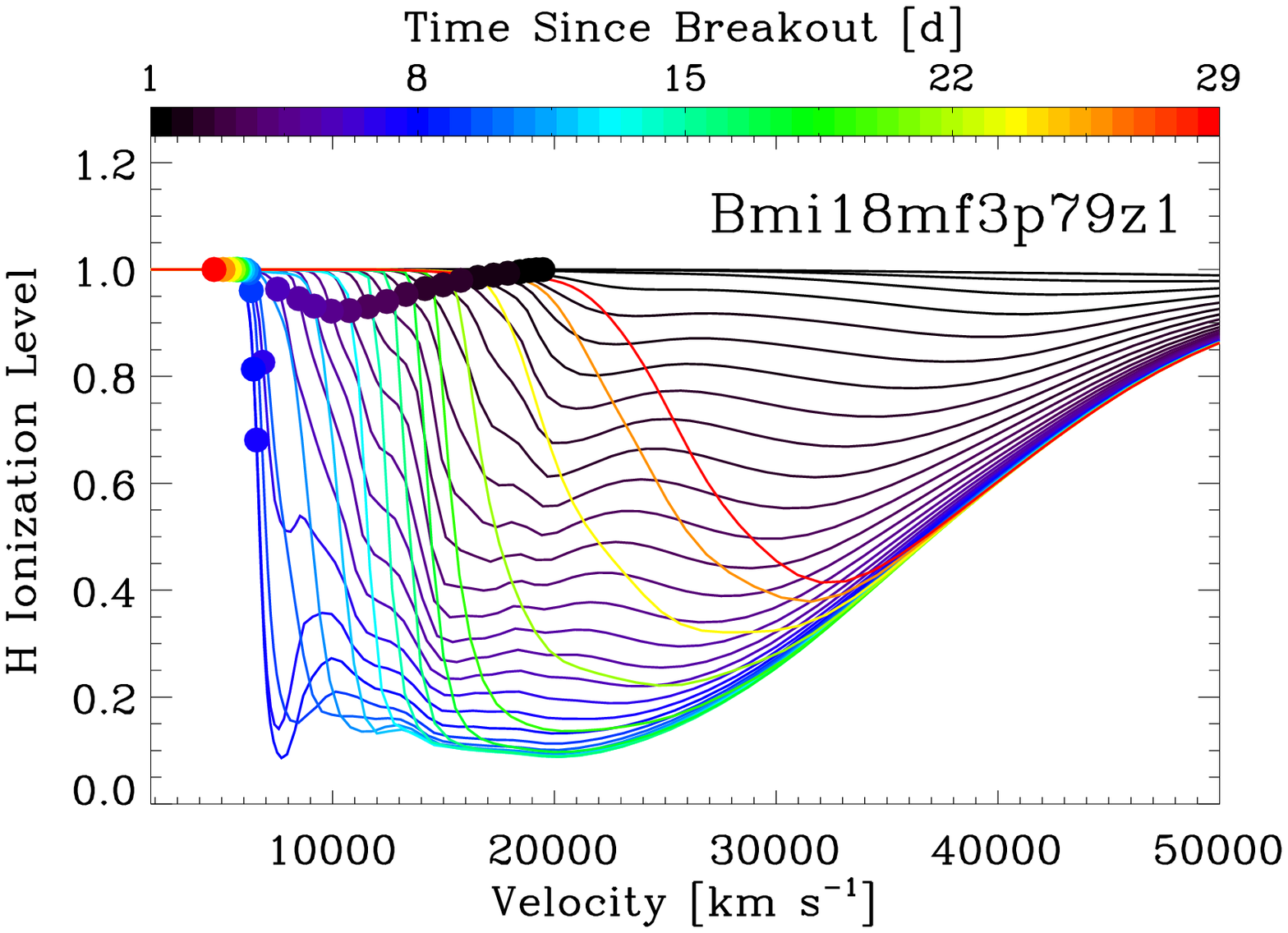,width=8.5cm}
\caption{{\it Top and Middle:} Evolution of H$\beta$ (top) and H$\alpha$ (middle) line fluxes  versus Doppler velocity
from 1.5 to 13.7\,d after shock breakout for model Bmi18mf3p79z1.
More specifically, the quantity plotted is obtained by subtracting from the total flux ($F_{\lambda}$)
the flux obtained by ignoring all bound-bound transitions due to H\one\ ($F'_{\lambda}$) ,
and then normalizing by the continuum flux ($F_{\rm c}$; i.e., the model flux obtained by accounting only for continuum processes).
This approach illustrates the H\one-line flux contribution to the total flux and its evolution with time.
{\it Bottom:} Illustration of the mean hydrogen ionisation level for model  Bmi18mf3p79z1.
The coloured dots refer to this ionisation level at the photosphere and its evolution with time.
Notice the re-ionization of hydrogen above the photosphere after $\gtrsim$7\,d in model Bmi18mf3p79z1
caused by outward-diffusing decay energy released at depth, and which causes a kink in the H$\alpha$ absorption profile (middle panel).
\label{fig_H}}
 \end{figure}

    The presence of H\one\ lines (and primarily H$\alpha$) calls for a Type II classification. The 7-day spectrum does not look
    quite like that of a SN II-Plateau at the same epoch \citep{dessart_etal_08}, and the corresponding model light curve even less so.
    The subsequent evolution exacerbates the differences as the strength of the H\one\ lines rapidly decreases as the photosphere
    recedes to H-deficient
    layers of the ejecta (Fig.~\ref{fig_phot_comp}). Both H$\beta$ and H$\alpha$ vanish at $\sim$10\,d after the explosion.
    Prior to that, H$\alpha$ shows a broad flat-top profile with a weak absorption feature at large velocity
    (Fig.~\ref{fig_H}). With this morphology, the line is often referred to as being ``detached". This stems here from
    chemical stratification, with hydrogen being absent in all ejecta mass shells with a velocity $\lesssim$10000\,\kms.
    In the present context, and if helium lines strengthened at later times, this event would accurately be
    classified as a Type IIb.\footnote{SN classification is, however, subject to errors. For example, SN 2008ax was
    originally classified by \citet{blondin_etal_08} as a Type II-pec and later revised by \citet{chornock_etal_08}, based on additional
    observations, as a type IIb.}

    The evolution of Balmer lines reflects in part the ionisation conditions at and above the photosphere
    (Fig.~\ref{fig_phot_comp}).
    As time progresses, the regions in the vicinity of the photosphere recombine and cool, photoionisation/recombination
    rates decrease,
    so that the Balmer lines eventually weaken. This is mitigated at large radii/velocities by an ionisation freeze-out which stems
    from time-dependent effects \citep{UC05_time_dep,DH08_time}.
    The arrival of the heat wave at the photosphere at $\sim$8\,d after explosion raises the hydrogen ionisation and
    causes the hydrogen-recombination front to travel outwards in velocity/mass. Because hydrogen is under-abundant in those
    ejecta regions, it does not reverse the recession of the photosphere to deeper ejecta layers, since helium, the
    dominant species there,
    continues to recombine. Spectroscopically, this ionisation shift causes the H$\beta$ absorption to plateau,
    while in H$\alpha$, it causes a kink in the P-Cygni absorption.
    The H$\alpha$-line survival time, strength, and width, could easily be modulated by varying, for example,
    the progenitor radius, forcing hydrogen to recombine earlier/later, or by mixing hydrogen to deeper layers
    (this coud result from stellar evolution or from fluid instabilities triggered after shock passage).

    From these simulations, which are based on realistic binary-star evolution models,
    we predict that H\one\ lines are produced for very low hydrogen abundances, corresponding here
    to $\gtrsim$0.01 mass fraction (if $\lesssim$10$^{-4}$, no H\one\ line is obtained) and  $\gtrsim$0.001\,\msun\
    cumulative mass (these values appear quite standard
    for the corresponding progenitor regions, which are at the base of the hydrogen envelope and thus show
    significant mixing of hydrogen and helium).
    The H\one\ lines are predicted for up to $\sim$10\,d {\it in the absence of non-thermal processes} and more generally
    without any need for \isoni.
    H$\alpha$, the strongest Balmer line we predict, is strong at early times but considerably weakens as time goes on, so that its unambiguous
    identification can only be done at early times. At later times, it may still be present but this will be conditioned
    by multiple, complicated, and sometimes largely unconstrained effects.
    These include mixing, \isoni\ production and distribution, ejecta mass, chemical-stratification etc...
    Further, its identification will be compromised by overlap with numerous and potentially stronger lines. Obviously, obtaining
    very early-time spectra  alleviates all these issues and can set a firm constraint on the properties of the progenitor star.

\subsection{Helium}
\label{sect_he}

  The production of He\one\ lines from ejecta arising from W-R-star explosions has been routinely associated with
non-thermal excitation processes \citep{harkness_etal_87,lucy_91,swartz_91,KF92,swartz_etal_93a,KF98a,
KF98b,swartz_etal_95}. This has been motivated by the inability of producing them by other physical
means. However, these early explorations employed radiative-transfer simulations that did not include all the relevant
physics, in particular lacked a rigorous treatment of non-LTE effects. In the context of SN 1987A,
\citet{EK89_87A} and \citet{SAR90_87A} were also unable to reproduce the He\one\ lines at 1-2\,d after explosion,
and obviously, non-thermal processes could not be the culprit.
This He\one\ discrepancy has now been resolved, although at a great computational cost,
with the use of a full non-LTE treatment \citep{DH10a}, demonstrating the need for a fully-consistent solution
of the radiative-transfer problem. More generally, it is interesting to investigate whether
non-LTE effects alone can lead to the production of He\one\ lines in W-R-star explosions. Using the present simulations,
we study this possibility at early-times when \isoni\ produced at depth has influenced the photospheric layers
neither directly nor indirectly. We also include in this discussion those models that do not contain \isoni\ or
other unstable nuclei.
Our results thus apply irrespective of \isoni\ production and therefore yield stronger and un-compromised
constraints on the progenitor star itself.

\begin{figure*}
\epsfig{file=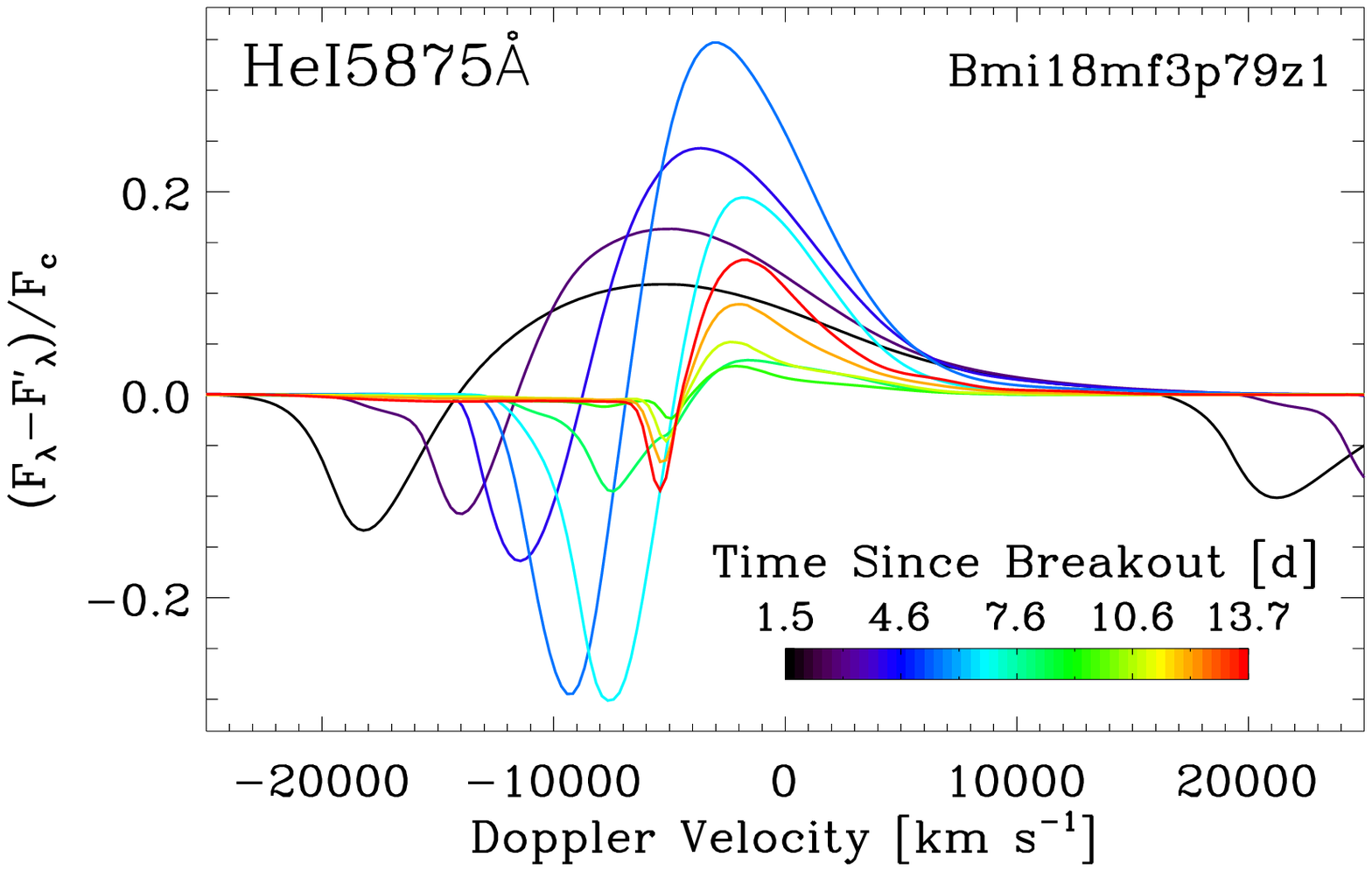,width=8.5cm}
\epsfig{file=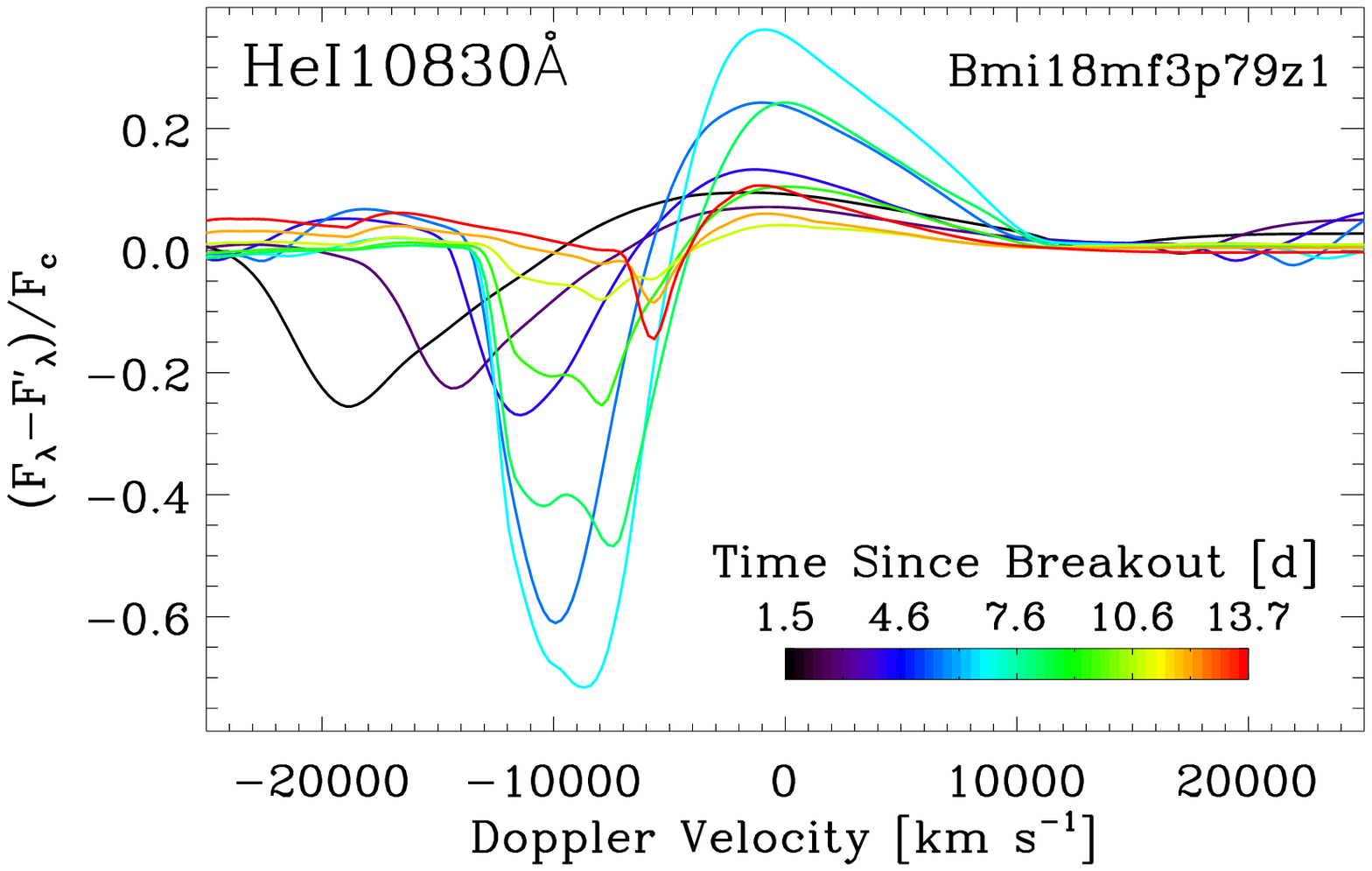,width=8.5cm}
\epsfig{file=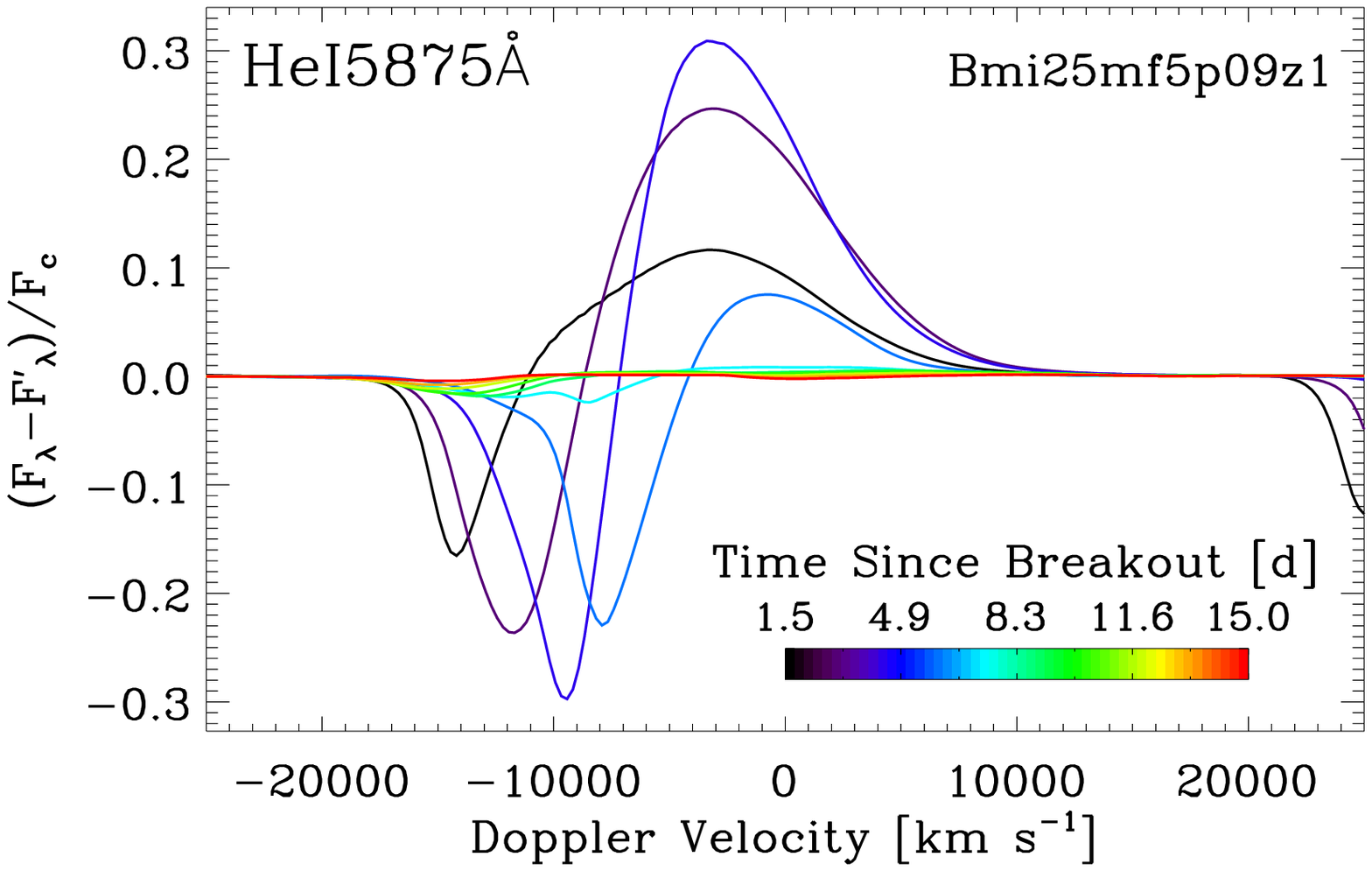,width=8.5cm}
\epsfig{file=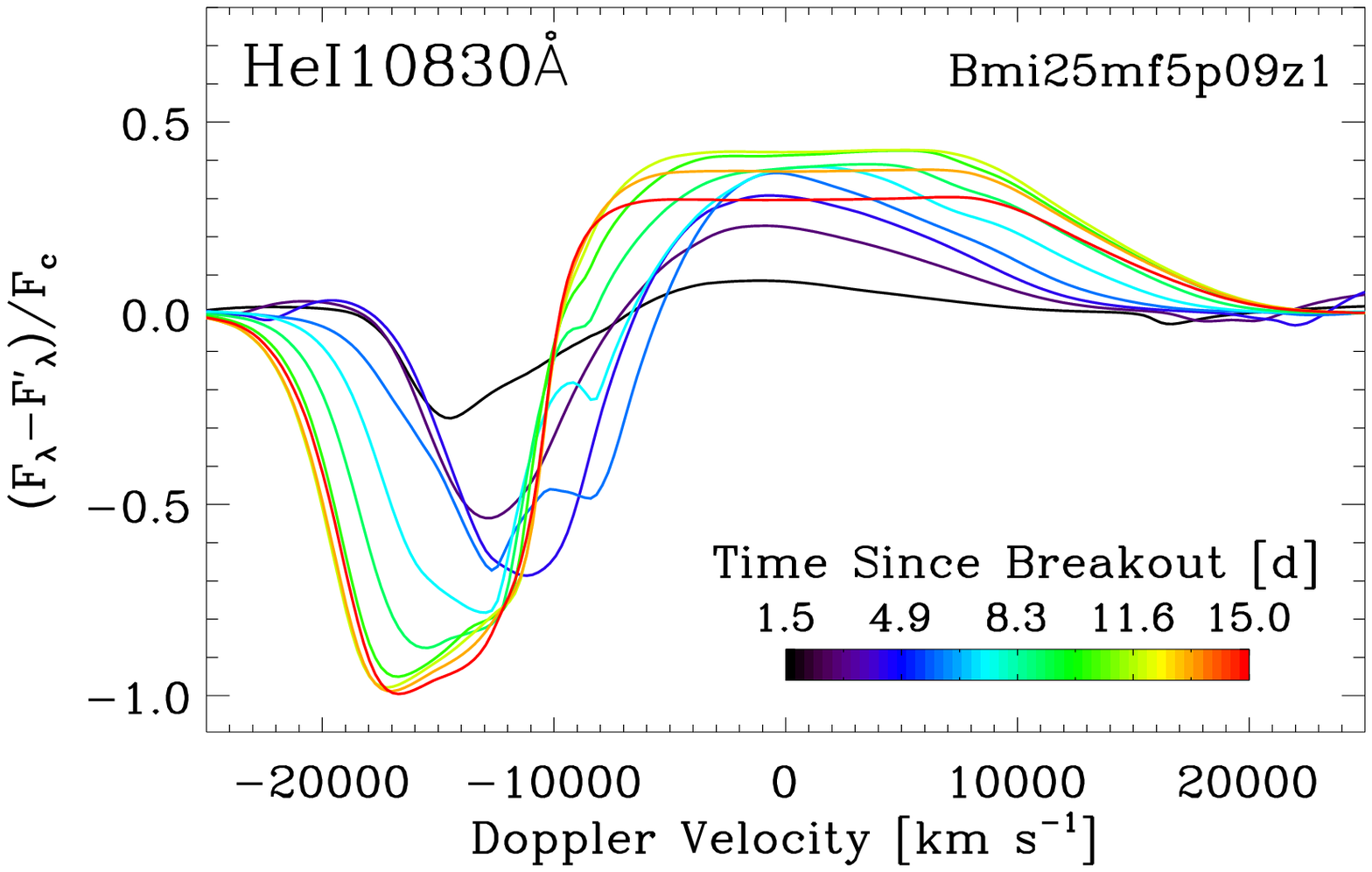,width=8.5cm}
\epsfig{file=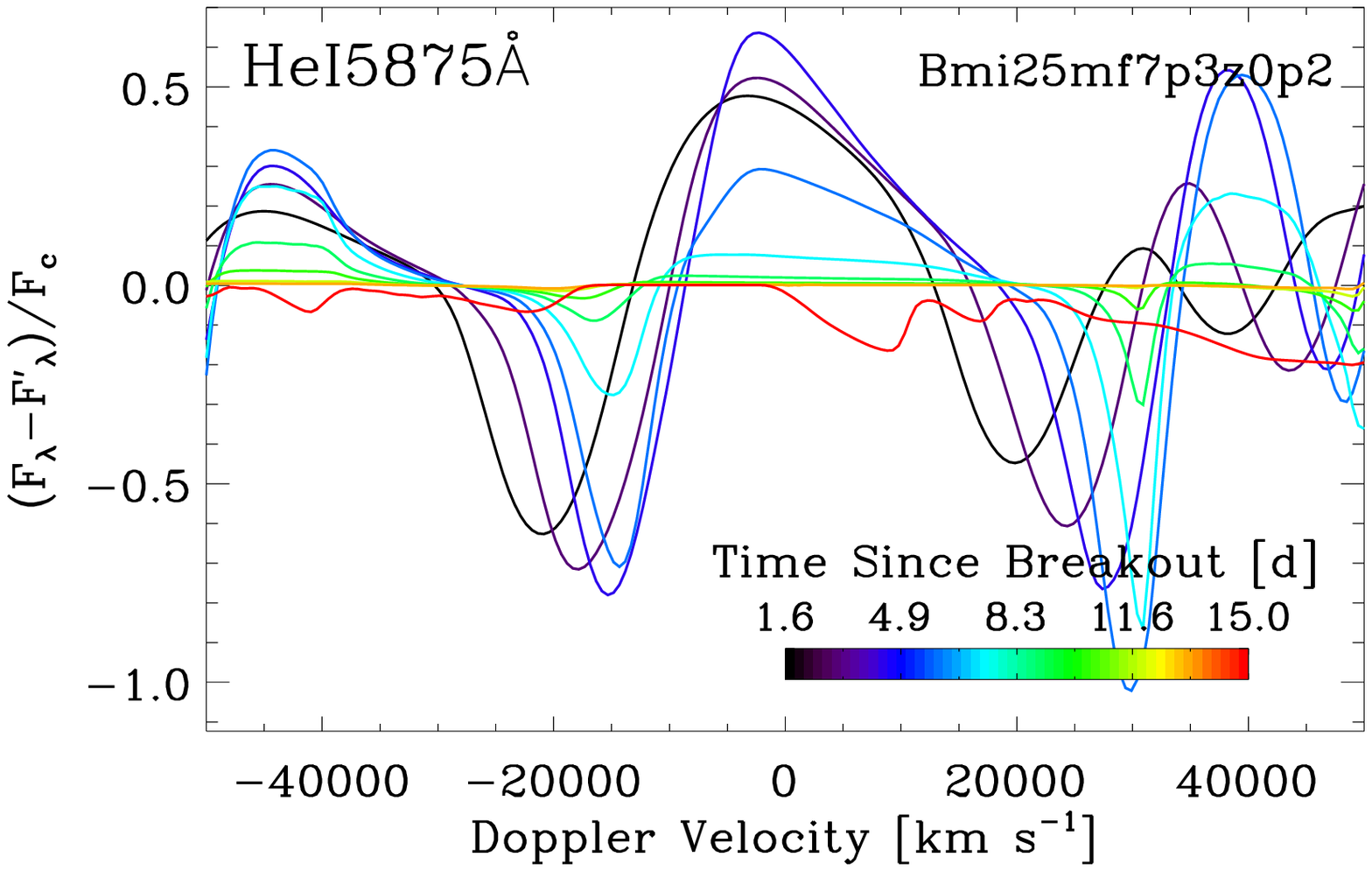,width=8.5cm}
\epsfig{file=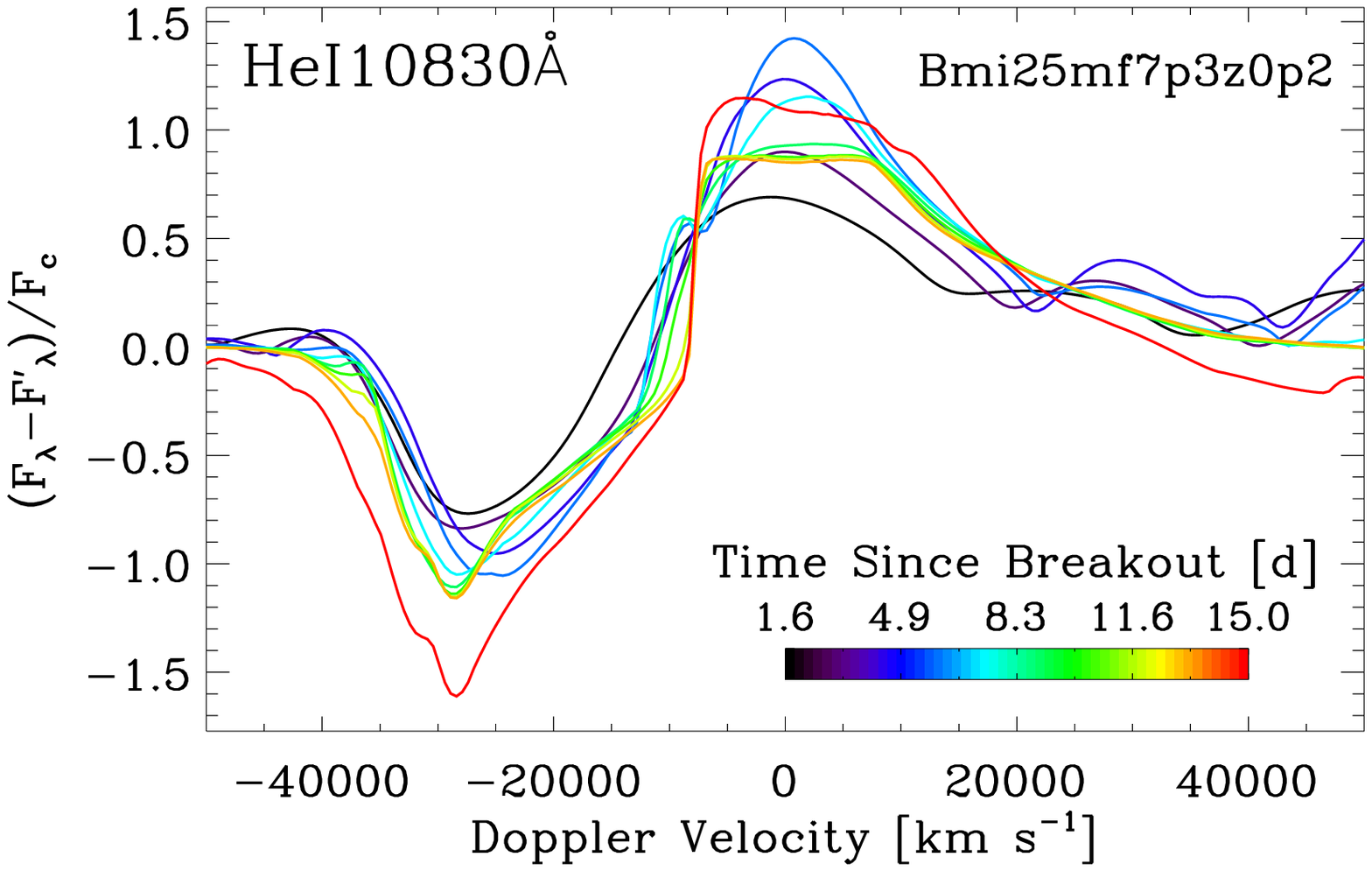,width=8.5cm}
\caption{{\it Top:} Same as Fig.~\ref{fig_H}, but now for the He{\sc i}\,5875\,\AA\ region (left; we also witness absorption
from the 6678\,\AA\ line to the red) and the He{\sc i}\,10830\,\AA\ region (right). The model depicted is Bmi18mf3p79z1.
{\it Middle:} Same as top, but now for model Bmi25mf5p09z1.
{\it Bottom:} Same as top, but now for model Bmi25mf7p3z0p2.
In this low-metallicity model, He{\sc i} lines are present at early times, and persist even for two weeks for the near-IR line,
even in the absence of non-thermal excitation by high-energy electrons.
Note that the abscissa scale now extends to 50000\,\kms.
\label{fig_He}}
 \end{figure*}

  The montage of spectra shown in Figs.~\ref{fig_spec1}--\ref{fig_spec2} for 1.5 and 7.0\,d post-explosion times
  suggest that He\one\ lines are present, although sometimes quite weak, in all models apart from
  Smi60mf7p08z1. In model Bmi25mf7p3z0p2, the synthetic spectrum at 1.56\,d
  is almost exclusively composed of He\one\ lines, specifically at 3705, 3819, 3888, 3964, 4026, 5875,
  6678, 7065, 7281, 10830, 10914, and 11969\,\AA\ (these are the He\one\ lines with a Sobolev optical depth
  greater than 50\,\AA\ in the range 3500--12000\,\AA).
  When He\one\ lines are rare and weak, He\one\,5875\,\AA\
  is the most apparent in the optical range. Compared to what they were at 1.5\,d, He\one\ lines at 7.0\,d
  are of similar (and weak) strength in models Bmi18mf3p79z1, Bmi18mf4p41z1 or Bmi25mf5p09z1,
  but of even lower strength in models Bmi25mf6p49z1 and Bmi25mf7p3z0p2.
  This evolution is more pronounced for He\one\,5875\,\AA\ than for He\one\,10830\,\AA, which can remain
  relatively strong both in absorption and in emission up to 15\,d after explosion (Fig.~\ref{fig_He}).

  Overall, the presence of He\one\ lines in our synthetic spectra suggests that He\one\ is thermally
  and radiatively excited in the corresponding models. The conditions are more favorable in progenitors
  with a sizeable and loosely-bound helium envelope (Fig.~\ref{fig_eb}), which tend to produce ejecta with a suitably
  larger internal energy (for a given ejecta kinetic energy).
  Note that in some of our models, the feature at 5900\,\AA\ is due to Na\,{\sc i}\,D rather than He\,{\sc i}\,5875\,\AA\
  (models Bmi25mf5p09z1, Bmi25mf6p49z1, and Bmi25mf7p3z0p2; Fig.~\ref{fig_spec2}).

  In model Smi25mf18p3z0p05, the only helium line present is He\one\,10830\,\AA, which appears
  as a weak (``detached") absorption at a Doppler velocity of $\sim$-26,000\,\kms,
  reminiscent of the line-profile structure found in
  time-dependent simulations displaying an ionization freeze-out (see Figs.~11 and 12 of \citealt{DH08_time}).
  In this model, both ionisation freeze-out and chemical stratification are the cause of this feature.

   In \isoni-rich models, although we neglect high-energy electrons and treat all decay energy as a local heat source,
  He\one\ lines can be present beyond the end of the post-breakout plateau. For example, He\one\ lines are
  present up until peak brightness at $\sim$30\,d in model Bmi18mf3p79z1 because at that time the
  helium mass fraction at the photosphere is still $\sim$0.8 and the diffusing heat from decay enhances the
  gas temperature at the photosphere to $\sim$10000\,K.
  In model Bmi25mf5p09z1, at 30\,d after explosion, the helium mass fraction at the photosphere is down to $\sim$0.5
  and the heat from decay, shared by a greater mass, is also diffusing out more slowly (Fig.~\ref{fig_phot_all}).
  This later point may be ciritical for the SN Ic classification: They may be genuinely helium-deficient, or may correspond
  to ejecta with a mixture of helium, carbon, and oxygen, and separated from the \isoni-rich layers by a sizeable mass buffer.
  In any case, we stress that, even as a pure heat source, decay energy diffusing out from greater depths may
  be sufficient to produce He\one\ lines at the light-curve peak, provided the ratio of ejecta mass to \isoni\ mass is not too large
  and the helium mass fraction at the photosphere is high (as in this model Bmi18mf3p79z1).

\begin{figure}
\epsfig{file=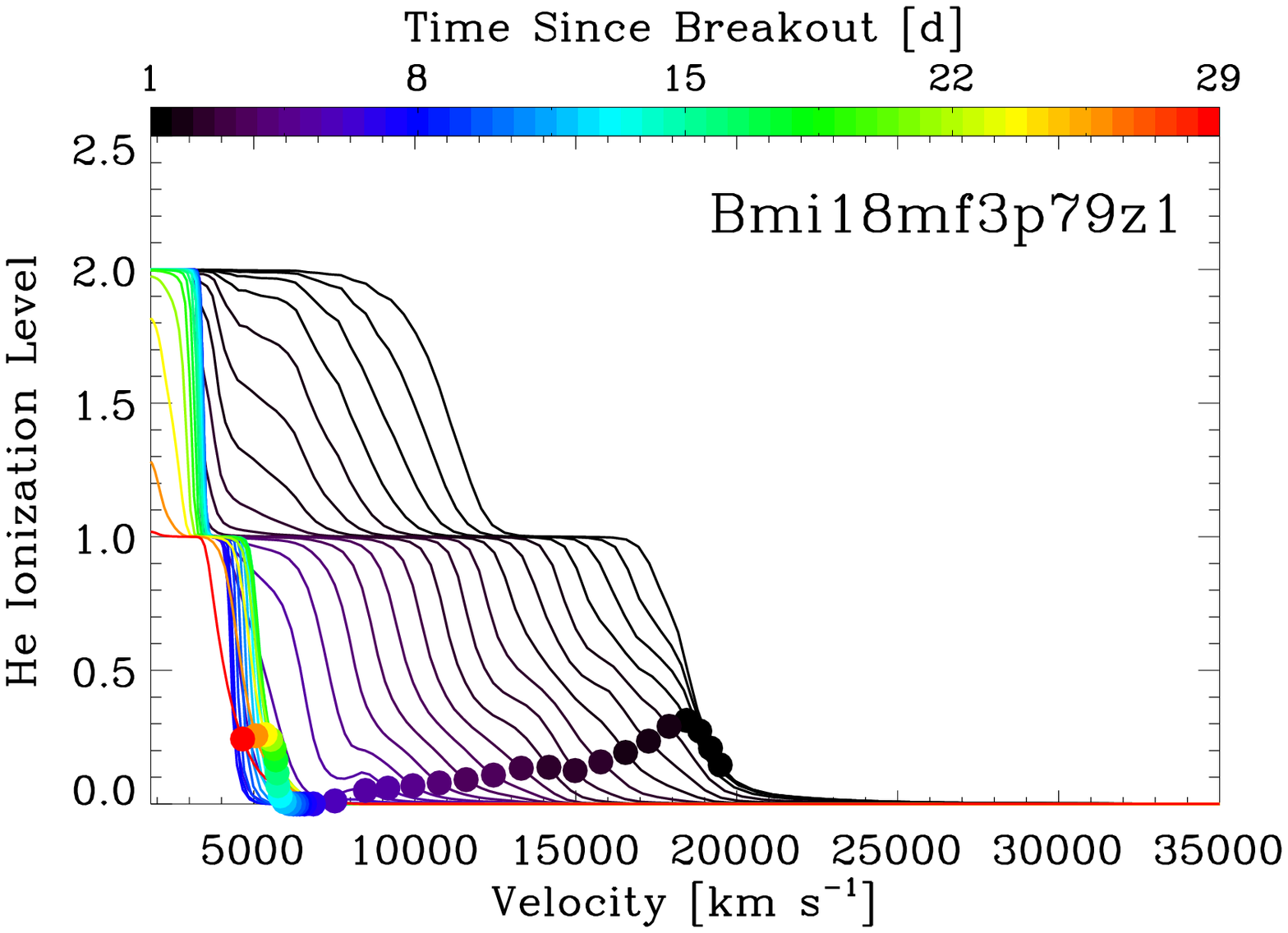,width=8.3cm}
\epsfig{file=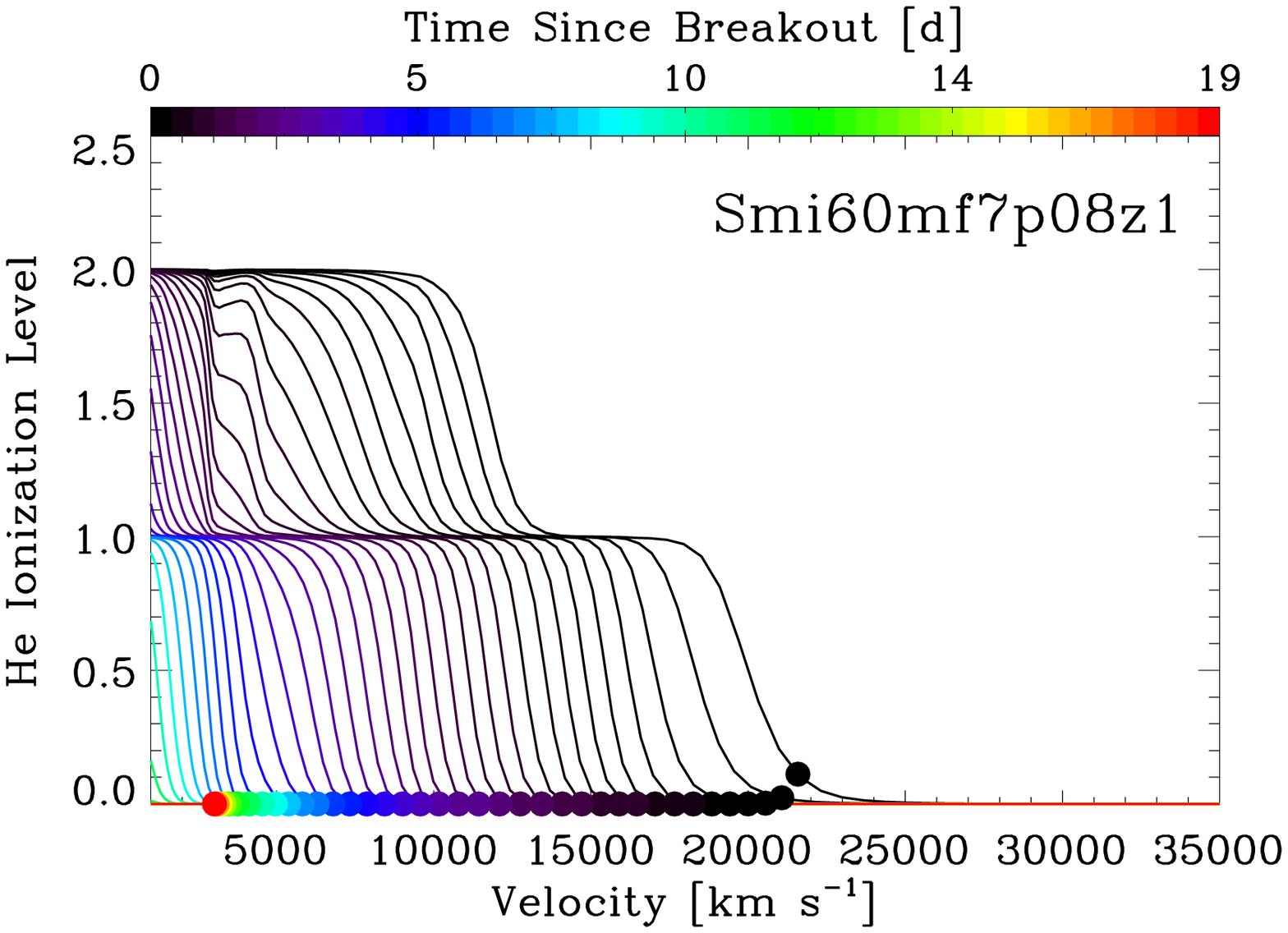,width=8.3cm}
\epsfig{file=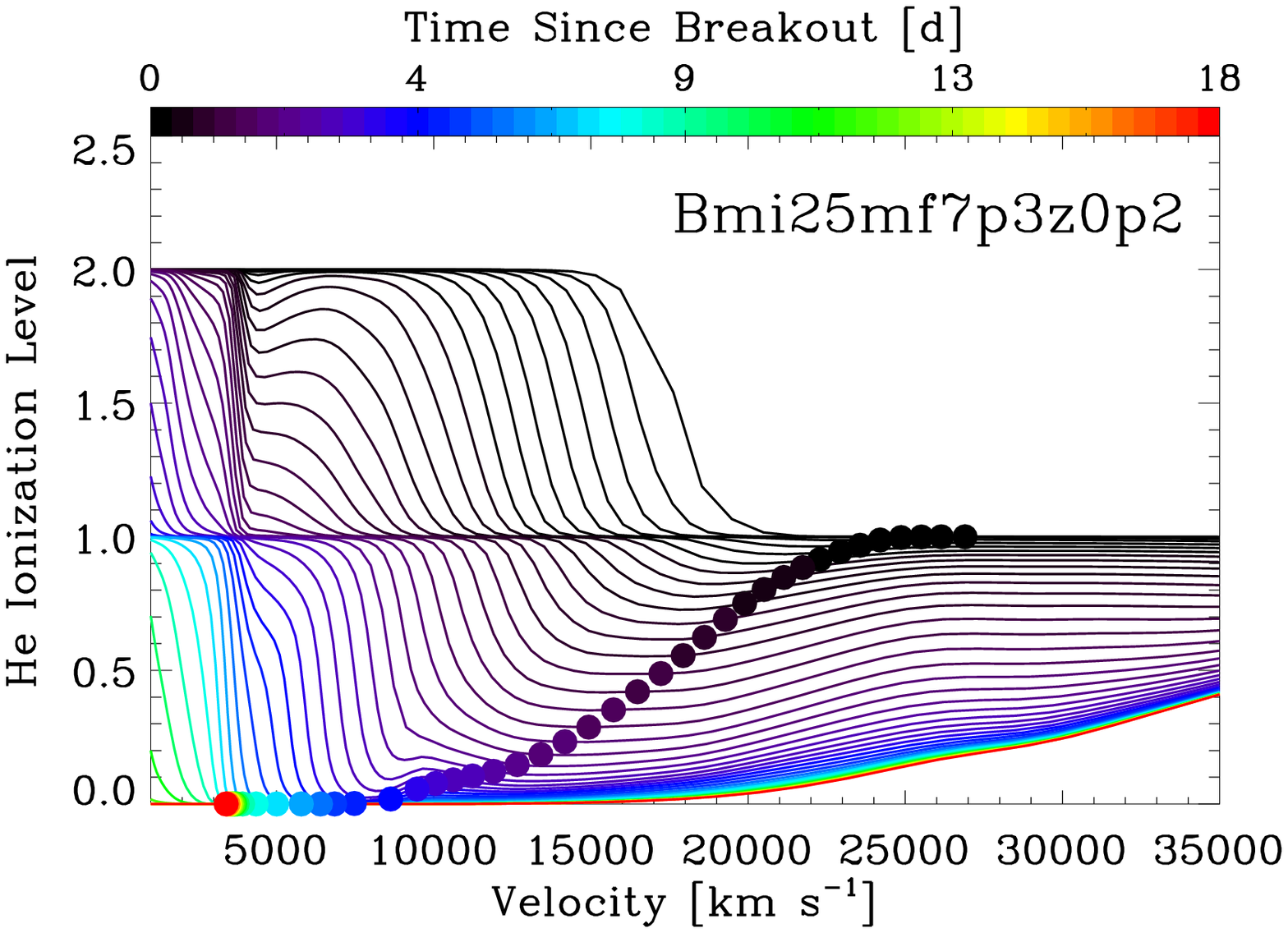,width=8.3cm}
\caption{Illustration of the mean helium ionisation level for models
Bmi18mf3p79z1 (top),  Smi60mf7p08z1 (middle), and Bmi25mf7p3z0p2 (bottom).
In each panel, the coloured dots refer to this ionisation level at the photosphere and its evolution with time.
An interesting feature is the systematic partial ionisation of hydrogen and helium in models
showing H\one\ and/or He\one\ lines at early times, even in the absence of \isoni/\isoco;
the counterexample here is model Smi60mf7p08z1. [See text for discussion.]
\label{fig_ion_He}}
\end{figure}

  The notion that one needs He\one\,10830\,\AA\ to unambiguously assess the presence or absence
  of helium is partially supported by our simulations. At early times, whenever He\one\,10830\,\AA\ is present,
  He\one\,5875\,\AA\  is present too and should be identifiable. At $\sim$10\,d, however, He\one\,5875\,\AA\ tends to
  become very weak while He\one\,10830\,\AA\ may still be strong and more easily identifiable.
  He\one\,10830\,\AA\ is systematically broader and stronger, both in absorption and emission, than
  other He\one\ lines, and it suffers less from line blending.

  The disappearance of He\one\ lines is directly connected to the decreasing helium mass fraction
  at the photosphere (Fig.~\ref{fig_phot_comp}), as it migrates
  towards the originally more tightly-bound inner regions of the ejecta.
  This phase also corresponds to the rapid luminosity fading that ends the post-breakout plateau (Fig.~\ref{fig_lbol_all}).
  As for hydrogen discussed above, the helium ionisation is a critical component for the production of He\one\
  lines. In model Bmi25mf7p3z0p2, which shows the strongest He\one\ lines at 1.5\,d, helium is partially ionised
  above the photosphere, while in other simulations with weaker He\one\ lines, it is very nearly neutral (Fig.~\ref{fig_ion_He}).

  There is something quite spectacular here though. While we find that hydrogen present at a few percent
  by mass is sufficient to produce a very strong H$\alpha$ line, helium present at $\sim$90\% by mass
  may produce feeble He\one\ lines, as in simulations Bmi18mf3p79z1, Bmi18mf4p41z1 or Bmi25mf5p09z1.
  It is not clear at what threshold helium mass fraction He\one\ lines would disappear; in our restricted set
  of simulations, He\one\ lines are invisible for helium mass fractions of $\lesssim$50\%.
  In these models, the lines we do predict stem from species that have a mass fraction at the per cent level
  or less. This is a dramatic illustration of the direct effects of ionisation and excitation on the emergent radiation,
  which can easily compensate for even large variations in abundances \citep{DH08_time}.
  Accounting for such ``dark matter" is a challenge.
  Of course, the presence (and dominance) of helium does condition the radiative-transfer solution (for example, helium
  ionisation controls the location of the photosphere). It also conditions the expansion rate since helium
  is associated with a sizeable mass, but its total mass cannot be derived easily in a direct manner.

   All our binary-star evolution models are characterised by very high helium-surface mass fractions of $\gtrsim$85\% and
   their associated ejecta produce spectra that contain He\one\ lines (potentially difficult to identify in the optical
   but not with He\one\,10830\,\AA). The identification of He\one\ lines {\it prior
   to the SN re-brightening phase} is a critical signature of a very high helium content, potentially favouring a binary-star
   evolution channel.  Our results also imply that a significant amount of helium may be present in the outer regions
   of the ejecta without any clear radiative signature.
   Recall that we are ignoring non-thermal processes, which could help produce He\one\ lines  when
   we do not predict them. However, in those cases where we do predict them, such processes would only make
   He\one\ lines stronger.

\subsection{CNO elements}
\label{sect_cno}

  With more erosion of the progenitor star and/or helium core, deeper shells are revealed and one expects a
  progression from a nitrogen-rich (and potentially hydrogen-rich) to a carbon-rich, and then to an oxygen-rich
  W-R star.
  In the same sequence the helium mass fraction should steadily decrease.
  Our set of progenitor models reflects this dichotomy. The binary-star models correspond to
  WN stars since they show enhancements of helium and nitrogen (with possible traces of hydrogen), and depletion
  of carbon and oxygen, as expected for the CNO cycle.
  The single-star models correspond to WC or WO stars since they show depletion of helium and nitrogen (with no trace of hydrogen)
  and enhancements of carbon and oxygen, as expected from helium-core burning.
  Hence, variations in helium mass fraction are accompanied by corresponding variations in CNO abundances.
  It is thus important to search for radiative signatures of such CNO elements.

   As expected, those (single-star) models that do not show any He\one\ line for the first 10\,d show a wealth of lines
  from C\one\ and O\one\ in the red part of the optical range, while only very few and weak lines are associated
  with nitrogen (i.e., N\one; Figs.~\ref{fig_spec1}--\ref{fig_spec2}).
  There are numerous C\one\ lines, but the stronger ones are at 8335, 9405, 9658,
  and 10691\,\AA. For O\one, the stronger lines are at 7774, 7987--7995,  8446, 9261, and
  11287\,\AA. Some of these overlap with He\one\,10830\,\AA, but He\one\,5875\,\AA\ should
  then be searched for in early-time observations to help resolve the ambiguity.
  There are also lines from once-ionised species such as C\two, for example in model Bmi25mf18p3z0p05 at 1.51\,d,
  with the strongest features at 5890 and 6580\,\AA\ (doublet lines; not shown here). All these C/O lines have similar strength in models
  Smi60mf7p08z1 and Smi25mf18p3z0p05. The appearance of such C/O lines arise from the large C/O mass fraction
  at the photosphere in both models, each element having a mass fraction of 0.2--0.5 (Table~\ref{tab_presn}).
  Such single-star progenitors would be classified WC stars and the lack of helium lines in their optical spectra at 1-10\,d after explosion,
  caused by an ejecta helium mass fraction $\lesssim$50\%, would make the corresponding explosion look like a Type Ic SN.

  In contrast, models for which we predict He\one\ lines do not show strong C/O lines. This occurs because of the
  lower photospheric C/O mass fraction. The nitrogen mass fraction is also low, but not as low, and so we
  predict some N\one\ lines in the red part of the spectrum. In simulation Bmi18mf3p79z1 at 7.0\,d, there
  are numerous C\one, N\one, and O\one\ lines of weak/moderate strength. The main N\one\ line is at 8680\,\AA,
  and as we shall see further below, and from what we discussed above, all these lines/species contributions overlap
  and challenge an accurate identification.

\subsection{Intermediate-mass elements}
\label{sect_ime}

   At the early times, the photospheric composition reflects only the
   results of (steady) core burning, causing modest enhancements of sodium, neon or magnesium abundances compared to solar.
   In contrast, titanium or silicon are at the environmental abundance, just like iron. Despite their
   generally low abundance, a number of IMEs have associated line features,
   in particular Na, Mg, Si, and Ti. This is analogous to what is observed in Type II SNe where species at the envrionmental
   metallicity cause significant blanketing (e.g., Ti\two), or produce obvious features associated with resonance lines (e.g., Na\one\,
   or Ca\two).
   If we evolved these ejecta to later times, lines from such IME
   would be favoured by the increasing mass fraction of such elements at the photosphere.

    Na\one\,D is weak or absent at 1.5\,d, but strengthens as the photosphere cools. At 7.0\,d, it appears as a strong line in
    all our models, whether at solar or sub-solar metallicity, and thus even for an Na mass fraction of 10$^{-5}$.
    In models showing helium lines, we find that it overlaps with He\one\,5875\,\AA\ and contributes significantly
    to the feature.

    Our single-star models, which show numerous and strong C\one\ and O\one\ lines, also show strong Mg\two\ lines, primarily
    at 4481, 7877--7896, 8213--8234, 9218--9244, 9632, and 10915\,\AA.

    Irrespective of abundance issues, Ca
    always shows strong Ca\two\ lines as soon as it becomes once ionised. In all our simulations, we predict a strong
    Ca\two\,H\&K as well as a strong Ca\two\ triplet at 8500\,\AA. This holds even in low-metallicity models once they possess the
    required ionisation.

    We also predict a few lines of Si, in particular Si\two\ at 6347--6371\,\AA.
    In the model Bmi18mf3p79z1, it contributes a line that overlaps quite closely with H$\alpha$, although
    it is weaker than H$\alpha$. In model Bmi25mf5p09z1, this Si\two\ doublet appears
    as a strong P-Cygni profile at 1.5\,d, and could fuel some controversy about the presence of hydrogen in the ejecta
    (present with a mass fraction $\lesssim$10$^{-5}$).
    As we demonstrated before, early-time observations can help lift this ambiguity, since even very small amounts of hydrogen
    would be sufficient to produce a strong, unambiguous, H$\alpha$ line. Models in which H$\alpha$ is strong at 1.5\,d also
    show strong H$\alpha$ at 7\,d. In contrast, the Si\two\ doublet in Bmi25mf5p09z1 is strong at 1.5\,d but
    has vanished by 7.0\,d after explosion.

\subsection{Iron-group elements}
\label{sect_z}

  The main iron-group element that contributes strong line-blanketing is iron (the IME titanium also causes significant
  blanketing, primarily in the B band).
  Due to the low photospheric temperatures, Fe\two\ is the dominant ion and causes numerous
  absorption (and some emission) lines throughout the optical range. This occurs in all solar-metallicity models
  presented [i.e., a large iron abundance is not required to produce significant iron line-blanketing \citep{harkness_etal_87}].
  However, a small iron abundance, as in the low-metallicity model  Smi25mf18p3z0p05, considerably reduces the strength
  of line blanketing. In that model at 6.93\,d, the spectrum shows only
  very weak Fe\two\ lines, in contrast with the dominating C\one/O\one/Na\one/Mg\two/Ca\two\ lines.

  This is a very interesting result which suggests that quantitative spectroscopic analyses of SN spectra may, when
  they have become accurate enough, be a tool for the determination of the SN environmental metallicity. This will require early-time
  observations since it is only at such epochs that we are confident the photosphere is only a probe of the progenitor
  surface layers, uncorrupted by elements synthesised explosively at depth. However, it may be limited to events in which external
  disturbances are negligible (i.e., cases where interaction with the pre-SN wind or with a companion star is weak).

\section{Comparison with observations and discussion}
\label{sect_disc}

  In the preceding section, we have presented a number of spectroscopic
  signatures for ejecta stemming from core-collapse explosions of W-R progenitor stars.
  Let us now discuss the main results and confront them to observations.

\begin{figure}
\epsfig{file=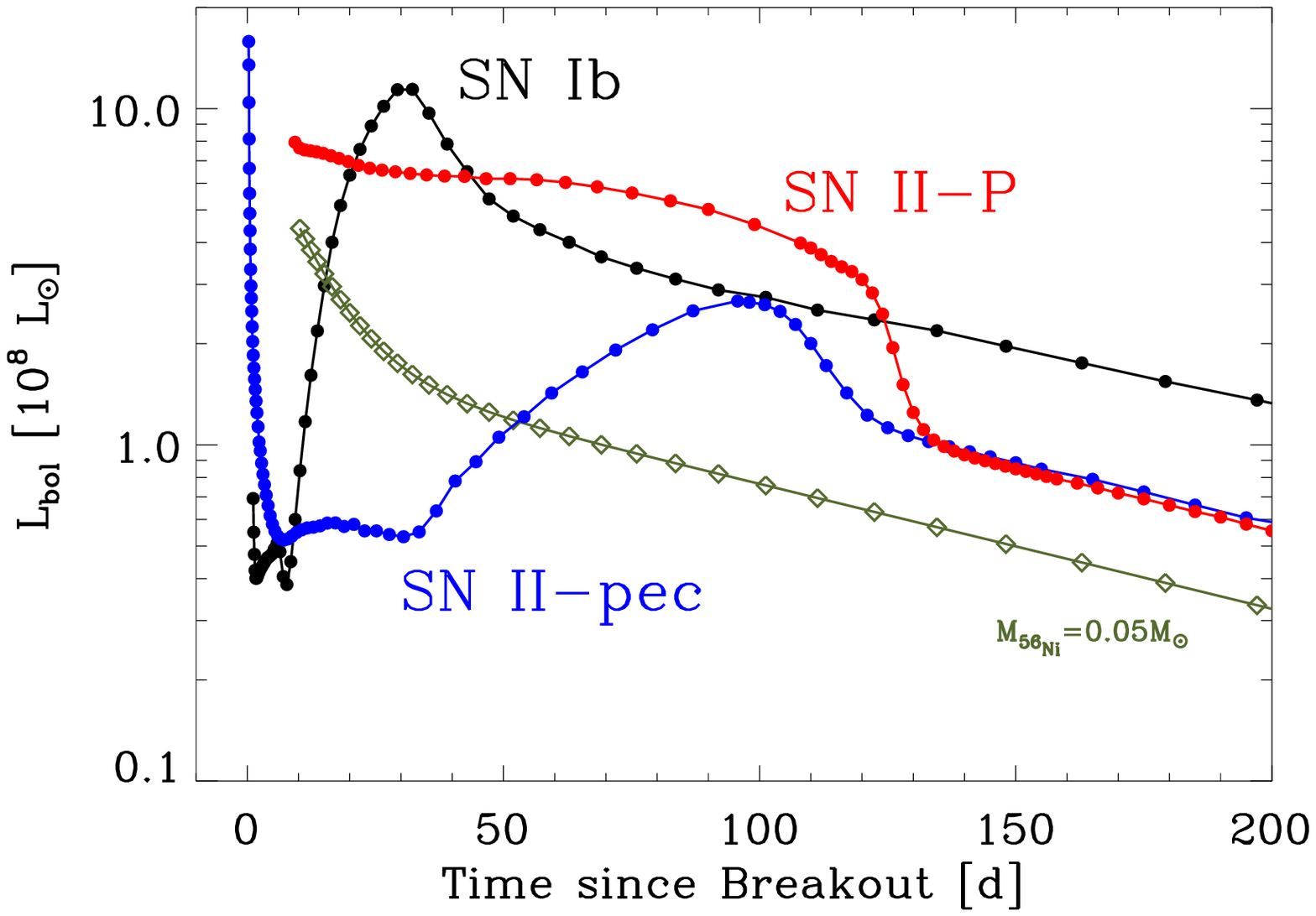,width=8.75cm}
\caption{Comparison of the bolometric-luminosity evolution
of SN Ib model Bmi18mf3p79z1 (0.184\,\msun\ of \isoni), SN II-pec model ``lm18Af" of SN 1987A
(0.084\,\msun\ of \isoni; \citealt{DH10a}), and SN II-P model s15e12 (0.086\,\msun\ of \isoni; \citealt{DH11}).
These three models have a similar ejecta kinetic energy of 1-1.2\,B, and full \gray\ trapping is assumed (it does not
hold for model Bmi18mf3p79z1 past $\gtrsim$70\,d; see Fig.~\ref{fig_lbol_ldecay}).
We also overplot the power associated with the decay of (initially) 0.05\,\msun\ of \isoni\ (olive curve).
All models show an early-time luminosity plateau whose magnitude is function of the progenitor radius
and the H/He abundance ratio, and a late-time luminosity behaviour reflecting the initial amount of \isoni.
The distinct evolution is primarily conditioned by the progenitor-envelope properties, i.e., in particular whether
the progenitor radius is $\sim$10\,\rsun\ (Bmi18mf3p79z1), $\sim$50\,\rsun\ (lm18a7Ad), or $\sim$800\,\rsun\ (s15e12).
\label{fig_lbol_snall}
 }
\end{figure}

\begin{figure}
\epsfig{file=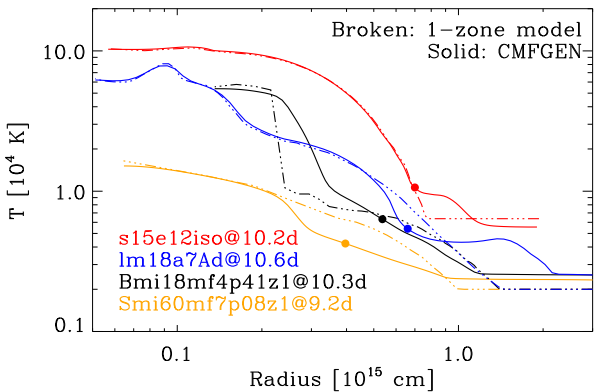,width=8.75cm}
\caption{Ejecta-temperature distribution versus radius at $\sim$10\,d after explosion for the SN Ic model Smi60mf7p08z1,
SN Ib model Bmi18mf4p41z1, SN II-pec model lm18a7Ad \citep{DH10a}, and SN II-P model s15e12iso \citep{DH11}.
We show {\sc cmfgen} results (solid lines; the photosphere location is shown as a dot), as well as results from a separate code
that assumes {\it no diffusion of heat and pure adiabatic cooling} (broken line; our ``1-zone model").
The contrast in brightness between models (Figures~\ref{fig_lbol_all} and \ref{fig_lbol_snall}) is reflected by the larger
radii/temperatures of the corresponding ejecta and epoch, and nearly exclusively reflects modulations in (adiabatic) expansion losses.
\label{fig_temp}
 }
\end{figure}

\subsection{Post-breakout plateau luminosity}

   A generic feature of all synthetic bolometric light curves presented in this work is the
   presence of a post-breakout plateau brightness that typically lasts for $\sim$10\,d (Section~\ref{sect_lbol};
   see also \citealt{EW88}).
   The origin of this plateau is analogous to what is seen in Type II SNe, but differs in brightness because
   of the small progenitor radius, the smaller ejecta mass, and progenitor composition.
   In Fig.~\ref{fig_lbol_snall},  we illustrate the diversity in SN light curves with different ejecta models
   all having 1-1.2\,B kinetic energy. We consider three cases ---
    progenitor stars having a small size ($\sim$12.3\,\rsun; model Bmi18mf4p41z1, this work), a modest size
   ($\sim$50\,\rsun; model ``lm18a7Ad" for SN 1987A from \citealt{DH10a}, but now evolved to nebular times),
   and a large size ($\sim$800\,\rsun;
   model s15e12 of \citealt{DH11}). All show a post-breakout plateau which tends to be brighter for larger
   progenitor objects, and in the range 4$\times$10$^7$ to 6$\times$10$^8$\,\lsun. Its duration is conditioned by the
   progenitor-envelope binding energy, ejecta mass, composition, as well as the mass and distribution of
   \isoni. Here, it varies from a few days in model Bmi18mf4p41z1, to about 20\,d in model ``lm18a7Ad"
   (compatible with the observations of SN 1987A, \citealt{phillips_etal_88}), and about 100\,d in the SN II-P model s15e12.

   This light-curve diversity is also directly connected to the evolution of the ejecta internal energy with time,
which is drastically different between such ejecta. To emphasize the importance of adiabatic cooling through expansion, we have
evolved a number of SN ejecta with a separate program that assumes {\it no diffusion of heat, pure adiabatic cooling},
and treats radioactive-decay energy as a pure local heating source.
In this respect, it evolves each ejecta mass-shell individually (i.e., one-zone model),
updating its energy according to $\Delta e = - p \Delta (1/\rho) + \delta e_{\rm decay}$,
where $e$, $\delta e_{\rm decay}$, $p$, and $\rho$ are the specific energy, the decay energy, the pressure and the density
associated with that mass shell. We solve this equation using the Newton-Raphson technique and employ an
equation of state, developed as part of the work shown in \citet{DLW10a}, which is function of density, temperature, and composition.

As shown in Fig.~\ref{fig_temp} for a post-explosion time of $\sim$10\,d,
the {\it inner-ejecta} temperature obtained with {\sc cmfgen} agrees very well with the one computed
assuming no diffusion and pure adiabatic cooling (this also serves as an independent and additional check on {\sc cmfgen} predictions).
Because of the huge range in progenitor radii, these inner-ejecta temperatures
vary by up to a factor of ten between the \isoni-deficient originally-compact SN Ic model Smi60mf7p08z1 and the originally-extended SN II-P
model s15e12iso \citep{DH11}.
Closer to the photosphere (shown as a dot), the temperature is influenced by non-LTE effects (s15e12iso), radiative cooling
(lm18a7Ad), and the heat wave generated by decay heating at depth (Bmi18mf4p41z1).
The contrast in brightness between models (shown in Figures~\ref{fig_lbol_all} and \ref{fig_lbol_snall}) is reflected by the larger
radii/temperatures of the corresponding model and epoch, and nearly exclusively reflects modulations in (adiabatic) expansion losses.

   The post-breakout plateau is generally only observed in SNe II, and in particular those of the plateau type because
   it is bright and lasts for a long time. In contrast, SNe Ib/c are generally discovered during the re-brightening phase, around peak light,
   or even later, so that any plateau phase taking place earlier is missed, and the shock-breakout time is poorly constrained.
  There are a few exceptions to this. For example, the fading from the breakout phase was observed in
  SN 1993J, which transitioned after a few days to a re-brightening, but by only $\sim$1\,mag \citep{richmond_etal_94}.
  SN 2005bf was a peculiar SN Ib in that it showed a double-peak bolometric light curve. No plateau is visible prior to the first
  peak \citep{folatelli_etal_06} and the explosion time is not known, although this SN was not discovered
  early.\footnote{The first peak shows a rise time and a maximum luminosity that is in fact very comparable to standard SNe Ib.
  The ``anomaly"  with SN 2005bf is the bright second peak, which does not seem to stem
  from decay energy but seems more consistent with the birth of a magnetar \citep{maeda_etal_07}.}
  In contrast, SNe associated with \gray-bursts (GRBs), e.g., SN 1998bw \citep{patat_etal_01} or SN 2010bh \citep{chornock_etal_10b},
  have a well defined explosion time, but in these, the object seems to be always brightening at $\gtrsim$1\,d after the GRB signal.
SN 2008D was caught as the shock broke out of the progenitor star \citep{soderberg_etal_08,chevalier_fransson_08,modjaz_etal_09}.
  By 1\,d after breakout, the luminosity plateaus, but already by 5\,d the SN re-brightens. Interestingly, it brightens by merely 1\,mag
  to reach a peak before fading again.

  The post-breakout plateau that we discuss here seems to have been seen in SN 2008D, but lasts for a few days only  and has a
  a brightness that is merely 1\,mag below the peak.
  This small post-breakout-plateau/peak brightness  contrast is in part caused by the faint light-curve peak, supporting a
  smaller-than-average \isoni\ mass of
  $\sim$0.07\,\msun\ \citep{tanaka_etal_09,drout_etal_10}, compared to $\sim$0.2\,\msun\  in our models.
  The short post-breakout plateau suggests that the ejecta mass is small and/or mixing quite efficient. The huge plateau brightness of
  $\sim$\,1.5$\times$10$^8$\,\lsun\  \citep{soderberg_etal_08, modjaz_etal_09}  is, however, perplexing.
  This plateau brightness is thrice that obtained in our simulations. It cannot be due to a large progenitor radius
  since analysis of the early X-ray light curve \citep{soderberg_etal_08, modjaz_etal_09} supports a progenitor radius
  of $R_{\ast}\lesssim1\,$\rsun. Efficient outward mixing of \isoni\
  would enhance the brightness, but it would likely induce a brightening rather than a plateau, and is therefore not the likely explanation.
Some light contamination could come from the galaxy host, although it appears unlikely
here since SN 2008D was followed up to very late times when the SN was visually much fainter than immediately
after breakout. The contamination could come from the SN light itself, but emitted earlier and scattered
by the surrounding CSM or pre-SN mass loss. The lack of narrow lines from photo-ionised and/or shocked CSM
gas in SN 2008D early-time spectra does not strongly support this. Furthermore, \citet{chevalier_fransson_08}
argued for a low mass rate for the SN 2008D progenitor W-R star.
Following upon the argument that the SN 2008D progenitor star
may be in a binary system, the large early post-breakout luminosity could in part arise from the collision of the SN ejecta with the companion star
\citep{kasen_10}. The same argument could also explain the early light curve of SN 1993J, which in current models requires a low-mass
H-poor and He-rich progenitor with a surprisingly large radius of $\gtrsim$630\,\rsun\ \citep{blinnikov_etal_98}.
Early-time light curves, starting immediately after breakout, may therefore provide an important  clue to the single/binary status
of the progenitor star, and complement independent constraints set by the outer-ejecta composition (Section~\ref{sect_spec}).

  Compared to the progenitor W-R stars, which emit the bulk of the radiation in the UV,
  the post-breakout plateau should be $\gtrsim$10\,mag brighter in the $V$-band (\citealt{wr_cat,crowther_07};
  this estimate accounts for the large bolometric correction of W-R stars). This enormous visual brightening might still be difficult
  to detect if the SN is located in a crowded region or if it falls on the galaxy light. For SNe located at large galactocentric
  radii, it should however be possible, provided the search is deep and performed on a daily cadence.

\subsection{Peak luminosity and width}

  In Nature, \isoni\ mixing within SN IIb/Ib/Ic ejecta will be more efficient  than in our set of
  \isoni-rich models, which assume no mixing at all.
  Our neglect of non-thermal processes will affect the gas ionisation and compromise our synthetic colors
   at peak, which are thus not discussed in this comparison to observations.
   We thus discuss instead the bolometric properties.

    Our \isoni-rich models Bmi18mf3p79z1 and Bmi18mf4p41z1 show a 20-d rise time from the
    post-breakout plateau to  the peak, whose bolometric magnitude is -17.89 and -17.64
    and bolometric luminosity of 1.14$\times$10$^9$ and 9.0$\times$10$^8$\,\lsun, respectively.
    These trends reflect the larger \isoni\ mass in the former model (0.184 compared to 0.170\,\msun) combined
    with the larger ejecta mass in the latter (2.39 compared to 2.91\,\msun). Importantly, \isoni\ is present out to the
    same ejecta velocity in both.

    In contrast, the bolometric light curve for model  Bmi25mf5p09z1 has a longer rise time of 40\,d from the
    end of the post-breakout plateau, with a peak bolometric magnitude (luminosity) of -17.49 (7.9$\times$10$^8$\,\lsun).
    Here, despite the larger \isoni\ mass of 0.24\,\msun, the peak is fainter than for the other two \isoni-rich models
    because of the larger ejecta mass and the significantly deeper location of \isoni, i.e. below 1250\,\kms\ compared
    to below $\gtrsim$2500\,\kms.

    Furthermore, a larger ejecta mass and a deeper location of \isoni\ broadens the peak width significantly.
    Fifteen days after the light curve peak, we obtain a magnitude fading of  0.82, 0.57, and 0.15 for models
    Bmi18mf3p79z1, Bmi18mf4p41z1, and Bmi25mf5p09z1. Modulations in ejecta mass and \isoni\ mixing
    can be interchanged to yield similar results. In a separate set of simulations for these three progenitors
    but characterised by a \isoni\ extent out to 3500\,\kms, all three models yielded similar light curve widths.
    In SNe Ibc light-curve modelling, mixing is usually adjusted freely to obtain a good match but we see here
    that there is a fundamental degeneracy in the problem. This requires an alternate constraint on mixing,
    which can only come from spectra.

    Overall, our results agree with those of \citet{EW88} when adopting similar ejecta characteristics.
    Our main characteristics for models Bmi18mf3p79z1 and Bmi18mf4p41z1 are in good agreement
    with the observations and estimates recently published by \citet{drout_etal_10} and based on a large
    sample of SNe Ibc multi-color light curves (see also \citealt{richardson_etal_02,richardson_etal_06}).

\subsection{Post-peak luminosity fading}
\label{sect_lbol_ldecay}

  The SN post-peak/nebular luminosity is understood to be powered by radioactive decay, and primarily that of \isoco\ at
  the times of interest here (we consider explosions that synthesise primarily \isoni).
  In SNe II, the large ejecta mass of typically 10-15\,\msun\ ensures the full trapping of \grays\ for
  a few hundred days. However, in Type I SNe, the lower ejecta mass is less efficient at trapping \grays, which can escape
  earlier (for a similar phenomenon in the context of SNe Ia, see, e.g., \citealt{hoeflich_khokhlov_96}).
  How much earlier, and how efficiently, are a very important question, since it connects to the ejecta mass and location of \isoni.
  The post-peak/nebular-phase luminosity decline rate represents an important diagnostic for the explosion and the progenitor star.
   Because \grays\ are insensitive to the gas ionisation state and composition (provided only elements heavier than
   helium are present), \gray\ transport is considerably simpler than optical-photon transport and less subject to uncertainty.

  The majority of SNe IIb/Ib/Ic show post-peak luminosity decline rates that are steeper than the expected rate of 0.01\,mag/d
  for full \gray\ trapping \citep{richardson_etal_06}.
  \citet{EW88} invoked the effect of clumping. Departures from sphericity or efficient intrinsic mixing may also propel \isoni\
  further out (see, for example, \citealt{tominaga_etal_05,folatelli_etal_06}), which would then increase \gray\ escape during the
  nebular phase.
  Unfortunately,  the emblematic SN 1987A, for which there is strong evidence of an asymmetric explosion and for which early \gray\
  escape is inferred,
  shows a nebular-luminosity fading rate that exactly matches the 0.01\,mag/d decline rate that implies full trapping \citep{arnett_etal_89}.
  As alluded to above, a systematically low ejecta mass combined with moderate mixing may alone explain the generically fast decline
  rate of SNe IIb/Ib/Ic.

\begin{figure*}
\epsfig{file=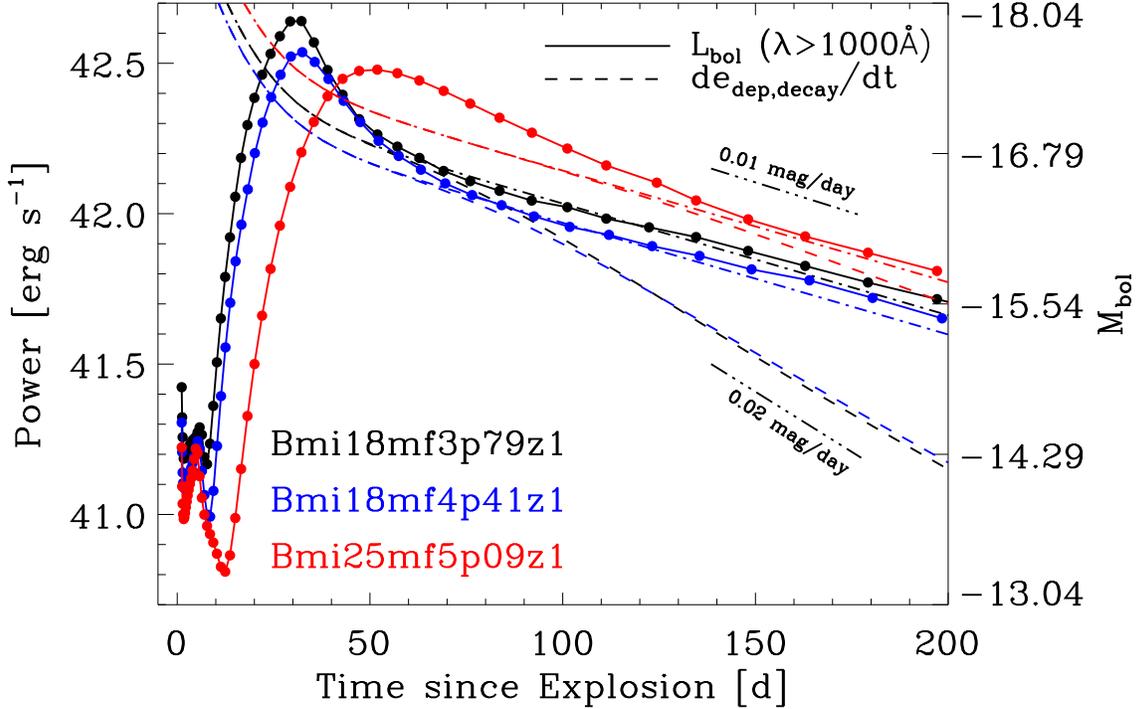,width=15cm}
\caption{Evolution of the bolometric luminosity of our non-LTE time-dependent
radiative-transfer simulations  ($L_{\rm bol}$; solid lines), which assume full-trapping of \gray\ photons, and based on the
\isoni-rich models  Bmi18mf3p79z1(black), Bmi18mf4p41z1 (blue), and Bmi25mf5p09z1 (red).
The dots correspond to the actual epochs computed.
If we had instead used in {\sc cmfgen} the \gray-energy deposition function output by our \gray-Monte-Carlo transport code,
the bolometric luminosity would have turned over at $\sim$70\,d in models Bmi18mf3p79z1 and Bmi25mf4p41z1
and fallen-off more steeply than the
0.01\,mag/d rate for full-trapping of \isoco-decay photons (it would have followed the dashed curve representing the energy
effectively deposited per unit time in the ejecta $de_{\rm dep,decay}/dt$).
The dot-dash line corresponds to the total energy released, in the form of \grays\ and positrons,
by radioactive decay in the Monte Carlo simulations (it is offset by $\lesssim$10\% from the {\sc cmfgen} curves at very late
times because of the numerical diffusion inherent to the necessary regridding at each time step in {\sc cmfgen}).
The smaller the mass overlying the \isoni-production site, the steeper is the luminosity-decline rate.
A low ejecta mass may thus be one way of explaining naturally the rapidly fading light curves of most SNe IIb/Ib/Ic, without
invoking the intervention of clumping, extreme mixing, or jets.
\label{fig_lbol_ldecay}
 }
\end{figure*}

  In the {\sc cmfgen} simulations we have presented so far, \isoni\ decay energy, when included, was fully deposited at the site
  of release in the ejecta. With this full-trapping assumption, our corresponding models fade at a 0.01\,mag/d at nebular times.
  To check how the light curve may be affected by this approximation,
  we have developed a \gray-transport Monte Carlo code that computes the fraction of \grays\ that escape, and of those
  that are trapped, how they distribute their energy within the ejecta (Hillier \& Dessart, in preparation).
  In Fig.~\ref{fig_lbol_ldecay}, we show results based on our \isoni-rich ejecta models Bmi18mf3p79z1, Bmi18mf4p41z1,
  and Bmi25mf5p09z1 (and assuming spherical symmetry).
  We find that the smaller the ejecta mass (for the same ejecta kinetic energy, here of 1.2\,B),
  the earlier the fading rate departs from the rate for full-trapping and the steeper the luminosity fades subsequently.

  In models Bmi18mf3p79z1 and Bmi25mf4p41z1, the post-peak luminosity curve smoothly joins the steep nebular decline, because the
  \grays\ start escaping immediately at the fall off from the peak.\footnote{Our \gray-transport Monte Carlo simulations for SN 1987A
  based on model ``lm18Af" \citep{DH10a} predict the onset of escape for the bulk of \grays\ at $\gtrsim$500\,d after explosion. The steeper
  decline of the SN 1987A bolometric luminosity occurs at that time \citep{arnett_etal_89}, but unfortunately this is concomitant
  with dust formation in the SN 1987A ejecta \citep{colgan_etal_94}, which may be responsible for part of the inferred attenuation.}
  The subsequent fading is very steep, and compatible with the decline rate of SNe Ib/Ic \citep{richardson_etal_06}. However,
  in model Bmi25mf5p09z1, full-trapping persists at the onset of the nebular phase and it is only later that \gray\ escape
  occurs and becomes noticeable. There is thus a point of inflection at 130\,d in the bolometric luminosity when full-trapping ceases
  to hold. Observations do not seem to support such late trapping, but instead suggest at least some moderate level of mixing.
  In model Bmi25mf5p09z1, adopting some mixing of \isoni\ out to 3000\,\kms\ (instead of 1250\,\kms) moves the onset
  of \gray\ leakage to 70\,d after explosion.

   In the observations of SNe 1993J \citep{utrobin_94,woosley_etal_94,young_etal_95},
   1994I \citep{young_etal_95,iwamoto_etal_94}, or 1998bw \citep{woosley_etal_99}, the post-breakout re-brightening
   occurred very soon after breakout,
   the peak was reached very early, the rise and fall from the peak occurred over a short time,
   and the decline rate was faster than for full-trapping. All these independently support
   the notion, which is not new, that SNe IIb/Ib/Ic are associated with low-mass ejecta, and quite generally low-mass
   progenitor stars. However, it may be that rather than invoking extreme clumping or mixing configurations, one may be
   able to accommodate such observations with moderate mixing and low-mass ejecta exclusively. In the context of hypernovae, associated or
   not with a \gray\ signal, the larger ejecta expansion rate reduces the diffusion time and enhances the probability
   of \gray\ escape, so that a larger mass may still be compatible with the observations. In the future, to address this issue,
   we will model simultaneously, in the fashion described here, the spectra and light curves of these rare events.

\subsection{Early-time spectra}

  Our early-time synthetic SEDs look like that of a cool thermal emitter, with a peak distribution around 5000\,\AA, and
  affected by numerous weak/strong lines from H/He/C/N/O and IME as well as forests of lines from blanketing species
  and in particular iron.
  The main results from our work are that hydrogen and/or helium lines are predicted {\it even in the absence of non-thermal
  processes}, or heat diffusion, caused by radioactive decay.
  H\one\ Balmer lines are seen for a hydrogen mass fraction as low as 0.01 in the corresponding ejecta layers, while
  He\one\ lines seem guaranteed only if the helium mass fraction is close to 100\%. During the rise to peak, decay heating
  may be strong enough to produce He\one\ lines, which may persist until the peak of the light curve in spite of the much
  lower helium mass fraction at the photosphere at that time.
  Such early-time observations (spectra and light curves) are critical since during the
  post-breakout plateau, up to $\lesssim$50\% of the ejecta mass may have gone through the photosphere, and thus
  potentially be unaccounted for.

     There is a scarcity of SN IIb/Ib/Ic spectroscopic observations at early times, prior to the post-breakout re-brightening.
Indeed, the first spectra are typically taken during the rise to peak, which is a phase influenced indirectly by radioactive decay
(an outward-diffusing heat wave).  The only exceptions with early-time observations are SNe 1993J (IIb) and 2008D (Ib). Numerous investigations
have studied the light curve or the spectra, but none have performed the full non-LTE time-dependent radiative-transfer simulation
of the ejecta, with the approach presented here. It is critical to do both at the same time, to test that the \isoni\ mixing adopted to fit
the light curve is also compatible with the excitation/ionisation seen in spectra. This has never been done.

   Our models were not tailored to match any specific observation so we cannot be quantitative. Qualitatively speaking,
our model Bmi18mf3p79z1 has some analogy with the early observations of SN 1993J (e.g., the presence of a strong H$\alpha$ line).
We note that the early-time SN 1993J spectra are nearly featureless and rather blue \citep{matheson_etal_00}. As discussed above,
this has been associated in former studies with the explosion of an extended low-mass progenitor in a binary system
\citep{blinnikov_etal_98}. Alternatively, the large early-time
flux could in a large part stem from the collision of the SN ejecta with the identified companion of SN 1993J's progenitor star
\citep{maund_etal_04}, which would call for a significant revision of current models of the event.
In model  Bmi18mf3p79z1, the pre-SN star has a radius of 10\,\rsun\ and the early-time spectra are consequently rather red
(the photosphere is cool and recombined), not blue, and line blanketing is strong. Irrespective of the mismatches with observations,
we do concur with previous investigations that low-hydrogen
mass fractions and low-hydrogen integral mass at the progenitor surface are sufficient to produce a strong (and broad) H$\alpha$ line
in the corresponding early-time SN IIb spectra, as observed in 93J \citep{utrobin_94,woosley_etal_94,young_etal_95,baron_etal_95,
blinnikov_etal_98}.

   None of our models match the observations of SN 2008D during the first week after explosion.
   Given our results for hydrogen, it is clear that
   08D is hydrogen-deficient. However, at early times, the observed spectra are nearly featureless, as for SN 1993J,
while our simulations predict He\one\ or Fe\two\ lines for such a SN Ib event. One striking property of SN 2008D is that
the early post-breakout plateau is about three times as bright as that of SN 1987A and only $\lesssim$1\,mag
fainter than the light-curve peak. SN 2008D is much brighter at those early times than expected for a rather
compact W-R-star progenitor. Together with the nearly featureless spectra, these properties suggest that
the SN light is indeed ``corrupted". As discussed above, an attractive solution to this is emission from the collision of
the SN ejecta with a companion star \citep{kasen_10}.

   At the time of re-brightening in  model Bmi18mf3p79z1, the heat wave causes the appearance of a kink in the H$\alpha$ absorption.
   If stronger, the corresponding change in ionisation could have led to a reversal of the Doppler velocity of maximum absorption in H$\alpha$.
   In model Bmi25mf5p09z1, the effects of the heat wave is very pronounced and causes the photosphere to move out mass, or velocity, space.
   However, it occurs too late to leave any imprint onto He\one\ lines.
   The velocity reversal in the He\one\,5875\,\AA\ line of SN 2005bf \citep{modjaz_etal_09} may stem from a similar circumstance,
   but in this case at a time when the photosphere is close to helium-rich layers. The origin of this unique observation may be the birth of
   a magnetar \citep{maeda_etal_07}. In general, such reversals/kinks/notches are not observed because SNe IIb/Ib/Ic are discovered too late
   (during the rise to the peak), at a time when the photosphere is already in the vicinity of the core. It is a dramatic change in heating/excitation/ionisation
   conditions that permits this reversal, such as when the photosphere starts feeling the energy from decay or, in exceptional circumstances,
   from the magnetar radiation.

\subsection{\isoni-deficient W-R-star explosions}

   Our \isoni-deficient models cannot be compared to observations since all SNe IIb/Ib/Ic are discovered by the
   virtue of being \isoni\ rich. However, this could stem from the current bias for the discovery of brighter objects in bright galaxies,
   making objects devoid of a post-breakout re-brightening impossible to detect. Future deep and un-targeted surveys
   may in fact discover  \isoni-deficient core-collapse SN explosions, and if they do so, it will be either through the breakout
   signal or the post-breakout plateau. As discussed in this paper, the radiative signatures during that phase are conditioned
   by the properties of the progenitor outer envelope. If detected, they would provide important information on the progenitor
   and perhaps clues on what distinguishes progenitors that explode as bright or faint supernovae, despite having
   a standard ejecta-kinetic energy of 1\,B.

\section{Conclusions}
\label{sect_concl}

  We have presented non-LTE time-dependent radiative-transfer simulations of SN ejecta, with and without
  \isoni-decay products, resulting from the core-collapse explosions of single and binary W-R stars
  evolved at solar and sub-solar metallicity.
  Our approach allows the simultaneous computation of the spectra and light curves, and is comparable to
  our earlier studies of SN 1987A \citep{DH10a} and SNe II-P \citep{DH11}.
  We pay particular attention to
  binary-star models for the production of SNe IIb/Ib/Ic, using the physically-consistent calculations
  of \citet{yoon_etal_10}. This work contains about 400 separate calculations performed in separate time sequences,
  each time taking typically 1-2 days of computing, from which we infer the following:

  \begin{enumerate}

\item All our SNe go through a $\sim$10-d-long post-breakout plateau with a luminosity of 1--5$\times$10$^7$\,\lsun,
which is optically $\sim$10\,mag brighter than its progenitor.
In models endowed with 0.17--0.24\,\msun\ of \isoni\ initially, this plateau is followed by a 20-30\,d re-brightening phase up
 to a peak luminosity of 8 to 10$\times$10$^8$\,\lsun.
 Assuming full \gray\ trapping, the nebular-luminosity decline rate of such ejecta agrees with the expected rate
 for \isoco\ decay. However, \gray\ escape in our low-mass ejecta with moderate \isoni\ mixing becomes effective at $\sim$70\,d
 after explosion and lead to a much faster fading rate of $\sim$0.02\,mag/d.

 \item The absence of a sizeable post-breakout plateau, the narrow peak width, and the fast nebular-luminosity
 decline rate of most observed SNe IIb/Ib/Ic light curves may be explained by a revision downward of their associated
 ejecta mass (as well as of the ejecta kinetic energy to retain the same ejecta kinetic energy per unit mass).\footnote{A
 lower mass ejecta may help resolve the discrepant fitting of the SN 1998bw light curve \citep{woosley_etal_99},
 although it might then compromise our current understanding of the progenitors of long-soft GRBs.
 In the future, we will specifically study the case of high-energy W-R star explosions yielding $\sim$10\,B-kinetic-energy ejecta
 to address this issue.}
 This may be more physically justifiable than invoking strong clumping, efficient mixing to the progenitor surface,
 or extreme asymmetry of the explosion in the form of a ``jet". Observing the post-breakout plateau and the post-peak
 luminosity decline rate of all SNe IIb/Ib/Ic is key to further constrain this issue.

\item We find that by the time of re-brightening, up to half the ejecta mass has gone through the photosphere.
Studies on SNe IIb/Ib/Ic based on peak and post-peak observations exclusively are thus subject to major
shortcomings, since they miss important information on the outer-ejecta properties, and in particular the H/He
composition and the total ejecta mass.

\item We present results for our W-R explosions with standard ejecta kinetic-energy but no explosively synthesised \isoni.
These models go through a similar post-breakout plateau but irrevocably fade away as the photosphere recedes
to helium-deficient inner regions, at about 10\,d after explosion.

\item  Hydrogen present with a small mass fraction of $\sim$0.01 and a small cumulative mass of $\gtrsim$0.001\,\msun\
gives rise to H\one\ Balmer lines and in particular strong H$\alpha$ in our WN-star progenitor models over
the period 1-10\,d. As latter times only H$\alpha$ remains. However, it is weak and broad which,
combined with possible overlap by C\two\ and Si\two\ lines, makes its identification difficult.
Observed spectroscopically a few days after shock breakout, these events would eventually be classified as Type IIb SNe (and correspond
to the SNe cIIb of \citealt{chevalier_soderberg_10}), and probably as Type Ib SNe otherwise.
To address this issue, it is critical to obtain early-time spectra,
preferably before the time of re-brightening (which may be caused by the recession of the photosphere to deeper,
possibly H-deficient, ejecta layers). Using spectra taken at the peak of the light curve can hardly help resolve
this problem since the photosphere may be a solar mass below the progenitor surface at that time.

\item Simulations based on binary-star W-R (i.e. WN) progenitor models that have an helium mass fraction
of $\gtrsim$95\% in the outer $\sim$1\,\msun\
of the ejecta show He\one\ lines throughout their post-breakout plateau phase, {\it irrespective of the absence/presence
of  \isoni}. Past the end of the plateau, some models still show He\one\ lines, even though our decay energy is
treated as a pure local heat/thermal source. Non-thermal excitation/ionisation is however key to
strengthen He\one\ lines further, as observed. In general, the He\one\,10830\,\AA\ is stronger, broader, and longer lived
than other He\one\ lines like  He\one\,5875\,\AA.

\item  Simulations based on single-star W-R (i.e. WC/WO) progenitor models that have an helium mass fraction of $\lesssim$50\%
in the outer $\sim$1\,\msun\ of the ejecta do not show He\one\ lines during their post-breakout plateau phase,
but instead lines of C\one, O\one, Na\one, Mg\two, Ca\two, and Fe\two. This may occur
even if helium is present with a total mass of $\sim$1\,\msun, and thus comparable to the models characterised
by a surface helium mass fraction of $\gtrsim$85\,\%.
Early-time spectra are thus an important tool to determine whether the progenitor star is a WN (in a binary system) or a WC/WO (single) star.

\item Our simulations of W-R-star explosions at low metallicity show weaker metal lines, and in particular
those from Fe\two.
Because spectra of SNe IIb/Ib/Ic taken prior to or at the time of re-brightening are unaffected by the explosively nucleo-synthesised
metals, quantitative spectroscopy of early-time SN observations may offer a powerful and accurate means to
directly infer the metallicity of the primordial gas from which the progenitor of the SNt formed.
This would also be a direct way of constraining the effect of metallicity on stellar evolution, by passing alternate
measurements performed on the galaxy environment of the SN \citep{kewley_etal_08,modjaz_etal_08,levesque_etal_10}.

\item SN classification suffers from a great bias due to the time of observation. Many events classified as Ib
could have been IIb's if discovered earlier, when H$\alpha$ appears as a strong P-Cygni profile.
Most SNe IIb/Ib/Ic are detected during the brightening to peak, or sometimes so late that even the peak
is missed. Failing early detection, seeking He\one\,10830\,\AA\ is key to address the presence
of Helium in the progenitor star, and discriminate between a type Ib and a type Ic SN.
We are currently in an embarrassing situation, with SNe Ib (Ic) showing signs of hydrogen (helium).

\item Our set of binary models, systematically characterised by low-mass ejecta, \isoni\ present out to $\sim$2500\,\kms,
and a  very large helium surface abundance,
seem to reproduce the key properties of SNe IIb and Ib.
Our single star models, which give rise to SNe Ic owing to the lack of helium lines in their spectra, have
lower but still significant ($X$(He)$\lesssim$50\%) surface He mass-fractions. Not included in our sample here,
the lowest-mass massive stars in binary systems  are expected to produce pre-SN objects
with very little surface helium, likely producing Type Ic SNe \citep{wellstein_langer_99}.
In this context, SNe IIb and Ib could stem from binary-star evolution, and SNe Ic could stem from both single- and binary-star
evolution with an obvious bimodal distribution in ejecta mass (the recent analysis of \citealt{drout_etal_10} reflects this in part).
In a forthcoming study, we will include such progenitors, as well as allow for the presence of \isoni\ in all ejecta to compare
the light-curve morphology for all resulting SNe.

\item Owing to mass loss through mass transfer to a companion star, massive stars evolving in a binary system
can become W-R stars at a much lower main-sequence mass, perhaps as low as 10\,\msun\
 \citep{vanbeveren_etal_98,yoon_etal_10}.
A single star of the same mass would die as a RSG (it would also give rise to a Type II-P SN).
Because the initial-mass function favors the formation of lower-mass stars, we
expect a large number of low-mass low-luminosity W-R stars.
Paradoxically, most W-R stars are inferred to stem from stars with a main-sequence mass in excess of 25\,\msun,
potentially extending to huge masses of a few 100\,\msun\ \citep{crowther_etal_10},
generically characterised by a large luminosity of 10$^5$--10$^6$\,\lsun\ and large mass-loss rates
of $\sim$10$^{-5}$\msunyr\ \citep{crowther_07}.

One striking example of this mismatch is the hydrogen-rich WN progenitor star corresponding to model Bmi18mf3p79z1.
It has a luminosity of $\sim$70000\,\lsun, a surface temperature of 29000\,K,  an escape velocity  of 380\,\kms,
is at 57\% of the Eddington luminosity, and is expected to have a feeble wind mass loss rate of $\sim$5$\times$10$^{-7}$\,\msunyr.
In contrast, known WNh stars are amongst the most luminous and most massive W-R stars observed today.
Hence, the higher-mass higher-luminosity W-R stars we know today seem to have little in common with
the progenitors of currently observed SNe Ib/c.
This issue is in fact not incompatible with observations. Low-luminosity low-mass W-R stars in binary systems are
sitting next to a bright and more massive companion, to which the primary's envelope was largely transferred.
The W-R luminosity is dwarfed by that of the OB-star companion and is consequently not easily identifiable. Their longer-term
descendants, the Be/X-ray binaries, are substantial evidence of this scenario \citep{wellstein_etal_01,petrovic_etal_05}.

\item If the majority of SNe IIb/Ib/Ic progenitors are from low-mass massive stars in binary systems, the
notion that the sequence II-P $\rightarrow$ II-L $\rightarrow$ IIb $\rightarrow$ Ib $\rightarrow$ Ic
represents {\it one of increasing main-sequence mass and increasing mass loss needs revision}.
Instead, one could understand this trend merely by invoking single- or binary-star evolution.
Namely, stars with a main-sequence mass $\lesssim$20\,\msun\ would make
SNe II-P and II-L if single, and SNe IIb/Ib/Ic if part of a binary system (this does not exclude that
a subset of SNe Ic could come from single massive stars).
Numerous studies have focused on understanding how such massive objects could lose sufficient
amount of mass and produce a successful explosion despite their highly-bound high-density cores.
If most SNe IIb/Ib/Ic stem from lower-mass massive stars, these issues are largely irrelevant.
In this context, high-mass W-R stars would collapse to a black hole without any associated SN or GRB signal.

\item The possibility that most, if not all, SNe IIb/Ib/Ic progenitors stem from low-mass massive stars in binary systems
would modify the scene for the core-collapse explosion mechanism. Such low-mass progenitors have small and
loosely-bound cores that will be no more difficult to explode than those of SNe II-P progenitors. In contrast, if
SNe IIb/Ib/Ic came from high-mass massive stars, a very powerful explosion mechanism, which is presently lacking,
would be needed to successfully eject their massive highly-bound envelopes. If all core-collapse SNe come from low-mass
massive stars, either single of binary, the challenge for the explosion mechanism is to be viable for
10-20\,\msun\ main-sequence mass stars, something that may be more easily attained.
Such SNe would also experience little or no fallback, corroborating
the observation that low-luminosity \isoni-poor SNe II-P come from 8-10\,\msun\ massive stars
(e.g., SN 2005cs, \citealt{maund_etal_05,li_etal_06}; SN 2008bk, \citealt{mattila_etal_08}).
Such binary-star progenitors  would thus eventually make an important contribution toward the oxygen enrichment of
the interstellar medium.

\end{enumerate}

   Despite the recent observational efforts, our view of SNe IIb/Ib/Ic (and Ia) is compromised by the lack of
   early-time {\it spectroscopic and photometric} observations, many days prior to peak when the SN
   is faint and the photosphere still resides
   in the outer layers of the progenitor star. Such observations would offer useful constraints on the
   progenitor surface composition, the metallicity of the progenitor molecular cloud, and consolidate
   the determination of the ejecta mass.
   Forthcoming blind deep full-sky survey will allow us to resolve these biases, by capturing a near-complete
   sample of stellar explosions covering from a very early post-explosion time up to late times.

Non-LTE time-dependent radiative-transfer modelling of the type presented here
allows the simultaneous computation of spectra
and light curves with the same high level of physical consistency. In the future, with the added treatment
of heating/excitation/ionisation from non-thermal electrons, this approach will permit a better determination
of the level of \isoni-mixing in SN ejecta and a more robust modelling of SNe IIb/Ib/Ic, concerning both
the progenitor as well as the explosion properties.

\section*{Acknowledgments}

LD acknowledges financial support from the European Community through an
International Re-integration Grant, under grant number PIRG04-GA-2008-239184.
DJH acknowledges support from STScI theory grant HST-AR-11756.01.A and NASA theory grant NNX10AC80G.
At UCSC this work was supported by NASA (NNX09AK36G) and the DOE SciDAC Program under contract DE-FC02-06ER41438.
Calculations presented in this work were performed in part at the French National Super-computing Centre  (CINES)
on the Altix ICE JADE machine.

\label{lastpage}


\begin{thebibliography}{101}
\expandafter\ifx\csname natexlab\endcsname\relax\def\natexlab#1{#1}\fi

\bibitem[{{Arnett} {et~al.}(1989){Arnett}, {Bahcall}, {Kirshner}, \&
  {Woosley}}]{arnett_etal_89}
{Arnett}, W.~D., {Bahcall}, J.~N., {Kirshner}, R.~P., \& {Woosley}, S.~E. 1989,
  \araa, 27, 629

\bibitem[{{Baron} {et~al.}(1995){Baron}, {Hauschildt}, {Branch}, {Austin},
  {Garnavich}, {Ann}, {Wagner}, {Filippenko}, {Matheson}, \&
  {Liebert}}]{baron_etal_95}
{Baron}, E., {Hauschildt}, P.~H., {Branch}, D., {Austin}, S., {Garnavich}, P.,
  {Ann}, H.~B., {Wagner}, R.~M., {Filippenko}, A.~V., {Matheson}, T., \&
  {Liebert}, J. 1995, \apj, 441, 170

\bibitem[{{Blinnikov} {et~al.}(1998){Blinnikov}, {Eastman}, {Bartunov},
  {Popolitov}, \& {Woosley}}]{blinnikov_etal_98}
{Blinnikov}, S.~I., {Eastman}, R., {Bartunov}, O.~S., {Popolitov}, V.~A., \&
  {Woosley}, S.~E. 1998, \apj, 496, 454

\bibitem[{{Blondin} {et~al.}(2008){Blondin}, {Filippenko}, {Foley}, {Li},
  {Dessart}, \& {Vaz}}]{blondin_etal_08}
{Blondin}, S., {Filippenko}, A.~V., {Foley}, R.~J., {Li}, W., {Dessart}, L., \&
  {Vaz}, A. 2008, Central Bureau Electronic Telegrams, 1285, 1

\bibitem[{{Bouret} {et~al.}(2005){Bouret}, {Lanz}, \&
  {Hillier}}]{bouret_etal_05}
{Bouret}, J., {Lanz}, T., \& {Hillier}, D.~J. 2005, \aap, 438, 301

\bibitem[{{Branch} {et~al.}(2002){Branch}, {Benetti}, {Kasen}, {Baron},
  {Jeffery}, {Hatano}, {Stathakis}, {Filippenko}, {Matheson}, {Pastorello},
  {Altavilla}, {Cappellaro}, {Rizzi}, {Turatto}, {Li}, {Leonard}, \&
  {Shields}}]{branch_etal_02}
{Branch}, D., {Benetti}, S., {Kasen}, D., {Baron}, E., {Jeffery}, D.~J.,
  {Hatano}, K., {Stathakis}, R.~A., {Filippenko}, A.~V., {Matheson}, T.,
  {Pastorello}, A., {Altavilla}, G., {Cappellaro}, E., {Rizzi}, L., {Turatto},
  M., {Li}, W., {Leonard}, D.~C., \& {Shields}, J.~C. 2002, \apj, 566, 1005

\bibitem[{{Branch} {et~al.}(2006){Branch}, {Jeffery}, {Young}, \&
  {Baron}}]{branch_etal_06}
{Branch}, D., {Jeffery}, D.~J., {Young}, T.~R., \& {Baron}, E. 2006, \pasp,
  118, 791

\bibitem[{{Burrows} {et~al.}(2007){Burrows}, {Livne}, {Dessart}, {Ott}, \&
  {Murphy}}]{burrows_etal_07a}
{Burrows}, A., {Livne}, E., {Dessart}, L., {Ott}, C.~D., \& {Murphy}, J. 2007,
  \apj, 655, 416

\bibitem[{{Castor} {et~al.}(1975){Castor}, {Abbott}, \& {Klein}}]{cak}
{Castor}, J.~I., {Abbott}, D.~C., \& {Klein}, R.~I. 1975, \apj, 195, 157

\bibitem[{{Chevalier} \& {Fransson}(2008)}]{chevalier_fransson_08}
{Chevalier}, R.~A. \& {Fransson}, C. 2008, \apjl, 683, L135

\bibitem[{{Chevalier} \& {Soderberg}(2010)}]{chevalier_soderberg_10}
{Chevalier}, R.~A. \& {Soderberg}, A.~M. 2010, \apjl, 711, L40

\bibitem[{{Chornock} {et~al.}(2010){Chornock}, {Berger}, {Levesque},
  {Soderberg}, {Foley}, {Fox}, {Frebel}, {Simon}, {Bochanski}, {Challis},
  {Kirshner}, {Podsiadlowski}, {Roth}, {Rutledge}, {Schmidt}, {Sheppard}, \&
  {Simcoe}}]{chornock_etal_10b}
{Chornock}, R., {Berger}, E., {Levesque}, E.~M., {Soderberg}, A.~M., {Foley},
  R.~J., {Fox}, D.~B., {Frebel}, A., {Simon}, J.~D., {Bochanski}, J.~J.,
  {Challis}, P.~J., {Kirshner}, R.~P., {Podsiadlowski}, P., {Roth}, K.,
  {Rutledge}, R.~E., {Schmidt}, B.~P., {Sheppard}, S.~S., \& {Simcoe}, R.~A.
  2010, ArXiv e-prints

\bibitem[{{Chornock} {et~al.}(2008){Chornock}, {Filippenko}, {Li}, {Foley},
  {Stockton}, {Moran}, {Hodge}, \& {Merriman}}]{chornock_etal_08}
{Chornock}, R., {Filippenko}, A.~V., {Li}, W., {Foley}, R.~J., {Stockton}, A.,
  {Moran}, E.~C., {Hodge}, J., \& {Merriman}, K. 2008, Central Bureau
  Electronic Telegrams, 1298, 1

\bibitem[{{Colgan} {et~al.}(1994){Colgan}, {Haas}, {Erickson}, {Lord}, \&
  {Hollenbach}}]{colgan_etal_94}
{Colgan}, S.~W.~J., {Haas}, M.~R., {Erickson}, E.~F., {Lord}, S.~D., \&
  {Hollenbach}, D.~J. 1994, \apj, 427, 874

\bibitem[{{Conti}(1976)}]{conti_76}
{Conti}, P.~S. 1976, Memoires of the Societe Royale des Sciences de Liege, 9,
  193

\bibitem[{{Crowther}(2007)}]{crowther_07}
{Crowther}, P.~A. 2007, \araa, 45, 177

\bibitem[{{Crowther} {et~al.}(2002){Crowther}, {Dessart}, {Hillier}, {Abbott},
  \& {Fullerton}}]{crowther_etal_02}
{Crowther}, P.~A., {Dessart}, L., {Hillier}, D.~J., {Abbott}, J.~B., \&
  {Fullerton}, A.~W. 2002, \aap, 392, 653

\bibitem[{{Crowther} {et~al.}(2010){Crowther}, {Schnurr}, {Hirschi}, {Yusof},
  {Parker}, {Goodwin}, \& {Kassim}}]{crowther_etal_10}
{Crowther}, P.~A., {Schnurr}, O., {Hirschi}, R., {Yusof}, N., {Parker}, R.~J.,
  {Goodwin}, S.~P., \& {Kassim}, H.~A. 2010, \mnras, 408, 731

\bibitem[{{Crowther} {et~al.}(1995){Crowther}, {Smith}, {Hillier}, \&
  {Schmutz}}]{crowther_etal_95}
{Crowther}, P.~A., {Smith}, L.~J., {Hillier}, D.~J., \& {Schmutz}, W. 1995,
  \aap, 293, 427

\bibitem[{{Deng} {et~al.}(2000){Deng}, {Qiu}, {Hu}, {Hatano}, \&
  {Branch}}]{deng_etal_00}
{Deng}, J.~S., {Qiu}, Y.~L., {Hu}, J.~Y., {Hatano}, K., \& {Branch}, D. 2000,
  \apj, 540, 452

\bibitem[{{Dessart} {et~al.}(2008){Dessart}, {Blondin}, {Brown}, {Hicken},
  {Hillier}, {Holland}, {Immler}, {Kirshner}, {Milne}, {Modjaz}, \&
  {Roming}}]{dessart_etal_08}
{Dessart}, L., {Blondin}, S., {Brown}, P.~J., {Hicken}, M., {Hillier}, D.~J.,
  {Holland}, S.~T., {Immler}, S., {Kirshner}, R.~P., {Milne}, P., {Modjaz}, M.,
  \& {Roming}, P.~W.~A. 2008, \apj, 675, 644

\bibitem[{{Dessart} {et~al.}(2000){Dessart}, {Crowther}, {Hillier}, {Willis},
  {Morris}, \& {van der Hucht}}]{DCH00_WC_neon}
{Dessart}, L., {Crowther}, P.~A., {Hillier}, D.~J., {Willis}, A.~J., {Morris},
  P.~W., \& {van der Hucht}, K.~A. 2000, \mnras, 315, 407

\bibitem[{{Dessart} \& {Hillier}(2008)}]{DH08_time}
{Dessart}, L. \& {Hillier}, D.~J. 2008, \mnras, 383, 57

\bibitem[{{Dessart} \& {Hillier}(2010)}]{DH10a}
---. 2010, \mnras, 405, 2141

\bibitem[{{Dessart} \& {Hillier}(2011)}]{DH11}
---. 2011, \mnras, 410, 1739

\bibitem[{{Dessart} {et~al.}(2010{\natexlab{a}}){Dessart}, {Livne}, \&
  {Waldman}}]{DLW10b}
{Dessart}, L., {Livne}, E., \& {Waldman}, R. 2010{\natexlab{a}}, \mnras, 408,
  827

\bibitem[{{Dessart} {et~al.}(2010{\natexlab{b}}){Dessart}, {Livne}, \&
  {Waldman}}]{DLW10a}
---. 2010{\natexlab{b}}, \mnras, 405, 2113

\bibitem[{{Drout} {et~al.}(2010){Drout}, {Soderberg}, {Gal-Yam}, {Cenko},
  {Fox}, {Leonard}, {Sand}, {Moon}, {Arcavi}, \& {Green}}]{drout_etal_10}
{Drout}, M.~R., {Soderberg}, A.~M., {Gal-Yam}, A., {Cenko}, S.~B., {Fox},
  D.~B., {Leonard}, D.~C., {Sand}, D.~J., {Moon}, D., {Arcavi}, I., \& {Green},
  Y. 2010, ArXiv:1011.4959

\bibitem[{{Eastman} \& {Kirshner}(1989)}]{EK89_87A}
{Eastman}, R.~G. \& {Kirshner}, R.~P. 1989, \apj, 347, 771

\bibitem[{{Ensman} \& {Woosley}(1988)}]{EW88}
{Ensman}, L.~M. \& {Woosley}, S.~E. 1988, \apj, 333, 754

\bibitem[{{Folatelli} {et~al.}(2006){Folatelli}, {Contreras}, {Phillips},
  {Woosley}, {Blinnikov}, {Morrell}, {Suntzeff}, {Lee}, {Hamuy},
  {Gonz{\'a}lez}, {Krzeminski}, {Roth}, {Li}, {Filippenko}, {Foley},
  {Freedman}, {Madore}, {Persson}, {Murphy}, {Boissier}, {Galaz},
  {Gonz{\'a}lez}, {McCarthy}, {McWilliam}, \& {Pych}}]{folatelli_etal_06}
{Folatelli}, G., {Contreras}, C., {Phillips}, M.~M., {Woosley}, S.~E.,
  {Blinnikov}, S., {Morrell}, N., {Suntzeff}, N.~B., {Lee}, B.~L., {Hamuy}, M.,
  {Gonz{\'a}lez}, S., {Krzeminski}, W., {Roth}, M., {Li}, W., {Filippenko},
  A.~V., {Foley}, R.~J., {Freedman}, W.~L., {Madore}, B.~F., {Persson}, S.~E.,
  {Murphy}, D., {Boissier}, S., {Galaz}, G., {Gonz{\'a}lez}, L., {McCarthy},
  P.~J., {McWilliam}, A., \& {Pych}, W. 2006, \apj, 641, 1039

\bibitem[{{Fryer} {et~al.}(2007){Fryer}, {Mazzali}, {Prochaska}, {Cappellaro},
  {Panaitescu}, {Berger}, {van Putten}, {van den Heuvel}, {Young},
  {Hungerford}, {Rockefeller}, {Yoon}, {Podsiadlowski}, {Nomoto}, {Chevalier},
  {Schmidt}, \& {Kulkarni}}]{fryer_etal_07}
{Fryer}, C.~L., {Mazzali}, P.~A., {Prochaska}, J., {Cappellaro}, E.,
  {Panaitescu}, A., {Berger}, E., {van Putten}, M., {van den Heuvel}, E.~P.~J.,
  {Young}, P., {Hungerford}, A., {Rockefeller}, G., {Yoon}, S.,
  {Podsiadlowski}, P., {Nomoto}, K., {Chevalier}, R., {Schmidt}, B., \&
  {Kulkarni}, S. 2007, \pasp, 119, 1211

\bibitem[{{Georgy} {et~al.}(2009){Georgy}, {Meynet}, {Walder}, {Folini}, \&
  {Maeder}}]{georgy_etal_09}
{Georgy}, C., {Meynet}, G., {Walder}, R., {Folini}, D., \& {Maeder}, A. 2009,
  \aap, 502, 611

\bibitem[{{Guzik}(2005)}]{guzik_05}
{Guzik}, J.~A. 2005, in Astronomical Society of the Pacific Conference Series,
  Vol. 332, The Fate of the Most Massive Stars, ed. {R.~Humphreys \&
  K.~Stanek}, 204--+

\bibitem[{{Harkness} {et~al.}(1987){Harkness}, {Wheeler}, {Margon}, {Downes},
  {Kirshner}, {Uomoto}, {Barker}, {Cochran}, {Dinerstein}, {Garnett}, \&
  {Levreault}}]{harkness_etal_87}
{Harkness}, R.~P., {Wheeler}, J.~C., {Margon}, B., {Downes}, R.~A., {Kirshner},
  R.~P., {Uomoto}, A., {Barker}, E.~S., {Cochran}, A.~L., {Dinerstein}, H.~L.,
  {Garnett}, D.~R., \& {Levreault}, R.~M. 1987, \apj, 317, 355

\bibitem[{{Heger} {et~al.}(2003){Heger}, {Fryer}, {Woosley}, {Langer}, \&
  {Hartmann}}]{heger_etal_03}
{Heger}, A., {Fryer}, C.~L., {Woosley}, S.~E., {Langer}, N., \& {Hartmann},
  D.~H. 2003, \apj, 591, 288

\bibitem[{{Hillier} \& {Miller}(1998)}]{HM98_lb}
{Hillier}, D.~J. \& {Miller}, D.~L. 1998, \apj, 496, 407

\bibitem[{{Hoeflich} \& {Khokhlov}(1996)}]{hoeflich_khokhlov_96}
{Hoeflich}, P. \& {Khokhlov}, A. 1996, \apj, 457, 500

\bibitem[{{Iwamoto} {et~al.}(1994){Iwamoto}, {Nomoto}, {Hoflich}, {Yamaoka},
  {Kumagai}, \& {Shigeyama}}]{iwamoto_etal_94}
{Iwamoto}, K., {Nomoto}, K., {Hoflich}, P., {Yamaoka}, H., {Kumagai}, S., \&
  {Shigeyama}, T. 1994, \apjl, 437, L115

\bibitem[{{James} \& {Baron}(2010)}]{james_baron_10}
{James}, S. \& {Baron}, E. 2010, \apj, 718, 957

\bibitem[{{Jeffery} {et~al.}(1991){Jeffery}, {Branch}, {Filippenko}, \&
  {Nomoto}}]{jeffery_etal_91}
{Jeffery}, D.~J., {Branch}, D., {Filippenko}, A.~V., \& {Nomoto}, K. 1991,
  \apjl, 377, L89

\bibitem[{{Kasen}(2010)}]{kasen_10}
{Kasen}, D. 2010, \apj, 708, 1025

\bibitem[{{Ketchum} {et~al.}(2008){Ketchum}, {Baron}, \&
  {Branch}}]{ketchum_etal_08}
{Ketchum}, W., {Baron}, E., \& {Branch}, D. 2008, \apj, 674, 371

\bibitem[{{Kewley} \& {Ellison}(2008)}]{kewley_etal_08}
{Kewley}, L.~J. \& {Ellison}, S.~L. 2008, \apj, 681, 1183

\bibitem[{{Kingsburgh} {et~al.}(1995){Kingsburgh}, {Barlow}, \&
  {Storey}}]{kingsburgh_etal_95}
{Kingsburgh}, R.~L., {Barlow}, M.~J., \& {Storey}, P.~J. 1995, \aap, 295, 75

\bibitem[{{Kozma} \& {Fransson}(1992)}]{KF92}
{Kozma}, C. \& {Fransson}, C. 1992, \apj, 390, 602

\bibitem[{{Kozma} \& {Fransson}(1998{\natexlab{a}})}]{KF98a}
---. 1998{\natexlab{a}}, \apj, 496, 946

\bibitem[{{Kozma} \& {Fransson}(1998{\natexlab{b}})}]{KF98b}
---. 1998{\natexlab{b}}, \apj, 497, 431

\bibitem[{{Langer} {et~al.}(1994){Langer}, {Hamann}, {Lennon}, {Najarro},
  {Pauldrach}, \& {Puls}}]{langer_etal_94}
{Langer}, N., {Hamann}, W., {Lennon}, M., {Najarro}, F., {Pauldrach}, A.~W.~A.,
  \& {Puls}, J. 1994, \aap, 290, 819

\bibitem[{{Leonard}(2010)}]{leonard_10}
{Leonard}, D.~C. 2010, in Astronomical Society of the Pacific Conference
  Series, Vol. 425, Astronomical Society of the Pacific Conference Series, ed.
  {C.~Leither, P.~Bennet, P.~Morris, \& J.~van Loon}, 79--+

\bibitem[{{Levesque} {et~al.}(2010){Levesque}, {Berger}, {Kewley}, \&
  {Bagley}}]{levesque_etal_10}
{Levesque}, E.~M., {Berger}, E., {Kewley}, L.~J., \& {Bagley}, M.~M. 2010, \aj,
  139, 694

\bibitem[{{Li} {et~al.}(2006){Li}, {Van Dyk}, {Filippenko}, {Cuillandre},
  {Jha}, {Bloom}, {Riess}, \& {Livio}}]{li_etal_06}
{Li}, W., {Van Dyk}, S.~D., {Filippenko}, A.~V., {Cuillandre}, J., {Jha}, S.,
  {Bloom}, J.~S., {Riess}, A.~G., \& {Livio}, M. 2006, \apj, 641, 1060

\bibitem[{{Livne}(1993)}]{livne_93}
{Livne}, E. 1993, \apj, 412, 634

\bibitem[{{Lucy}(1991)}]{lucy_91}
{Lucy}, L.~B. 1991, \apj, 383, 308

\bibitem[{{Maeda} {et~al.}(2007){Maeda}, {Tanaka}, {Nomoto}, {Tominaga},
  {Kawabata}, {Mazzali}, {Umeda}, {Suzuki}, \& {Hattori}}]{maeda_etal_07}
{Maeda}, K., {Tanaka}, M., {Nomoto}, K., {Tominaga}, N., {Kawabata}, K.,
  {Mazzali}, P.~A., {Umeda}, H., {Suzuki}, T., \& {Hattori}, T. 2007, \apj,
  666, 1069

\bibitem[{{Maeder} \& {Meynet}(1994)}]{maeder_meynet_94}
{Maeder}, A. \& {Meynet}, G. 1994, \aap, 287, 803

\bibitem[{{Maeder} \& {Meynet}(2000)}]{maeder_meynet_00}
---. 2000, \aap, 361, 159

\bibitem[{{Matheson} {et~al.}(2000){Matheson}, {Filippenko}, {Barth}, {Ho},
  {Leonard}, {Bershady}, {Davis}, {Finley}, {Fisher}, {Gonz{\'a}lez}, {Hawley},
  {Koo}, {Li}, {Lonsdale}, {Schlegel}, {Smith}, {Spinrad}, \&
  {Wirth}}]{matheson_etal_00}
{Matheson}, T., {Filippenko}, A.~V., {Barth}, A.~J., {Ho}, L.~C., {Leonard},
  D.~C., {Bershady}, M.~A., {Davis}, M., {Finley}, D.~S., {Fisher}, D.,
  {Gonz{\'a}lez}, R.~A., {Hawley}, S.~L., {Koo}, D.~C., {Li}, W., {Lonsdale},
  C.~J., {Schlegel}, D., {Smith}, H.~E., {Spinrad}, H., \& {Wirth}, G.~D. 2000,
  \aj, 120, 1487

\bibitem[{{Mattila} {et~al.}(2008){Mattila}, {Smartt}, {Eldridge}, {Maund},
  {Crockett}, \& {Danziger}}]{mattila_etal_08}
{Mattila}, S., {Smartt}, S.~J., {Eldridge}, J.~J., {Maund}, J.~R., {Crockett},
  R.~M., \& {Danziger}, I.~J. 2008, \apjl, 688, L91

\bibitem[{{Maund} {et~al.}(2005){Maund}, {Smartt}, \&
  {Danziger}}]{maund_etal_05}
{Maund}, J.~R., {Smartt}, S.~J., \& {Danziger}, I.~J. 2005, \mnras, 364, L33

\bibitem[{{Maund} {et~al.}(2004){Maund}, {Smartt}, {Kudritzki},
  {Podsiadlowski}, \& {Gilmore}}]{maund_etal_04}
{Maund}, J.~R., {Smartt}, S.~J., {Kudritzki}, R.~P., {Podsiadlowski}, P., \&
  {Gilmore}, G.~F. 2004, \nat, 427, 129

\bibitem[{{Millard} {et~al.}(1999){Millard}, {Branch}, {Baron}, {Hatano},
  {Fisher}, {Filippenko}, {Kirshner}, {Challis}, {Fransson}, {Panagia},
  {Phillips}, {Sonneborn}, {Suntzeff}, {Wagoner}, \&
  {Wheeler}}]{millard_etal_99}
{Millard}, J., {Branch}, D., {Baron}, E., {Hatano}, K., {Fisher}, A.,
  {Filippenko}, A.~V., {Kirshner}, R.~P., {Challis}, P.~M., {Fransson}, C.,
  {Panagia}, N., {Phillips}, M.~M., {Sonneborn}, G., {Suntzeff}, N.~B.,
  {Wagoner}, R.~V., \& {Wheeler}, J.~C. 1999, \apj, 527, 746

\bibitem[{{Modjaz} {et~al.}(2008){Modjaz}, {Kewley}, {Kirshner}, {Stanek},
  {Challis}, {Garnavich}, {Greene}, {Kelly}, \& {Prieto}}]{modjaz_etal_08}
{Modjaz}, M., {Kewley}, L., {Kirshner}, R.~P., {Stanek}, K.~Z., {Challis}, P.,
  {Garnavich}, P.~M., {Greene}, J.~E., {Kelly}, P.~L., \& {Prieto}, J.~L. 2008,
  \aj, 135, 1136

\bibitem[{{Modjaz} {et~al.}(2009){Modjaz}, {Li}, {Butler}, {Chornock},
  {Perley}, {Blondin}, {Bloom}, {Filippenko}, {Kirshner}, {Kocevski},
  {Poznanski}, {Hicken}, {Foley}, {Stringfellow}, {Berlind}, {Barrado y
  Navascues}, {Blake}, {Bouy}, {Brown}, {Challis}, {Chen}, {de Vries},
  {Dufour}, {Falco}, {Friedman}, {Ganeshalingam}, {Garnavich}, {Holden},
  {Illingworth}, {Lee}, {Liebert}, {Marion}, {Olivier}, {Prochaska},
  {Silverman}, {Smith}, {Starr}, {Steele}, {Stockton}, {Williams}, \&
  {Wood-Vasey}}]{modjaz_etal_09}
{Modjaz}, M., {Li}, W., {Butler}, N., {Chornock}, R., {Perley}, D., {Blondin},
  S., {Bloom}, J.~S., {Filippenko}, A.~V., {Kirshner}, R.~P., {Kocevski}, D.,
  {Poznanski}, D., {Hicken}, M., {Foley}, R.~J., {Stringfellow}, G.~S.,
  {Berlind}, P., {Barrado y Navascues}, D., {Blake}, C.~H., {Bouy}, H.,
  {Brown}, W.~R., {Challis}, P., {Chen}, H., {de Vries}, W.~H., {Dufour}, P.,
  {Falco}, E., {Friedman}, A., {Ganeshalingam}, M., {Garnavich}, P., {Holden},
  B., {Illingworth}, G., {Lee}, N., {Liebert}, J., {Marion}, G.~H., {Olivier},
  S.~S., {Prochaska}, J.~X., {Silverman}, J.~M., {Smith}, N., {Starr}, D.,
  {Steele}, T.~N., {Stockton}, A., {Williams}, G.~G., \& {Wood-Vasey}, W.~M.
  2009, \apj, 702, 226

\bibitem[{{Nomoto} {et~al.}(1995){Nomoto}, {Iwamoto}, \&
  {Suzuki}}]{nomoto_etal_95}
{Nomoto}, K.~I., {Iwamoto}, K., \& {Suzuki}, T. 1995, \physrep, 256, 173

\bibitem[{{Owocki} {et~al.}(2004){Owocki}, {Gayley}, \&
  {Shaviv}}]{owocki_etal_04}
{Owocki}, S.~P., {Gayley}, K.~G., \& {Shaviv}, N.~J. 2004, \apj, 616, 525

\bibitem[{{Parrent} {et~al.}(2007){Parrent}, {Branch}, {Troxel}, {Casebeer},
  {Jeffery}, {Ketchum}, {Baron}, {Serduke}, \& {Filippenko}}]{parrent_etal_07}
{Parrent}, J., {Branch}, D., {Troxel}, M.~A., {Casebeer}, D., {Jeffery}, D.~J.,
  {Ketchum}, W., {Baron}, E., {Serduke}, F.~J.~D., \& {Filippenko}, A.~V. 2007,
  \pasp, 119, 135

\bibitem[{{Patat} {et~al.}(2001){Patat}, {Cappellaro}, {Danziger}, {Mazzali},
  {Sollerman}, {Augusteijn}, {Brewer}, {Doublier}, {Gonzalez}, {Hainaut},
  {Lidman}, {Leibundgut}, {Nomoto}, {Nakamura}, {Spyromilio}, {Rizzi},
  {Turatto}, {Walsh}, {Galama}, {van Paradijs}, {Kouveliotou}, {Vreeswijk},
  {Frontera}, {Masetti}, {Palazzi}, \& {Pian}}]{patat_etal_01}
{Patat}, F., {Cappellaro}, E., {Danziger}, J., {Mazzali}, P.~A., {Sollerman},
  J., {Augusteijn}, T., {Brewer}, J., {Doublier}, V., {Gonzalez}, J.~F.,
  {Hainaut}, O., {Lidman}, C., {Leibundgut}, B., {Nomoto}, K., {Nakamura}, T.,
  {Spyromilio}, J., {Rizzi}, L., {Turatto}, M., {Walsh}, J., {Galama}, T.~J.,
  {van Paradijs}, J., {Kouveliotou}, C., {Vreeswijk}, P.~M., {Frontera}, F.,
  {Masetti}, N., {Palazzi}, E., \& {Pian}, E. 2001, \apj, 555, 900

\bibitem[{{Petrovic} {et~al.}(2005){Petrovic}, {Langer}, \& {van der
  Hucht}}]{petrovic_etal_05}
{Petrovic}, J., {Langer}, N., \& {van der Hucht}, K.~A. 2005, \aap, 435, 1013

\bibitem[{{Phillips} {et~al.}(1988){Phillips}, {Heathcote}, {Hamuy}, \&
  {Navarrete}}]{phillips_etal_88}
{Phillips}, M.~M., {Heathcote}, S.~R., {Hamuy}, M., \& {Navarrete}, M. 1988,
  \aj, 95, 1087

\bibitem[{{Richardson} {et~al.}(2006){Richardson}, {Branch}, \&
  {Baron}}]{richardson_etal_06}
{Richardson}, D., {Branch}, D., \& {Baron}, E. 2006, \aj, 131, 2233

\bibitem[{{Richardson} {et~al.}(2002){Richardson}, {Branch}, {Casebeer},
  {Millard}, {Thomas}, \& {Baron}}]{richardson_etal_02}
{Richardson}, D., {Branch}, D., {Casebeer}, D., {Millard}, J., {Thomas}, R.~C.,
  \& {Baron}, E. 2002, \aj, 123, 745

\bibitem[{{Richmond} {et~al.}(1994){Richmond}, {Treffers}, {Filippenko},
  {Paik}, {Leibundgut}, {Schulman}, \& {Cox}}]{richmond_etal_94}
{Richmond}, M.~W., {Treffers}, R.~R., {Filippenko}, A.~V., {Paik}, Y.,
  {Leibundgut}, B., {Schulman}, E., \& {Cox}, C.~V. 1994, \aj, 107, 1022

\bibitem[{{Sauer} {et~al.}(2006){Sauer}, {Mazzali}, {Deng}, {Valenti},
  {Nomoto}, \& {Filippenko}}]{sauer_etal_06}
{Sauer}, D.~N., {Mazzali}, P.~A., {Deng}, J., {Valenti}, S., {Nomoto}, K., \&
  {Filippenko}, A.~V. 2006, \mnras, 369, 1939

\bibitem[{{Schmutz} {et~al.}(1990){Schmutz}, {Abbott}, {Russell}, {Hamann}, \&
  {Wessolowski}}]{SAR90_87A}
{Schmutz}, W., {Abbott}, D.~C., {Russell}, R.~S., {Hamann}, W.-R., \&
  {Wessolowski}, U. 1990, \apj, 355, 255

\bibitem[{{Shaviv}(2000)}]{shaviv_00}
{Shaviv}, N.~J. 2000, \apjl, 532, L137

\bibitem[{{Smith} {et~al.}(2010){Smith}, {Li}, {Filippenko}, \&
  {Chornock}}]{smith_etal_10}
{Smith}, N., {Li}, W., {Filippenko}, A.~V., \& {Chornock}, R. 2010,
  ArXiv:1006.3899

\bibitem[{{Soderberg} {et~al.}(2008){Soderberg}, {Berger}, {Page}, {Schady},
  {Parrent}, {Pooley}, {Wang}, {Ofek}, {Cucchiara}, {Rau}, {Waxman}, {Simon},
  {Bock}, {Milne}, {Page}, {Barentine}, {Barthelmy}, {Beardmore}, {Bietenholz},
  {Brown}, {Burrows}, {Burrows}, {Byrngelson}, {Cenko}, {Chandra}, {Cummings},
  {Fox}, {Gal-Yam}, {Gehrels}, {Immler}, {Kasliwal}, {Kong}, {Krimm},
  {Kulkarni}, {Maccarone}, {M{\'e}sz{\'a}ros}, {Nakar}, {O'Brien}, {Overzier},
  {de Pasquale}, {Racusin}, {Rea}, \& {York}}]{soderberg_etal_08}
{Soderberg}, A.~M., {Berger}, E., {Page}, K.~L., {Schady}, P., {Parrent}, J.,
  {Pooley}, D., {Wang}, X., {Ofek}, E.~O., {Cucchiara}, A., {Rau}, A.,
  {Waxman}, E., {Simon}, J.~D., {Bock}, D., {Milne}, P.~A., {Page}, M.~J.,
  {Barentine}, J.~C., {Barthelmy}, S.~D., {Beardmore}, A.~P., {Bietenholz},
  M.~F., {Brown}, P., {Burrows}, A., {Burrows}, D.~N., {Byrngelson}, G.,
  {Cenko}, S.~B., {Chandra}, P., {Cummings}, J.~R., {Fox}, D.~B., {Gal-Yam},
  A., {Gehrels}, N., {Immler}, S., {Kasliwal}, M., {Kong}, A.~K.~H., {Krimm},
  H.~A., {Kulkarni}, S.~R., {Maccarone}, T.~J., {M{\'e}sz{\'a}ros}, P.,
  {Nakar}, E., {O'Brien}, P.~T., {Overzier}, R.~A., {de Pasquale}, M.,
  {Racusin}, J., {Rea}, N., \& {York}, D.~G. 2008, \nat, 453, 469

\bibitem[{{Swartz}(1991)}]{swartz_91}
{Swartz}, D.~A. 1991, \apj, 373, 604

\bibitem[{{Swartz} {et~al.}(1993{\natexlab{a}}){Swartz}, {Clocchiatti},
  {Benjamin}, {Lester}, \& {Wheeler}}]{swartz_etal_93b}
{Swartz}, D.~A., {Clocchiatti}, A., {Benjamin}, R., {Lester}, D.~F., \&
  {Wheeler}, J.~C. 1993{\natexlab{a}}, \nat, 365, 232

\bibitem[{{Swartz} {et~al.}(1993{\natexlab{b}}){Swartz}, {Filippenko},
  {Nomoto}, \& {Wheeler}}]{swartz_etal_93a}
{Swartz}, D.~A., {Filippenko}, A.~V., {Nomoto}, K., \& {Wheeler}, J.~C.
  1993{\natexlab{b}}, \apj, 411, 313

\bibitem[{{Swartz} {et~al.}(1995){Swartz}, {Sutherland}, \&
  {Harkness}}]{swartz_etal_95}
{Swartz}, D.~A., {Sutherland}, P.~G., \& {Harkness}, R.~P. 1995, \apj, 446, 766

\bibitem[{{Tanaka} {et~al.}(2009){Tanaka}, {Tominaga}, {Nomoto}, {Valenti},
  {Sahu}, {Minezaki}, {Yoshii}, {Yoshida}, {Anupama}, {Benetti}, {Chincarini},
  {Della Valle}, {Mazzali}, \& {Pian}}]{tanaka_etal_09}
{Tanaka}, M., {Tominaga}, N., {Nomoto}, K., {Valenti}, S., {Sahu}, D.~K.,
  {Minezaki}, T., {Yoshii}, Y., {Yoshida}, M., {Anupama}, G.~C., {Benetti}, S.,
  {Chincarini}, G., {Della Valle}, M., {Mazzali}, P.~A., \& {Pian}, E. 2009,
  \apj, 692, 1131

\bibitem[{{Tominaga} {et~al.}(2005){Tominaga}, {Tanaka}, {Nomoto}, {Mazzali},
  {Deng}, {Maeda}, {Umeda}, {Modjaz}, {Hicken}, {Challis}, {Kirshner},
  {Wood-Vasey}, {Blake}, {Bloom}, {Skrutskie}, {Szentgyorgyi}, {Falco},
  {Inada}, {Minezaki}, {Yoshii}, {Kawabata}, {Iye}, {Anupama}, {Sahu}, \&
  {Prabhu}}]{tominaga_etal_05}
{Tominaga}, N., {Tanaka}, M., {Nomoto}, K., {Mazzali}, P.~A., {Deng}, J.,
  {Maeda}, K., {Umeda}, H., {Modjaz}, M., {Hicken}, M., {Challis}, P.,
  {Kirshner}, R.~P., {Wood-Vasey}, W.~M., {Blake}, C.~H., {Bloom}, J.~S.,
  {Skrutskie}, M.~F., {Szentgyorgyi}, A., {Falco}, E.~E., {Inada}, N.,
  {Minezaki}, T., {Yoshii}, Y., {Kawabata}, K., {Iye}, M., {Anupama}, G.~C.,
  {Sahu}, D.~K., \& {Prabhu}, T.~P. 2005, \apjl, 633, L97

\bibitem[{{Utrobin}(1994)}]{utrobin_94}
{Utrobin}, V. 1994, \aap, 281, L89

\bibitem[{{Utrobin} \& {Chugai}(2005)}]{UC05_time_dep}
{Utrobin}, V.~P. \& {Chugai}, N.~N. 2005, \aap, 441, 271

\bibitem[{{van der Hucht}(2001)}]{wr_cat}
{van der Hucht}, K.~A. 2001, New Astronomy Review, 45, 135

\bibitem[{{Vanbeveren} {et~al.}(1998){Vanbeveren}, {De Loore}, \& {Van
  Rensbergen}}]{vanbeveren_etal_98}
{Vanbeveren}, D., {De Loore}, C., \& {Van Rensbergen}, W. 1998, \aapr, 9, 63

\bibitem[{{Weaver} {et~al.}(1978){Weaver}, {Zimmerman}, \&
  {Woosley}}]{weaver_etal_78}
{Weaver}, T.~A., {Zimmerman}, G.~B., \& {Woosley}, S.~E. 1978, \apj, 225, 1021

\bibitem[{{Wellstein} \& {Langer}(1999)}]{wellstein_langer_99}
{Wellstein}, S. \& {Langer}, N. 1999, \aap, 350, 148

\bibitem[{{Wellstein} {et~al.}(2001){Wellstein}, {Langer}, \&
  {Braun}}]{wellstein_etal_01}
{Wellstein}, S., {Langer}, N., \& {Braun}, H. 2001, \aap, 369, 939

\bibitem[{{Wheeler} {et~al.}(1994){Wheeler}, {Harkness}, {Clocchiatti},
  {Benetti}, {Brotherton}, {Depoy}, \& {Elias}}]{wheeler_etal_94}
{Wheeler}, J.~C., {Harkness}, R.~P., {Clocchiatti}, A., {Benetti}, S.,
  {Brotherton}, M.~S., {Depoy}, D.~L., \& {Elias}, J. 1994, \apjl, 436, L135+

\bibitem[{{Woosley} \& {Bloom}(2006)}]{woosley_bloom_06}
{Woosley}, S.~E. \& {Bloom}, J.~S. 2006, \araa, 44, 507

\bibitem[{{Woosley} {et~al.}(1999){Woosley}, {Eastman}, \&
  {Schmidt}}]{woosley_etal_99}
{Woosley}, S.~E., {Eastman}, R.~G., \& {Schmidt}, B.~P. 1999, \apj, 516, 788

\bibitem[{{Woosley} {et~al.}(1994{\natexlab{a}}){Woosley}, {Eastman}, {Weaver},
  \& {Pinto}}]{woosley_etal_94}
{Woosley}, S.~E., {Eastman}, R.~G., {Weaver}, T.~A., \& {Pinto}, P.~A.
  1994{\natexlab{a}}, \apj, 429, 300

\bibitem[{{Woosley} {et~al.}(1994{\natexlab{b}}){Woosley}, {Eastman}, {Weaver},
  \& {Pinto}}]{WEW94_SN1993A}
---. 1994{\natexlab{b}}, \apj, 429, 300

\bibitem[{{Woosley} {et~al.}(2002){Woosley}, {Heger}, \& {Weaver}}]{WHW02}
{Woosley}, S.~E., {Heger}, A., \& {Weaver}, T.~A. 2002, Reviews of Modern
  Physics, 74, 1015

\bibitem[{{Woosley} {et~al.}(1995){Woosley}, {Langer}, \&
  {Weaver}}]{woosley_etal_95}
{Woosley}, S.~E., {Langer}, N., \& {Weaver}, T.~A. 1995, \apj, 448, 315

\bibitem[{{Yoon} {et~al.}(2006){Yoon}, {Langer}, \& {Norman}}]{yoon_etal_06}
{Yoon}, S., {Langer}, N., \& {Norman}, C. 2006, \aap, 460, 199

\bibitem[{{Yoon} {et~al.}(2010){Yoon}, {Woosley}, \& {Langer}}]{yoon_etal_10}
{Yoon}, S., {Woosley}, S.~E., \& {Langer}, N. 2010, \apj, 725, 940

\bibitem[{{Young} {et~al.}(1995){Young}, {Baron}, \& {Branch}}]{young_etal_95}
{Young}, T.~R., {Baron}, E., \& {Branch}, D. 1995, \apjl, 449, L51+

\end{thebibliography}
\end{document}